%% file: main.tex
\documentclass{article}

\PassOptionsToPackage{numbers, compress}{natbib}
\usepackage[eandd, final]{neurips_2026}

\usepackage[utf8]{inputenc} 
\usepackage[T1]{fontenc}    
\usepackage{hyperref}       
\usepackage{url}            
\usepackage{booktabs}       
\usepackage{amsfonts}       
\usepackage{nicefrac}       
\usepackage{microtype}      
\usepackage{pifont}
\usepackage{xcolor}         
\usepackage{graphicx}
\usepackage{float}
\usepackage{fvextra}
\usepackage{titletoc}
\usepackage{tikz}
\usetikzlibrary{arrows.meta,calc}
\usepackage{pgfplots}
\usepgfplotslibrary{groupplots}
\pgfplotsset{compat=1.18}

\fvset{
    breaklines,
    breakanywhere,
    breakautoindent,
    breaksymbolleft={},
    fontsize=\footnotesize,
    frame=single,
    baselinestretch=1.2,
    numbers=left,
    numbersep=5pt
}

\title{HierSVA: A Data Synthesis Pipeline, Dataset, and Benchmark for LLM-Generated SystemVerilog Assertions on Hierarchical RTL}

\author{%
  \textbf{Maohua Nie}\textsuperscript{1}\thanks{Equal Contribution.}\hspace{4pt}
  \quad
  \textbf{Jiang Zhu}\textsuperscript{2}\footnotemark[1]\hspace{3pt}\thanks{This work does not relate to Jiang Zhu's position at Amazon.}\hspace{4pt}
  \quad
  \textbf{Jingqun Zhang}\textsuperscript{1}
  \quad
  \textbf{Zhichen Zeng}\textsuperscript{1}\\
  \textbf{Jiayi Wang}\textsuperscript{1} 
  \quad
  \textbf{Sibo Zhang}\textsuperscript{1}
  \quad
  \textbf{Jialin Wang}\textsuperscript{1}
  \quad
  \textbf{C.-J. Richard Shi}\textsuperscript{1}
  \\
  \\
  \textsuperscript{1}University of Washington 
  \quad
  \textsuperscript{2}Annapurna Labs, Amazon
  \quad
}

\begin{document}

\maketitle

\begin{abstract}
We present HierSVA, an integrated suite that combines a data synthesis pipeline, a dataset, and a benchmark for LLM-generated SystemVerilog Assertions (SVA) on hierarchical RTL. HierSVA-SP pairs an RTL preprocessing toolchain with an LLM-in-the-loop formal verification flow to produce reference SVA on hierarchical RTL. Applying it to BaseJump STL yields HierSVA-DS, a dataset of 342 modules with hierarchy metadata, accompanied by a deep subset of 28 module-bug pairs with natural-language specifications and bug variants. HierSVA-B decomposes assertion quality into six metric axes: syntax correctness, assertion proof success rate, vacuity, specification faithfulness, mutation coverage, and formal core coverage. Applying HierSVA-B to twelve recent LLMs reveals three findings. First, averaged across models, the module-level compile rate is 66.6\%, and 82.1\% of generated assertions in evaluable runs prove non-vacuously, but the assertion sets detect only 70.2\% of eligible injected faults and cover 36.2\% of the formal core. Second, on 211 evaluable model--module entries in the deep subset, generated assertions flag buggy RTL with 0.87 recall, but 40\% of predicted-buggy outcomes are false positives on correct RTL, limiting precision to 0.60. Third, agentic mode improves provability, but strength metrics do not improve consistently and oscillate across iterations. We release dataset at \href{https://huggingface.co/datasets/AnonymousHierSVA/HierSVA}{https://huggingface.co/datasets/AnonymousHierSVA/HierSVA}, and also codes and artifacts at \href{https://github.com/HierSVAAnon/HierSVACodeAndArtifacts}{https://github.com/HierSVAAnon/HierSVACodeAndArtifacts}.
\end{abstract}

\input{sections/1_Introduction}

\input{sections/2_Background_and_Related_Work}

\input{sections/3_Synthesis_Methodology}

\input{sections/4_The_HierSVA_Dataset}

\input{sections/5_Evaluation_Framework}


\input{sections/7_Results}

\input{sections/9_Conclusion}

\begin{ack}
This work received no third-party funding. We thank the University of Washington for providing the EDA tool licenses used in this work. The authors have no other competing interests to declare.
\end{ack}

\bibliography{refs}
\bibliographystyle{plainnat}

\newpage
\appendix





\input{appendix/A_Datasheet_for_HierSVA}

\input{appendix/D_Full_Prompt_Templates}

\input{appendix/E_SVA_Primer_for_ML_Audience}

\input{appendix/F_Per_Model_Detailed_Results}

\input{appendix/G_Deep_Subset_Bug_Patterns_and_Specifications}

\input{appendix/I_Mutation_Operator_Breakdown}

\input{appendix/J_Comparison_with_Prior_LLM_for_SVA_Benchmarks}

\input{appendix/N_Compute_and_Cost_Accounting}

\newpage
\input{checklist.tex}

\end{document}

%% file: sections/1_Introduction.tex
\section{Introduction}
\label{sec:intro}
 
Functional verification, also known as design verification (DV), is a key step of integrated circuit design that determines whether a silicon tapeout succeeds~\cite{dvlsi}. Formal verification has become an established part of modern DV flows, supplementing simulation-based verification through the use of SystemVerilog Assertions and formal verification tools to prove design correctness~\cite{the_formal_book}. As LLM capabilities advance, recent studies have explored their use in hardware design and verification, including hardware formal verification~\cite{asserteval,assertionbench,assertionforge,assertllm,chiraag,fveval,harm,sangam,spec2assertion}.
 
Early work on using LLMs to generate SVA has shown that current models can produce assertions that pass simulation or formal checks on small designs, and that checker-guided iterative refinement improves these results further~\cite{chiraag,sangam}. However, the infrastructure supporting this line of research still falls short in three respects. First, existing benchmarks rely on limited sources of ground truth: automated mining tools with known scaling limitations~\cite{assertionbench}, expert-written or randomly generated assertions~\cite{fveval}, or small sets of designs accompanied by specifications~\cite{asserteval,assertllm}. These limitations reflect the absence of a reusable pipeline for synthesizing reference assertions directly from RTL. Second, existing reference datasets are correspondingly small, flat, or templated, with SoC modules excluded because of their sub-module dependencies~\cite{fveval}. Third, existing evaluations focus on syntax correctness and pass-or-fail metrics, leaving out vacuity, mutation coverage, and specification faithfulness (Section~\ref{sec:metrics})~\cite{assertionbench, asserteval, assertllm}. The three gaps are linked. Producing reference assertions by hand at scale requires too much labor, so constructing such a dataset requires an automated synthesis pipeline. Such a dataset, together with independent specifications for a deep subset (Section~\ref{sec:ds-deep}), enables a benchmark that evaluates LLM-generated assertions against formal-evidence criteria (Section~\ref{sec:dataset-qa}) rather than only pass-or-fail outcomes.
 
We address these gaps with an integrated suite built around industry practice. It uses standard SystemVerilog grammar, industrial formal verification tools, tapeout-validated RTL, industrial verification methodology, including assume-guarantee composition (Section~\ref{sec:sp-contracts}), fault injection, vacuity checking, and coverage analysis. The deep-subset artifacts are reviewed by industrial DV engineers. Appendix~\ref{app:sva-primer} introduces SVA and formal verification for readers outside hardware design.

\paragraph{Contributions.}
We present HierSVA as three independently reusable components:

\begin{enumerate}
\item \textbf{HierSVA-SP}, a \textbf{S}ynthesis \textbf{P}ipeline that produces reference SVA directly from RTL. It couples an open-source RTL preprocessing toolchain with an iterative loop in which an LLM proposes assertions and formal tools check them. Beyond this generate-and-check loop, it adds three hardware-specific extensions, namely hierarchical assume-guarantee composition, parameter cross-product expansion, and mutation and coverage analysis (Section~\ref{sec:sp}).

\item \textbf{HierSVA-DS}, a hierarchical SVA \textbf{D}ata\textbf{S}et of 342 modules drawn from the tapeout-validated BaseJump STL~\cite{basejump}, each retained only after passing a formal-evidence filter. A deep subset of 28 module-bug pairs further provides expert-authored natural-language specifications and bug variants from two sources, all reviewed by senior industrial DV engineers, to support spec-versus-RTL faithfulness evaluation (Section~\ref{sec:ds}).

\item \textbf{HierSVA-B}, a \textbf{B}enchmark that evaluates assertion quality along six axes (syntax correctness, assertion proof success rate, vacuity, specification faithfulness, mutation coverage, and formal core coverage) in both static and agentic modes (Section~\ref{sec:b}). Measuring these axes separately yields three findings that a single pass-or-fail rate would hide.
\end{enumerate}


\paragraph{Findings preview.}
Evaluating twelve recent LLMs yields three findings. First, averaged across models, 82.1\% of the generated assertions in evaluable runs (Section~\ref{sec:b-task}) prove non-vacuously, but the assertion sets detect only 70.2\% of eligible injected faults (Section~\ref{sec:dataset-qa}) and cover only 36.2\% of the formal core, which falls from about 75\% at depth~0 (Section~\ref{sec:sp-contracts}) to 14--26\% at depths~5--7. Second, on the deep subset, recall is 0.87 and precision 0.60. Third, agentic mode improves provability, but verification strength (Section~\ref{sec:metrics}) does not improve consistently.


%% file: sections/2_Background_and_Related_Work.tex
\section{Related Work}
\label{sec:related}

\paragraph{Benchmarks for LLM-generated assertions.}
A small number of benchmarks evaluate LLMs on SVA generation, and Table~\ref{tab:related-compact} shows them against HierSVA along the dimensions relevant to industrial verification practice. AssertionBench~\cite{assertionbench} pairs OpenCores~\cite{opencores} designs with assertions mined using GoldMine~\cite{goldmine} and HARM~\cite{harm_miner}, and classifies each generated assertion as Pass, Fail (counterexample), or Error under JasperGold. FVEval~\cite{fveval} introduces three sub-benchmarks built from expert-written, synthetically generated, and template-generated test cases. Its Design2SVA subset uses parameterized pipeline templates and FSM templates rather than industrial multi-module codebases. AssertEval~\cite{asserteval} and AssertLLM~\cite{assertllm} pair small sets of open-source designs with specifications and check assertions by FPV, VERT~\cite{vert} provides synthetic fine-tuning data, and CVDP~\cite{cvdp} checks assertions in simulation. None of these works reports a vacuity rate.

\paragraph{Automated and LLM-driven SVA generation methods.}
Before LLMs, AutoSVA~\cite{autosva} generated formal testbenches and properties from RTL interface annotations. LLM-based methods use prompt engineering, RAG, and iterative refinement, for example simulation feedback in ChIRAAG~\cite{chiraag}, multi-LLM spec-to-SVA pipelines in AssertLLM~\cite{assertllm}, knowledge graphs in AssertionForge~\cite{assertionforge}, pre-RTL specifications in Spec2Assertion~\cite{spec2assertion}, MCTS refinement in SANGAM~\cite{sangam}, and security properties in LASSO~\cite{lasso,lasa}. NSPG~\cite{nspg} and LLM4DV~\cite{llm4dv} target adjacent settings. LASSO filters vacuous properties inside the method, building on standard vacuity-detection theorems~\cite{beer2001efficient,kupferman2003vacuity}.

\begin{table}[htbp]
\centering
\small
\setlength{\tabcolsep}{4pt}
\caption{Comparison of HierSVA with prior LLM-for-SVA benchmarks. \#Items counts each benchmark's evaluation unit (training samples for VERT), so the counts are not directly comparable. For HierSVA, it counts 342 main-set modules and 28 deep-subset module-bug pairs. \emph{Hier.}: per-module evaluation along a module hierarchy. \emph{Cov.}: coverage metric (COI~=~cone of influence, FC~=~formal core, path~=~complete path coverage). \emph{Vac.} and \emph{Mut.}: whether vacuity and mutation coverage are reported (\emph{partial}~=~mutants only judge correctness). Appendix~\ref{app:prior-comparison} gives the full comparison.}
\label{tab:related-compact}
\begin{tabular}{lrlccccc}
\toprule
Benchmark & \#Items & RTL source & Hier. & Eval. & Cov. & Vac. & Mut. \\
\midrule
AssertionBench & 100 & OpenCores & \ding{55} & Formal & \ding{55} & \ding{55} & \ding{55} \\
FVEval & 79$+$300$+$192 & Expert$+$synthetic & \ding{55} & Formal & \ding{55} & \ding{55} & \ding{55} \\
AssertEval & 18 & Open-source & \ding{55} & Formal & COI & \ding{55} & \ding{55} \\
AssertLLM & 20 & Open-source & \ding{55} & Formal & \ding{55} & \ding{55} & \ding{55} \\
VERT & 20K & Synthetic & \ding{55} & Formal$+$Sim & path & \ding{55} & partial \\
CVDP cid14 & 98 & Human-authored & \ding{55} & Sim/harness & \ding{55} & \ding{55} & \ding{55} \\
\midrule
\textbf{HierSVA (ours)} & 342$+$28 & BaseJump STL & \ding{51} & Formal & \textbf{FC} & \ding{51} & \ding{51} \\
\bottomrule
\end{tabular}
\end{table}

%% file: sections/3_Synthesis_Methodology.tex

\section{HierSVA-SP: Dataset Synthesis Pipeline}
\label{sec:sp}

\subsection{Design Principles: Following Industry Practice}
\label{sec:sp-fidelity}

Industry verification practice translates into three principles of HierSVA-SP. First, with respect to assertion grammar, HierSVA-SP emits and validates IEEE-1800-compliant SVA accepted by VC~Formal~\cite{sva_standard}, rather than the SVA subset supported by open-source tools such as SymbiYosys~\cite{symbiyosys}. Second, with respect to tooling, validation runs on Synopsys VC Formal~\cite{vcf}, using its Formal Property Verification (FPV) and Formal Testbench Analyzer (FTA) app modes, with Formal Core (FC) analysis invoked inside FPV. Third, on methodology, HierSVA-SP integrates assume-guarantee composition (Section~\ref{sec:sp-contracts}), mutation analysis through FTA, and formal core coverage analysis through FC, mirroring the assertion quality criteria used in industrial signoff flows. HierSVA-SP accepts any RTL written in standard Verilog or SystemVerilog and treats the input RTL as correct, so its reference assertions describe the behavior of that RTL.

\subsection{RTL Preprocessing}
\label{sec:sp-prep}

A handful of scripts prepare raw RTL for downstream synthesis. The \emph{macro expander} expands \texttt{\textasciigrave define} macros and \texttt{\textasciigrave include} files. The \emph{module dependency graph builder} (Section~\ref{sec:sp-contracts}) recognizes the hierarchy of all RTL modules and emits a \emph{module dependency graph} used to determine processing order and identify leaf modules. The \emph{comment filter} is an optional pass that strips RTL code comments, reducing prompt noise. The \emph{syntax normalizer} rewrites whitespaces, port declarations, and parameter declarations into a uniform style. The \emph{design information extractor} script extracts each module's clock and reset signals, reset polarity, latch presence, and whether it is pure-combinational. The latter two properties propagate upward over the module dependency graph. A parent module inherits the latch-containing label from any instantiated child whose dependency cone contains a latch; the non-combinational label propagates likewise.

\subsection{Hierarchical Composition with Submodule Contracts}
\label{sec:sp-contracts}

\begin{figure}[t]
\centering
\input{figures/module_hierarchy}
\vspace{-0.4em}
\caption{(a) Instance tree with one node per instance. (b) Module dependency graph with one node per module definition and its depth (d). Orange marks a module instantiated with two values of $W$, and green marks two copies in one parent.}
\label{fig:hierarchy}
\vspace{-1.2em}
\end{figure}

\paragraph{Module hierarchy as a dependency graph.}
An RTL design is usually described by its instance hierarchy, a tree in which each node is a module instance (Figure~\ref{fig:hierarchy}a). Because HierSVA writes and verifies assertions once per module definition rather than once per instance, it operates on the \emph{module dependency graph} instead (Figure~\ref{fig:hierarchy}b). Each node in this graph is a module definition, and a directed edge from $P$ to $C$ means that $P$ instantiates $C$ at least once. A module may be instantiated by several parents, so the graph is not a tree. We define a module's \emph{depth} as the length of the longest chain of instantiations from it down to a leaf module, which instantiates no other module and has depth~0. Throughout this paper, \emph{hierarchy} and \emph{hierarchical} refer to this graph and this notion of \emph{depth}.

\paragraph{Assume-guarantee reasoning.}
Assume-guarantee reasoning is a classical formal verification pattern~\cite{assumeguarantee}. A module is verified against its environment under \emph{assumptions} on its inputs and produces \emph{guarantees} about its outputs. When the module is instantiated as a submodule, its guarantees become assumptions that the parent verification can rely on rather than properties the parent must re-prove. The pair (assumption set, guarantee set) is the module's \emph{contract}.
 
\paragraph{Contracts in HierSVA-SP.}
The pipeline verifies modules bottom-up along the module dependency graph. Leaf modules are verified first by the iterative loop (Section~\ref{sec:sp-loop}), and their assumptions and proven assertions are saved into an SVA file. When verifying a parent module, the SVA files of its direct submodules are loaded into the VC Formal. Each submodule's proven assertion becomes an \texttt{assume}, and each submodule's assumption becomes an \texttt{assert} that the parent must prove. Because each module definition is verified once, its contract applies to every instance of it. Modules further down the graph are blackboxed, so each parent reasons only about its own RTL plus the contracts of its direct submodules.
 
\paragraph{Why hierarchical contracts are necessary.}
Hierarchical proof through assume-guarantee contracts is the standard DV practice for verifying designs of nontrivial depth. A flattened verification is feasible, in which FPV reasons about the parent modules and all child modules together, but verification cost rises sharply with module-structure complexity and becomes intractable at deeper hierarchy.
 
\subsection{Iterative Synthesis with Formal Feedback}
\label{sec:sp-loop}

\begin{figure}[tp!]
    \centering
    \includegraphics[width=0.9\linewidth]{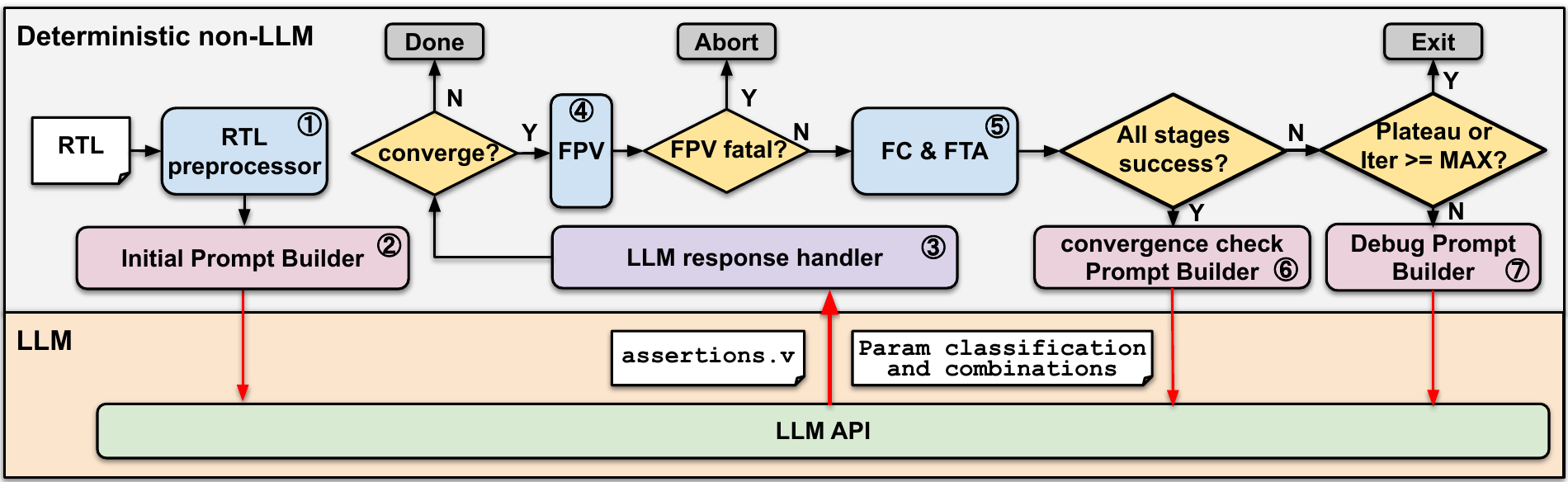}
    \vspace{-0.5em}
    \caption{Synthesis loop control flow. Boxes on the top are deterministic non-LLM code, and boxes on the bottom are LLM-controlled. The deterministic code decides whether the loop continues or aborts. The LLM produces the two artifacts of each iteration, which are \texttt{assertions.v} and the parameter classification and combinations.}
    \vspace{-1em}
    \label{fig:loop}
\end{figure}

For each module, HierSVA-SP takes the preprocessed RTL, the design information from Section~\ref{sec:sp-prep}, and SVA files of the direct submodules without their RTL, and optionally a natural-language specification, and it produces an assertion file. To produce them, the pipeline runs the iterative loop shown in Figure~\ref{fig:loop}. In each iteration, an LLM generates the SVA file, the EDA flow validates it, and the control logic decides whether to terminate, abort, or feed the results back to the LLM for revision.


At each iteration $t$, the LLM generates a JSON object containing the only two artifacts it is permitted to modify: the assertion file (\texttt{assertions.v}) and a set of parameter configurations (\texttt{param\_combinations}). To build these configurations, the LLM first sorts each module parameter into one of three types. Type-1 parameters set data sizes, such as widths and depths of a FIFO, and do not change the module's functional behavior. Type-2 parameters select behavioral modes, such as a priority policy of an arbiter, and therefore change what the module does. Type-N parameters are synthesis hints with no effect on RTL behavior. The LLM then constructs \texttt{param\_combinations} as the set of configurations that covers valid Type-2 values. Type-1 parameters take only their default and one minimal edge-case value, and Type-N parameters stay fixed at their defaults. The EDA tool verifies each resulting configuration independently, and the LLM uses the formal feedback to refine both its classification and the combinations in the next iteration.


The rest of the loop runs as deterministic, non-LLM code. In iteration $t$, the pipeline executes FPV, followed by the FTA and FC analyses. A static rule-based parser processes the VC Formal logs into structured per-stage and per-property statuses, ensuring the LLM never receives raw logs. To formulate the feedback prompt for iteration $t+1$, static templates extract information exclusively from unsuccessful stages. 
The loop terminates under one of three conditions: (1) convergence, where the three tool-based criteria in Section~\ref{sec:dataset-qa} are satisfied and the LLM, answering a convergence-check prompt (Appendix~\ref{app:prompt-convergence}), declares that its assertions sufficiently cover the intended functionality; (2) reaching \texttt{MAX\_ITERATIONS}; or (3) a plateau where the same single stage blocks convergence for three consecutive iterations.

%% file: figures/module_hierarchy.tex
\definecolor{mhNavy}{HTML}{1F2633}
\definecolor{mhSlate}{HTML}{4A6385}
\definecolor{mhFill}{HTML}{F3F5F8}
\definecolor{mhOrange}{HTML}{B5652B}
\definecolor{mhOrangeFill}{HTML}{F6E8DB}
\definecolor{mhGreen}{HTML}{3E7A5A}
\definecolor{mhGreenFill}{HTML}{E3F0E8}
\definecolor{mhEdge}{HTML}{6B7280}
\definecolor{mhMuted}{HTML}{5B6472}
\definecolor{mhOnDark}{HTML}{C8D0DB}
\begin{tikzpicture}[
  font=\footnotesize,
  mod/.style={draw=mhSlate, fill=mhFill, rounded corners=2pt, line width=0.5pt,
              minimum width=1.55cm, minimum height=0.42cm, inner sep=1.5pt},
  topmod/.style={mod, draw=mhNavy, fill=mhNavy, text=white},
  shared/.style={mod, draw=mhOrange, fill=mhOrangeFill},
  copies/.style={mod, draw=mhGreen, fill=mhGreenFill},
  edge/.style={-{Stealth[length=3.4pt, width=3pt]}, draw=mhEdge, line width=0.5pt, shorten >=0.3pt},
  tag/.style={font=\footnotesize\bfseries, text=mhNavy, anchor=north west, inner sep=0pt},
]
\newcommand{\inst}[2]{\textbf{#1}\,{\scriptsize\color{mhMuted}#2}}
\newcommand{\defn}[2]{\textbf{#1}\,{\scriptsize\color{mhMuted}d#2}}
\def\rowA{0}\def\rowB{-0.72}\def\rowC{-1.44}

\node[tag] at (-0.05, 0.22) {(a)};
\node[topmod] (it)  at (2.7, \rowA) {\textbf{top}};
\node[mod]    (ia)  at (0.9, \rowB) {\inst{A}{\texttt{u\_a}}};
\node[mod]    (ib)  at (4.5, \rowB) {\inst{B}{\texttt{u\_b}}};
\node[shared] (ix1) at (0.9, \rowC) {\inst{X}{$W{=}8$}};
\node[shared] (ix2) at (2.7, \rowC) {\inst{X}{$W{=}16$}};
\node[copies] (iy0) at (4.5, \rowC) {\inst{Y}{\texttt{u\_y0}}};
\node[copies] (iy1) at (6.3, \rowC) {\inst{Y}{\texttt{u\_y1}}};
\foreach \p/\c in {it/ia, it/ib, ia/ix1, ib/ix2, ib/iy0, ib/iy1}
  \draw[edge] (\p.south) -- (\c.north);

\draw[mhEdge!35, line width=0.4pt] (7.35, 0.25) -- (7.35, -1.7);

\node[tag] at (7.6, 0.22) {(b)};
\node[topmod] (gt) at (10.3, \rowA) {\textbf{top}\,{\scriptsize\color{mhOnDark}d2}};
\node[mod]    (ga) at (9.0, \rowB)  {\defn{A}{1}};
\node[mod]    (gb) at (11.6, \rowB) {\defn{B}{1}};
\node[shared] (gx) at (10.3, \rowC) {\defn{X}{0}};
\node[copies] (gy) at (12.9, \rowC) {\defn{Y}{0}};
\foreach \p/\c in {gt/ga, gt/gb, ga/gx, gb/gx, gb/gy}
  \draw[edge] (\p.south) -- (\c.north);
\end{tikzpicture}

%% file: sections/4_The_HierSVA_Dataset.tex
\section{HierSVA-DS: The Dataset}
\label{sec:ds}

\subsection{Source: BaseJump STL}
\label{sec:ds-source}

HierSVA-DS is built on BaseJump STL~\cite{basejump}, an open-source SystemVerilog standard library covering memory primitives, dataflow building blocks, network-on-chip components, and inter-chip communication links. BaseJump STL has been used across multiple tapeouts, making its modules reflect the actual design choices in real silicon, including clock-domain handling, parameterization, and module-level reuse patterns. We considered authoring custom RTL to better control bug rates, but chose BaseJump STL because its tapeout validation gives it both the production quality and a bug rate low enough for our purposes.

\subsection{Hierarchical Structure}
\label{sec:ds-hier}

In HierSVA-DS, each module entry records its top-level RTL files, its submodule RTL files, and its directly and indirectly instantiated submodules. Each module has its own reference SVA file containing its assertions and assumptions. Table~\ref{tab:ds-stats} summarizes HierSVA-DS. The main set contains 342 modules organized into the BaseJump STL category structure, with hierarchy depths from 0 to 9.

\begin{table}[h]
\centering
\small
\caption{HierSVA-DS at a glance. Categories follow the BaseJump STL directory structure.}
\label{tab:ds-stats}
\begin{tabular}{lr}
\toprule
Property & Value \\
\midrule
Main-set modules & 342 \\
Deep-subset bug pairs & 28 pairs over 27 unique modules \\
Categories (BaseJump STL) & bsg\_async, bsg\_cache, bsg\_dataflow, bsg\_link, \\
                          & bsg\_mem, bsg\_misc, bsg\_noc, bsg\_tag \\
Hierarchy depth (min, max) & (0, 9) \\
Modules per depth & L0:150, L1:61, L2:27, L3:36, L4:25, \\
                  & L5:10, L6:20, L7:10, L8:2, L9:1 \\
\bottomrule
\end{tabular}
\end{table}

\subsection{Deep Evaluation Subset}
\label{sec:ds-deep}



The faithfulness axis of HierSVA-B (Section~\ref{sec:metrics}) checks whether generated assertions follow the specification rather than mirror the RTL, which requires buggy RTL and an independently written specification. The deep evaluation subset provides both for 28 module-bug pairs, which are built on modules already in the main set rather than added to it. The bugs come from two sources, with 7 real bugs recovered by reverting confirmed fix commits in the BaseJump STL git history and 21 synthetic bugs injected by Claude Opus~4.6~\cite{anthropic_claude_opus_46} and Claude Opus~4.7~\cite{anthropic_claude_opus_47}. The subset is small because most bugs that LLMs generate in bulk are syntax-level mutations, such as replacing a logical AND with a logical OR, which the fault injection behind C4 (Section~\ref{sec:b-c4}) already covers systematically. 

Each pair contains the correct RTL, the buggy RTL, and a natural-language specification of the correct behavior that does not reveal the bug. Senior industrial DV engineers reviewed every pair for (i)~bug well-formedness, meaning that the buggy RTL compiles and differs from the correct RTL in a single semantic deviation, (ii)~patch isolation and effectiveness, meaning that the diff contains only that deviation and reverting it restores the correct behavior, and (iii)~specification faithfulness, meaning that the specification describes the correct module without revealing the bug.

We classify the bugs into five patterns (Appendix~\ref{app:g-taxonomy}). Protocol violations (P1, 10~pairs) break the module's contract at its boundary. FSM reachability errors (P2, 3~pairs) let the design enter a state the specification forbids or prevent it from reaching one the specification requires. Atomicity breakdowns (P3, 7~pairs) corrupt a multi-cycle transaction even though each individual cycle looks correct. Corner-case errors (P4, 1~pair) appear only at the edges of the parameter or input space. Parameter-specific errors (P5, 7~pairs) appear only under particular Type-2 parameter settings.


\subsection{Dataset Quality Assurance}
\label{sec:dataset-qa}

HierSVA-DS is the result of running HierSVA-SP with a mixture of LLMs on the pre-processed BaseJump STL modules, followed by two filters that determine which pipeline outputs are included in the released dataset. For the main set, outputs are retained only if they pass the formal evidence filter. For the deep subset, pairs are additionally retained only if they pass expert review.

\paragraph{Formal evidence filter.} This filter applies to all 342 modules in HierSVA-DS. For a module to be retained, its reference assertion set must satisfy all convergence criteria of the synthesis loop (Section~\ref{sec:sp-loop}) across every parameter combination. Specifically: (1) FPV must prove every assertion without counterexamples; (2) FC analysis must confirm that no design statements lie outside the formal core; and (3) FTA must achieve a 100\% fault-detection rate over the eligible fault set. Faults that SVA cannot meaningfully detect are excluded from the \emph{eligible fault set}. Each module includes a \texttt{limit.json} file documenting these audited exclusions and their rationale. During evaluation, HierSVA-B applies this exact exclusion list when computing the LLM's FTA (Section~\ref{sec:b-c4}), ensuring that the reference and LLM-generated assertion sets are measured against an identical denominator.

\paragraph{Expert review filter.} This filter applies to the deep evaluation subset of 28 module-bug pairs (Section~\ref{sec:ds-deep}). Each bug pair in the deep subset is reviewed by senior industrial DV engineers. This filter is not applied to the full dataset because production-grade review of 342 hierarchical modules would exceed the labor budget, and the formal evidence filter serves as the universal baseline.

%% file: sections/5_Evaluation_Framework.tex
\section{HierSVA-B: The Benchmark}
\label{sec:b}

\subsection{Task Formulation}
\label{sec:b-task}

In HierSVA-B, the LLM receives the target module's RTL together with the design information extracted by HierSVA-SP, and it returns a complete assertion file. We then run VC~Formal on the LLM's assertion file under every parameter combination defined in the reference dataset, and call each (module, parameter combination) pair an \emph{evaluation cell}. Syntax correctness is computed per module, and the remaining metrics are computed over \emph{evaluable runs}, which are the runs of evaluation cells in which the assertion file compiles and VC~Formal produces a report. Unless stated otherwise, the metrics below are \emph{micro-averaged}, which means that the counts of assertions, faults, or design elements are pooled over all evaluable runs of a model before the ratio is taken. \emph{Cross-model means} are unweighted means over the twelve models. Submodule contracts are loaded exactly as in HierSVA-SP (Section~\ref{sec:sp-contracts}). Each metric defined below is computed from parsed VC~Formal output. We evaluate LLMs in two modes. In \emph{Static} mode, the LLM produces a complete assertion file in one response. In \emph{Agentic} mode, the LLM iterates in response to VC~Formal feedback as in HierSVA-SP's synthesis loop.

\subsection{Metrics}
\label{sec:metrics}

\paragraph{Syntax Correctness C1.}
\label{sec:b-c1c2p}

For each module, C1 records whether the LLM's assertion file compiles in VC~Formal without errors under every parameter combination of the module, and is reported as the fraction of modules that pass. C1 captures whether the generated assertion file is syntactically well-formed under the IEEE 1800 SVA grammar.

\paragraph{Assertion Proof Success Rate C2-P.}

C2-P is the assertion-level rate of properties that VC~Formal \textbf{P}roves under FPV without vacuity (proven non-vacuously). Together, C1 and C2-P are the metrics that most prior LLM-for-SVA benchmarks stop at.

\paragraph{Vacuity C2-V.}
\label{sec:b-meaning}

C2-V is the assertion-level rate of properties that VC~Formal proves only \textbf{V}acuously. A property is \emph{vacuously proven} when its antecedent can never be satisfied on any reachable trace.

\paragraph{Faithfulness C3.}
\label{sec:b-c3}

C3 is evaluated on the deep subset (Section~\ref{sec:ds-deep}) at the module level, and each \emph{model--module entry} is one model evaluated on one module-bug pair. The LLM is prompted with the buggy RTL variant and the natural-language specification, with explicit instructions to follow the specification rather than mirror the RTL. Every assertion the LLM emits for an evaluable model--module entry is then run through FPV against both the correct RTL and the buggy RTL. Non-evaluable entries, where the assertion set does not compile or cannot be run on both variants, are excluded from the confusion matrix and reported separately. 

Let $b=1$ denote that at least one generated assertion is falsified on the buggy RTL, and let $c=1$ denote that at least one generated assertion is falsified on the correct RTL. The buggy RTL run is the positive example and the correct RTL run is the negative example: $b=1$ contributes a TP, $b=0$ contributes an FN, $c=1$ contributes an FP, and $c=0$ contributes a TN. These variant-level counts are used to compute precision, recall, F1, and accuracy in Table~\ref{tab:twodim}. 

For qualitative interpretation, we additionally group the paired outcomes $(b,c)$ into the module-level classes shown in Table~\ref{tab:c3-classes}. Spec-Aligned means the assertion set catches the bug while remaining valid on the correct RTL. Mirror-RTL means the assertion set accepts both the correct and buggy RTL, suggesting that it has followed the implementation rather than the specification. Broken means the assertion set fires on the correct RTL and is therefore not a valid specification check, regardless of whether it also fires on the buggy RTL.

\begin{table}[h]
\centering
\small
\caption{C3 paired-outcome classification for a model--module entry. Parenthesized terms indicate the corresponding variant-level confusion-matrix contributions used in Table~\ref{tab:twodim}.}
\label{tab:c3-classes}
\begin{tabular}{p{0.28\linewidth}p{0.31\linewidth}p{0.31\linewidth}}
\toprule
 & Buggy RTL: some assertion falsified ($b=1$) 
 & Buggy RTL: no assertion falsified ($b=0$) \\
\midrule
Correct RTL: no assertion falsified ($c=0$)
& Spec-Aligned (TP+TN)
& Mirror-RTL / accepts both (FN+TN) \\
Correct RTL: some assertion falsified ($c=1$)
& Broken: over-strong (TP+FP)
& Broken: spurious and missing (FN+FP) \\
\bottomrule
\end{tabular}
\vspace{-1em}
\end{table}
\paragraph{Mutation Coverage C4.}
\label{sec:b-c4}


C4 measures bug-catching ability with the FTA results. Specifically, the FTA tool injects faults into the RTL code and checks whether the assertions of the verification environment detect them. The C4 score, or mutation coverage, is the percentage of these injected faults that trigger an assertion failure. To ensure a fair comparison between the LLM and the reference baseline, we exclude faults that are not meaningfully detectable by SVA, as documented in the per-module limit.json (Section~\ref{sec:dataset-qa}). The injected fault categories are detailed in Appendix~\ref{app:mutation}.

\paragraph{Formal Core Coverage C5.}
\label{sec:b-c5}

C5 is the formal core (FC) coverage reported by VC~Formal. The \emph{formal core} of an assertion is the part of its cone of influence (COI) that its proof uses, and C5 covers the union of the formal cores of all assertions. FTA and FC run on the full verification environment of the module, which contains the LLM's assertions and, for depth $\geq 1$, the obligations inherited from submodule contracts (Section~\ref{sec:sp-contracts}); these obligations are identical for every model and for the reference assertions. Falsified properties do not count, so at depth~0 C4 and C5 come only from the LLM's assertions. We refer to C1 and C2-P as \emph{provability} and to C4 and C5 as \emph{verification strength}.



%% file: sections/7_Results.tex
\section{Evaluation and Results Analysis}
\label{sec:eval}

\subsection{Experimental Setup}
\label{sec:eval-setup}

\paragraph{Models.}
We evaluate twelve recent LLMs: Claude Opus~4.7 \cite{anthropic_claude_opus_47}, DeepSeek~V4~Pro \cite{deepseek_v4_pro}, Gemini~3.1~Pro \cite{google_gemini_31_pro}, GLM~5.1 \cite{zai_glm51}, GPT-5.5 \cite{openai_gpt55}, Kimi~K2.6 \cite{moonshot_kimi_k26}, MiniMax~M2.7 \cite{minimax_m27}, Qwen~3.6~Max \cite{qwen_qwen36_max}, Claude Haiku~4.5 \cite{anthropic_claude_haiku_45}, DeepSeek~V4~Flash \cite{deepseek_v4_flash}, GPT-5-mini \cite{openai_gpt5_mini}, and Qwen~3.6~Plus \cite{qwen_qwen36_plus}. All twelve are queried through a unified OpenRouter gateway with the same prompts and parameters (reasoning enabled, strict JSON-schema output, temperature \(=\) 0). Setting temperature to 0 does not guarantee identical outputs for identical prompts, because OpenRouter may route requests to different providers and provider-side inference can depend on request batching.

\paragraph{Evaluation settings.}
Three settings are reported. The \textbf{main-set static evaluation} runs all twelve models on the full HierSVA-DS in static mode and produces the primary metrics for C1, C2-P, C2-V, C4, and C5. The \textbf{deep-subset static evaluation} runs the same twelve models on the deep subset (Section~\ref{sec:ds-deep}) in static mode with each LLM prompted to follow the correct specification rather than mirror the buggy RTL, producing C3. The \textbf{agentic evaluation} runs five selected models on a fixed 40-module subset of HierSVA-DS in agentic mode, with an initial generation (iteration~0) followed by up to six rounds of VC~Formal feedback per module. Iteration~0 serves as the baseline of the agentic evaluation. The main-set static evaluation results on the same 40 modules come from a separate LLM call and are listed in Appendix~\ref{app:f-t4-iter} for reference only. The agentic evaluation reports C1, C2-P, C2-V, C4, and C5 and does not evaluate faithfulness (C3).

\paragraph{Prompts.}
The system prompt instructs the LLM to act as a formal verification engineer and specifies coding standards for SVA syntax, naming, parameterization, and bind commands (full template in Appendix ~\ref{app:prompts}). To simplify the VC Formal tool flow, the prompt tells the model not to assert or cover input-only behavior, while allowing assumptions on input protocols when needed to model the environment. It also tells the model not to verify reset behavior, while allowing reset guards such as \texttt{disable iff}. The same prompt is used across all twelve models with no per-model tuning.

\subsection{Main-Set Static Evaluation Results}
\label{sec:eval-cascade}

Current LLMs often produce syntactically valid SVA assertions (Table~\ref{tab:t1-summary}): eleven of the twelve evaluated models exceed 50\% C1, the fraction of modules whose assertion file compiles in VC~Formal under every parameter combination. Averaged across the twelve models, the module-level compile rate is 66.6\%. Among generated assertions in evaluable runs, the non-vacuous proof rate (C2-P) averages 82.1\%. Beyond provability, however, the assertion sets are much weaker. Only 70.2\% of eligible injected faults are detected (C4), and only 36.2\% of the design's formal core is covered (C5). C5 is the lowest of C2-P, C4, and C5 for every model. The 46-point gap from C2-P to C5 is the main result of this evaluation. Assertion sets whose assertions prove under FPV cover only a fraction of the design's formal core, so pass-rate evaluation overstates verification strength. C5 pools design elements over all evaluable runs, so large modules weigh more; averaged per module, it is 56.8\%, still 25 points below C2-P. Formal-core coverage also falls with hierarchy depth, although not monotonically. The cross-model mean is 75.5\% at depth~0 and 14.0--26.0\% at depths~5--7 (Table~\ref{tab:f-depth}), while depths~8 and~9 contain only two modules and one module.

\begin{table}[h]
\centering
\small

\caption{Per-model overall metrics in the main-set static evaluation. C1 is the strict module-level compile rate over all 342 main-set modules, where a module counts as compiled only if its assertion file compiles under every parameter combination; C2-P and C2-V are assertion-level micro-averages over generated assertions in evaluable runs. C4 is computed over eligible injected faults, and C5 reports union formal-core coverage over the covered design statements/signals. Per-depth and per-category breakdowns are in Appendix~\ref{app:per-model}.}

\label{tab:t1-summary}
\setlength{\tabcolsep}{6pt}
\begin{tabular}{lrrrrr}
\toprule
Model & C1\% & C2-P\% & C2-V\%~($\downarrow$) & C4\% & C5\% \\
\midrule
Claude Opus 4.7           & 87.7 & 87.5 & 1.8 & 70.5 & 34.7 \\
DeepSeek V4 Pro           & 67.8 & 86.3 & 0.3 & 84.4 & 37.4 \\
Gemini 3.1 Pro            & 84.2 & 90.3 & 1.5 & 74.7 & 33.8 \\
GLM 5.1                   & 76.0 & 91.0 & 1.3 & 78.1 & 43.3 \\
GPT-5.5                   & 84.5 & 76.8 & 0.5 & 84.5 & 51.4 \\
Kimi K2.6                 & 62.3 & 86.2 & 0.2 & 89.0 & 53.3 \\
MiniMax M2.7              & 19.0 & 58.2 & 15.4 & 37.8 & 26.5 \\
Qwen 3.6 Max              & 70.2 & 95.2 & 0.1 & 70.4 & 37.7 \\
\midrule
Claude Haiku 4.5          & 59.6 & 74.0 & 6.5 & 58.4 & 33.1 \\
DeepSeek V4 Flash         & 57.9 & 84.9 & 1.2 & 73.2 & 39.6 \\
GPT-5-mini                & 57.6 & 77.9 & 2.1 & 56.7 & 14.5 \\
Qwen 3.6 Plus             & 72.5 & 76.9 & 1.4 & 64.9 & 29.4 \\
\midrule
\emph{Cross-model mean} & 66.6 & 82.1 & 2.7 & 70.2 & 36.2 \\
\bottomrule
\end{tabular}
\end{table}

\subsection{Deep-Subset Static Evaluation Results}
\label{sec:eval-twodim}

C3 evaluates assertion sets by testing them on two RTL variants of each module: a buggy version (positive example) and a correct version (negative example). An assertion set predicts a ``bug'' if any of its assertions fail. Out of the 336 possible model--module entries (12 models $\times$ 28 pairs), 211 are evaluable on both variants and contribute to Table~\ref{tab:twodim}. Conditional on those 211 entries, the models achieve aggregate recall 0.87, but they frequently trigger false alarms on correct code. Aggregate precision is only 0.60, meaning that 40\% of predicted-buggy outcomes are false positives. In every evaluable entry, an assertion set that fails on the correct RTL also fails on the buggy RTL, so the recall includes the 122 Broken entries. Of the 211 entries, 62 (29\%) are Spec-Aligned, 27 (13\%) are Mirror-RTL, and 122 (58\%) are Broken (Table~\ref{tab:c3-classes}). For reference, a predictor that flags every variant as buggy reaches precision 0.50, recall 1.00, and F1 0.67.

\begin{table}[h]
\centering
\small

\caption{Per-model variant-level C3 confusion matrix and metrics on the deep subset. ``Eval'' is the number of evaluable model--module entries out of 28. Each evaluable entry contributes two predictions: one on the buggy RTL, treated as the positive example, and one on the correct RTL, treated as the negative example. The aggregate row pools the variant-level counts across models.}
\label{tab:twodim}
\setlength{\tabcolsep}{3pt}
\begin{tabular}{lrrrrrrrrr}
\toprule
Model & Eval & TP & FN & FP & TN & P & R & F1 & Acc \\
\midrule
Claude Opus 4.7         & 25 &  21 &  4 & 10 & 15 & 0.68 & 0.84 & 0.75 & 0.72 \\
DeepSeek V4 Pro         & 20 &  18 &  2 & 16 &  4 & 0.53 & 0.90 & 0.67 & 0.55 \\
Gemini 3.1 Pro          & 24 &  20 &  4 & 11 & 13 & 0.65 & 0.83 & 0.73 & 0.69 \\
GLM 5.1                 & 25 &  20 &  5 & 17 &  8 & 0.54 & 0.80 & 0.65 & 0.56 \\
GPT-5.5                 & 20 &  18 &  2 &  7 & 13 & 0.72 & 0.90 & 0.80 & 0.78 \\
Kimi K2.6               & 17 &  15 &  2 & 10 &  7 & 0.60 & 0.88 & 0.71 & 0.65 \\
MiniMax M2.7            &  1 &   1 &  0 &  1 &  0 & 0.50 & 1.00 & 0.67 & 0.50 \\
Qwen 3.6 Max            & 17 &  15 &  2 & 11 &  6 & 0.58 & 0.88 & 0.70 & 0.62 \\
\midrule
Claude Haiku 4.5        & 15 &  12 &  3 &  7 &  8 & 0.63 & 0.80 & 0.71 & 0.67 \\
DeepSeek V4 Flash       & 16 &  15 &  1 & 13 &  3 & 0.54 & 0.94 & 0.68 & 0.56 \\
GPT-5-mini              & 13 &  13 &  0 & 10 &  3 & 0.57 & 1.00 & 0.72 & 0.62 \\
Qwen 3.6 Plus           & 18 &  16 &  2 &  9 &  9 & 0.64 & 0.89 & 0.74 & 0.69 \\
\midrule
\emph{Aggregate}        &211 & 184 & 27 &122 & 89 & 0.60 & 0.87 & 0.71 & 0.65 \\
\bottomrule
\end{tabular}
\vspace{-1em}
\end{table}

\subsection{Model Behavior Patterns}
\label{sec:eval-taxonomy}

As an exploratory analysis, we compare macro- and micro-averaged C2-P through a \emph{dilution gap}, defined as C2-P macro minus C2-P micro. C2-P macro is the mean of the per-run non-vacuous proof rates over the evaluable runs, and C2-P micro, which Table~\ref{tab:t1-summary} reports as C2-P, pools all assertions in those runs. A positive gap means that runs with many weakly proven assertions pull the micro-average down, and a negative gap means that a few runs with many proven assertions dominate it.

\begin{table}[h]
\centering
\small
\caption{Exploratory grouping by dilution gap (C2-P macro $-$ C2-P micro). Models within a group are ordered by gap magnitude.}
\label{tab:taxonomy}
\begin{tabular}{llr}
\toprule
Group & Models & Gap range \\
\midrule
Positive gap & GPT-5.5 ($+14.1$) & $> +5$ \\
\midrule
Near zero     & Gemini 3.1 Pro ($-2.1$), Kimi K2.6 ($-2.1$) & $[-5,\, +5]$ \\
\midrule
Negative gap & Qwen 3.6 Max ($-17.0$), GLM 5.1 ($-15.5$), GPT-5-mini ($-14.2$), & $< -5$ \\
             & Claude Haiku 4.5 ($-12.9$), DeepSeek V4 Flash ($-12.3$), & \\
             & MiniMax M2.7 ($-11.0$), DeepSeek V4 Pro ($-8.8$), & \\
             & Qwen 3.6 Plus ($-7.1$), Claude Opus 4.7 ($-6.9$) & \\
\bottomrule
\end{tabular}
\vspace{-1em}
\end{table}

\subsection{Agentic Evaluation Results}
\label{sec:eval-agentic}

The agentic evaluation runs five selected models within the iterative loop (Section~\ref{sec:sp-loop}) on a 40-module subset. Iteration~0 is the initial assertion file generated in agentic mode and serves as the baseline, and iterations~1 to~6 are revisions after successive rounds of VC~Formal feedback. From iteration~0 to iteration~6, the cross-model mean C1 rises by 16.0~percentage points (pp) and C2-P rises by 28.0~pp, while C4 falls by 13.3~pp and C5 falls by 20.1~pp (Figure~\ref{fig:t4-iter}). C5 falls for four of the five models, and C4 falls for three. GPT-5.5 is the only model whose C5 rises, by 25.9~pp, and GPT-5.5 and Gemini 3.1 Pro gain 7.6~pp and 6.8~pp in C4, whereas GPT-5-mini loses 62.8~pp in C4 and 67.9~pp in C5.

Restricted to modules that compile under every parameter combination at both iteration~0 and iteration~6, the cross-model mean changes keep their signs with smaller magnitudes ($+9.4$~pp in C2-P, $-8.4$~pp in C4, $-5.7$~pp in C5), although the number of assertions grows for all five models. This suggests that the revised assertion sets are easier to prove but exercise less of the design. C4 and C5 also oscillate across iterations (Appendix~\ref{app:f-t4-iter}).
\begin{figure}[h]
\centering
\input{figures/t4_iter_trends}
\vspace{-0.5em}
\caption{Per-iteration trajectories of C1, C2-P, mutation coverage (C4), and formal core coverage (C5) in the agentic evaluation for the five evaluated models on the 40-module subset. Iteration~0 is the initial generation in agentic mode. Every parameter combination of each module is evaluated, and C2-P, C4, and C5 are micro-averaged over the evaluable runs.}
\label{fig:t4-iter}
\vspace{-1em}
\end{figure}

%% file: figures/t4_iter_trends.tex
\definecolor{mtBlue}{HTML}{1F77B4}
\definecolor{mtOrange}{HTML}{FF7F0E}
\definecolor{mtGreen}{HTML}{2CA02C}
\definecolor{mtRed}{HTML}{D62728}
\definecolor{mtPurple}{HTML}{9467BD}
\begin{tikzpicture}
\begin{groupplot}[
  group style={group size=4 by 1, horizontal sep=0.72cm},
  scale only axis, width=0.193\linewidth, height=2.5cm,
  xmin=-0.3, xmax=6.3, ymin=0, ymax=100,
  xtick={0,1,...,6}, ytick={0,25,50,75,100},
  tick label style={font=\scriptsize}, label style={font=\footnotesize}, title style={font=\footnotesize\bfseries, yshift=-4pt},
  xlabel={Iteration}, xlabel style={yshift=3pt},
  grid=major, grid style={draw=black!10},
  axis line style={draw=black!60}, tick style={draw=black!60},
  every axis plot/.append style={line width=0.9pt, mark=*, mark size=1.3pt},
  legend style={font=\footnotesize, draw=none, legend columns=5, /tikz/every even column/.append style={column sep=8pt}},
]
\nextgroupplot[title={C1 (\%)}, legend to name=t4legend]
\addplot[color=mtBlue] coordinates {(0,82.5) (1,87.5) (2,80.0) (3,92.5) (4,75.0) (5,82.5) (6,82.5)}; \addlegendentry{Claude Opus 4.7}
\addplot[color=mtOrange] coordinates {(0,55.0) (1,77.5) (2,82.5) (3,82.5) (4,77.5) (5,72.5) (6,75.0)}; \addlegendentry{DeepSeek V4 Pro}
\addplot[color=mtGreen] coordinates {(0,75.0) (1,85.0) (2,77.5) (3,85.0) (4,82.5) (5,75.0) (6,85.0)}; \addlegendentry{Gemini 3.1 Pro}
\addplot[color=mtRed] coordinates {(0,45.0) (1,80.0) (2,80.0) (3,72.5) (4,70.0) (5,80.0) (6,75.0)}; \addlegendentry{GPT-5-mini}
\addplot[color=mtPurple] coordinates {(0,77.5) (1,100.0) (2,90.0) (3,92.5) (4,95.0) (5,100.0) (6,97.5)}; \addlegendentry{GPT-5.5}
\nextgroupplot[title={C2-P (\%)}]
\addplot[color=mtBlue] coordinates {(0,64.9) (1,88.7) (2,85.5) (3,93.9) (4,69.3) (5,88.7) (6,98.8)};
\addplot[color=mtOrange] coordinates {(0,48.2) (1,48.9) (2,57.5) (3,49.3) (4,60.7) (5,73.0) (6,76.5)};
\addplot[color=mtGreen] coordinates {(0,62.4) (1,70.3) (2,45.2) (3,52.9) (4,54.5) (5,85.1) (6,90.5)};
\addplot[color=mtRed] coordinates {(0,45.2) (1,55.8) (2,85.9) (3,88.6) (4,80.6) (5,84.6) (6,81.9)};
\addplot[color=mtPurple] coordinates {(0,75.5) (1,72.4) (2,96.7) (3,80.6) (4,90.6) (5,84.4) (6,88.8)};
\nextgroupplot[title={C4 (\%)}]
\addplot[color=mtBlue] coordinates {(0,62.9) (1,46.4) (2,50.0) (3,74.7) (4,51.4) (5,49.6) (6,57.8)};
\addplot[color=mtOrange] coordinates {(0,94.3) (1,63.6) (2,85.9) (3,100.0) (4,95.2) (5,81.3) (6,81.2)};
\addplot[color=mtGreen] coordinates {(0,47.4) (1,90.2) (2,80.7) (3,78.0) (4,47.4) (5,55.2) (6,54.3)};
\addplot[color=mtRed] coordinates {(0,97.6) (1,49.4) (2,34.0) (3,32.6) (4,18.0) (5,23.0) (6,34.8)};
\addplot[color=mtPurple] coordinates {(0,78.4) (1,97.5) (2,98.7) (3,87.7) (4,89.4) (5,98.1) (6,86.0)};
\nextgroupplot[title={C5 (\%)}]
\addplot[color=mtBlue] coordinates {(0,54.0) (1,8.8) (2,33.6) (3,30.8) (4,34.2) (5,33.8) (6,39.8)};
\addplot[color=mtOrange] coordinates {(0,85.3) (1,48.9) (2,67.6) (3,61.5) (4,68.5) (5,57.3) (6,63.8)};
\addplot[color=mtGreen] coordinates {(0,29.3) (1,41.2) (2,25.4) (3,7.0) (4,5.9) (5,7.4) (6,6.5)};
\addplot[color=mtRed] coordinates {(0,74.9) (1,2.4) (2,14.6) (3,4.6) (4,2.0) (5,2.6) (6,7.0)};
\addplot[color=mtPurple] coordinates {(0,21.6) (1,38.2) (2,54.6) (3,69.6) (4,33.5) (5,41.6) (6,47.5)};
\end{groupplot}
\node[anchor=south] at ($(group c1r1.north west)!0.5!(group c4r1.north east) + (0,0.45cm)$) {\pgfplotslegendfromname{t4legend}};
\end{tikzpicture}

%% file: sections/9_Conclusion.tex
\section{Conclusion and Discussion}
\label{sec:conclusion}

HierSVA connects a synthesis pipeline, a dataset of 342 BaseJump STL modules with 28 expert-reviewed module-bug pairs, and a six-axis benchmark for LLM-generated SVA on hierarchical RTL. Across twelve models, provability overstates verification strength. On average across models, 82.1\% of the generated assertions in evaluable runs prove non-vacuously, but the assertion sets cover only 36.2\% of the formal core. On the deep subset, 29\% of the assertion sets catch the bug while remaining valid on the correct RTL, and agentic feedback does not consistently improve verification strength.

\paragraph{Interpretation.} The current experiments cannot separate limited RTL training data from more general reasoning limitations of LLMs. However, the results point to a distinction broader than RTL familiarity, between writing a provable assertion for a known requirement and deciding which requirements a module must satisfy. For example, a model may verify FIFO overflow, underflow, and occupancy yet omit first-in-first-out ordering. A model checker does not report a property that was never proposed, and only mutation and formal-core coverage give indirect evidence of such omissions. HierSVA therefore tests verification planning under an incomplete specification in addition to RTL and SVA knowledge.

\paragraph{Limitations and future work.} The evaluation prompts retain instructions about structural coverage from an earlier pipeline version. Structural coverage is disabled and not scored, but these instructions may have influenced generation. Also, the eligible-fault denominator of C4 depends on manually audited exclusions in \texttt{limit.json}. Future work includes a semantic audit of the main set, a larger deep subset with semantic-level bugs, multiple independently isolated bug variants per module, and standardization or automation of the eligible-fault denominator.

%% file: appendix/A_Datasheet_for_HierSVA.tex
\section{Datasheet for HierSVA-DS}
\label{app:datasheet}

This datasheet follows the questionnaire proposed by~\cite{gebru2021datasheets}.

\subsection{Motivation}

\paragraph{For what purpose was the dataset created?}
HierSVA-DS was created to support evaluation of LLMs on hierarchical SVA generation under industrial formal verification methodology. The Hugging Face release is a plain-file dataset containing preprocessed BaseJump STL RTL (\texttt{rtl/}), per-module reference SVA bundles and metadata (\texttt{golden/}), hierarchy manifests (\texttt{json/}), and deep-evaluation files (\texttt{buggy\_rtl/}, \texttt{buggy\_rtl2/}, and \texttt{spec/}). The full main set is retained through a formal-evidence filter, and the deep subset is additionally reviewed by senior industrial DV engineers.

\paragraph{Who created the dataset and on behalf of which entity?}
The authors of this paper created the dataset. The dataset was not created on behalf of any other entity.

\paragraph{Who funded the creation of the dataset?}
The creation of the dataset received no external funding. The authors paid all costs themselves.

\subsection{Composition}
\label{app:datasheet-composition}

\paragraph{What do the dataset instances represent?}
Each main-set instance is a preprocessed hierarchical RTL module from the BaseJump STL standard library~\cite{basejump} paired with a reference bundle under \texttt{golden/}. This bundle includes formally retained SVA plus module metadata used to interpret and score the module. The deep dataset contains module-bug pairs derived from the same preprocessed RTL corpus, together with bug variants, bug-pattern notes, and natural-language specifications for C3 evaluation.

\paragraph{How many instances are there in total?}
The dataset contains 342 modules in the main set, distributed as 342 \texttt{golden/} reference bundles over a preprocessed \texttt{rtl/} snapshot of BaseJump STL. It also contains eight category-level hierarchy manifests under \texttt{json/} and an auxiliary deep subset of 28 module-bug pairs over 27 unique modules. The deep subset includes 21 synthetically injected bugs in \texttt{buggy\_rtl/}, 7 real-history bugs in \texttt{buggy\_rtl2/}, and 27 module specifications plus a coverage map in \texttt{spec/}. Modules span hierarchy depths 0 to 9 across the BaseJump STL category structure (bsg\_async, bsg\_cache, bsg\_dataflow, bsg\_link, bsg\_mem, bsg\_misc, bsg\_noc, bsg\_tag).

\paragraph{Does the dataset contain all possible instances or is it a sample?}
The dataset contains every BaseJump STL module that the HierSVA-SP preprocessing toolchain admitted and that subsequently cleared the formal evidence layer of Section~\ref{sec:dataset-qa}. Modules excluded during preprocessing (parser failures on non-1800-compliant constructs or unresolved external dependencies) or by the convergence criteria (assertions not satisfying all three convergence criteria of the synthesis loop) are not included as dataset instances.

\paragraph{What data does each instance consist of?}
For each main-set module, the dataset contains preprocessed BaseJump STL RTL under \texttt{rtl/} and a reference bundle under \texttt{golden/basejump\_stl/rtl/basejump\_stl/<group>/<module>/}. The bundle contains \texttt{design\_info.json} and a \texttt{final/} directory with \texttt{assertions.v}, \texttt{param\_info.json}, \texttt{limit.json}, and, for some modules, a short \texttt{x.md} summary. The \texttt{json/<group>.json} files record top RTL paths, hierarchy depth, and direct and transitive submodule sets. The deep dataset additionally carries independent natural-language specifications, a spec-to-bug coverage map, buggy RTL variants, and per-bug markdown notes. Bug variants were either recovered from upstream fix commits or generated by Claude Opus~4.6 and Claude Opus~4.7; senior industrial DV engineers reviewed them for bug well-formedness, patch isolation and effectiveness, and specification faithfulness (Section~\ref{sec:ds-deep}) before inclusion in the released subset.

\paragraph{Is there a label or target associated with each instance?}
The reference assertion file functions as a formally retained reference specification rather than absolute ground truth. Most HierSVA-B metrics (C1, C2-P, C2-V, C4, and C5) are computed by submitting the LLM's output to VC~Formal directly and reading the engine's verdicts on the LLM, with the reference assertions never consulted.

\paragraph{Is any information missing from individual instances?}
None of the fields above are missing from any released module. Modules whose pipeline runs did not produce artifacts satisfying all three convergence criteria are excluded rather than released with partial fields.

\paragraph{Are relationships between individual instances made explicit?}
Yes. The preprocessed RTL preserves the BaseJump STL module hierarchy, and \texttt{json/<group>.json} records each module's top file, hierarchy depth, direct submodules, and transitive submodule set. Reference bundles are provided for retained modules in that hierarchy.

\paragraph{Are there recommended data splits?}
HierSVA-DS is an evaluation benchmark and is not intended to be split into train/validation/test. The full main set is the primary evaluation corpus, and the deep dataset is the target of C3 and bug-detection studies. The agentic evaluation (Section~\ref{sec:eval-setup}) uses a fixed 40-module subset of the main set.

\paragraph{Are there any errors, sources of noise, or redundancies in the dataset?}
BaseJump STL is tapeout-validated and carries a low residual bug rate but is not bug-free. Reference assertions therefore characterize the design as it appears in the tapeout-validated source tree rather than as an idealized specification, and the deep-subset bugs are deliberately injected relative to this baseline.

\paragraph{Is the dataset self-contained, or does it link to external resources?}
The dataset release contains the plain-file tree needed to inspect the benchmark inputs: \texttt{rtl/}, \texttt{golden/}, \texttt{json/}, \texttt{buggy\_rtl/}, \texttt{buggy\_rtl2/}, \texttt{spec/}, the dataset card, and repository metadata. It links to the original BaseJump STL repository for provenance. Experimental artifacts such as benchmark logs, model outputs, parsed verdicts, harness scripts, and the HierSVA-SP code are released separately and are not dataset instances.

\paragraph{Does the dataset contain data that might be considered confidential, offensive, or sensitive?}
No. The dataset contains only open-source hardware design files, derived reference assertions, generated metadata, hierarchy manifests, and deep-dataset bug/specification artifacts.

\subsection{Collection Process}

\paragraph{How was the data associated with each instance acquired?}
The RTL files were obtained by cloning the BaseJump STL repository at commit a43571d and preprocessing the admitted modules. The \texttt{design\_info.json}, \texttt{param\_info.json}, \texttt{limit.json}, hierarchy manifests, and reference assertions were produced by HierSVA-SP (Section~\ref{sec:sp}), which runs an LLM proposer in an iterative VC~Formal loop providing formal feedback. Deep-subset bug variants were either recovered from upstream fix commits or generated by Claude Opus~4.6 and Claude Opus~4.7. Senior industrial DV engineers reviewed the variants for bug well-formedness, patch isolation and effectiveness, and specification faithfulness. Natural-language specifications were authored or reviewed to describe the correct behavior without revealing the bug.

\paragraph{What mechanisms or procedures were used to collect the data?}
The data were collected using an automated software pipeline comprising \texttt{git clone}, Python preprocessing scripts (Section~\ref{sec:sp-prep}), LLM API calls through a unified OpenRouter gateway, and Python loop control with rule-based log parsing. Deep-subset specification authoring was a manual process performed by senior DV engineers.

\paragraph{If the dataset is a sample from a larger set, what was the sampling strategy?}
HierSVA-SP attempts every BaseJump STL module that the preprocessing toolchain accepts. Modules that fail preprocessing or fail to satisfy the three convergence criteria are excluded. This constitutes documented deterministic filtering rather than sampling.

\paragraph{Who was involved in the data collection process?}
The authors of this paper carried out the data collection. The expert review of the deep subset is described in Section~\ref{sec:dataset-qa}.

\paragraph{Over what timeframe was the data collected?}
2026.

\paragraph{Were any ethical review processes conducted?}
Not applicable. The dataset is built from open-source hardware design files and does not involve human subjects, personal data, or sensitive information.

\subsection{Preprocessing, Cleaning, and Labeling}

\paragraph{Was any preprocessing or cleaning of the data done?}
Yes. Section~\ref{sec:sp-prep} describes the preprocessing toolchain in detail: macro expansion for \texttt{`define} and \texttt{`include} directives, module dependency graph construction, optional comment stripping, syntax normalization of whitespace and declaration styles, and design information extraction (clocks, resets, latch and combinational flags propagated over the dependency graph).

\paragraph{Was the raw data saved in addition to the preprocessed data?}
The dataset release contains the preprocessed RTL under \texttt{rtl/} and a pointer to the original BaseJump STL files at the pinned commit, allowing downstream users to inspect provenance or rerun preprocessing under different settings.

\paragraph{Is the software used to preprocess the data available?}
Yes. The HierSVA-SP preprocessing and synthesis toolchain is released as code accompanying the paper artifact, but it is not part of HierSVA-DS.

\subsection{Uses}

\paragraph{Has the dataset been used for any tasks already?}
The twelve-LLM evaluation reported in Section~\ref{sec:eval} is the primary use to date.

\paragraph{What other tasks could the dataset be used for?}
Intended additional uses include evaluation of agentic-formal hybrid methods, studying the convergence behavior of LLM-formal feedback loops, evaluating new assertion-quality metrics, and supporting assertion-quality methodology research. Users wishing to extend the dataset with new RTL libraries should run HierSVA-SP on those libraries rather than training directly on HierSVA-DS itself.

\paragraph{Is there anything about the composition of the dataset that might affect future uses?}
The reference assertions are LLM-proposed, formally verified, and (for the deep subset) expert-reviewed. Users treating HierSVA-DS as fine-tuning supervision should be aware that the assertion style reflects the proposing LLMs' output distribution under the convergence criteria of HierSVA-SP, rather than a human-complete SVA distribution. The benchmark was designed for evaluation rather than training.

\paragraph{Are there tasks for which the dataset should not be used?}
HierSVA-DS should not be used to train an LLM that is then evaluated on HierSVA-DS, as this would create train/test leakage. The headline findings of Section~\ref{sec:eval} depend on the evaluated models not having been exposed to the reference assertions.

\subsection{Distribution}
\label{app:datasheet-distribution}

\paragraph{Will the dataset be distributed to third parties?}
Yes, publicly under an open-source license.

\paragraph{How will the dataset be distributed?}
The dataset will be distributed through GitHub for version-controlled releases and through Hugging Face for a packaged tarball with explicit version tags.

\paragraph{When will the dataset be distributed?}
The dataset is submitted alongside the paper.

\paragraph{Will the dataset be distributed under a copyright or IP license?}
The original BaseJump STL files retain their upstream license. The derived dataset artifacts (preprocessed RTL, reference assertions, \texttt{design\_info.json}, \texttt{param\_info.json}, \texttt{limit.json}, hierarchy manifests, and deep-dataset bug/specification files) are released under Solderpad Hardware License version 2.1, selected for compatibility with the BaseJump STL upstream license.

\paragraph{Have any third parties imposed IP-based or other restrictions?}
No. BaseJump STL is open-source, and the derived artifacts are subject to no further restrictions beyond the chosen license.

\subsection{Maintenance}

\paragraph{Who will be supporting, hosting, and maintaining the dataset?}
The authors of this paper support and maintain the dataset, which is hosted on Hugging Face at \url{https://huggingface.co/datasets/AnonymousHierSVA/HierSVA}.

\paragraph{How can the maintainers be contacted?}
Maintainers can be contacted through the release repository's issue tracker for technical questions.

\paragraph{Are there errata?}
Errata, version notes, and known issues will be published in the repository's CHANGELOG.

\paragraph{Will the dataset be updated?}
Yes. Updates may track upstream BaseJump STL changes and ship corrections to existing preprocessed RTL, reference SVA, metadata, hierarchy manifests, or deep-dataset entries when issues are reported. Users are also encouraged to run HierSVA-SP on their own RTL libraries to extend the dataset locally; the toolchain is released for this purpose.

\paragraph{Will older versions continue to be supported, hosted, and maintained?}
Yes. Each release is tagged in Git and archived on the dataset host, with the version evaluated in this paper designated as the canonical citable artifact.

\paragraph{If others want to extend, augment, build on, or contribute to the dataset, is there a mechanism for them to do so?}
Yes. HierSVA-SP is the recommended mechanism for producing compatible \texttt{rtl/}, \texttt{golden/}, and \texttt{json/} entries from new RTL libraries. Contributors can submit pull requests containing compatible dataset entries and, when applicable, deep-dataset review packages under the \texttt{buggy\_rtl/}, \texttt{buggy\_rtl2/}, and \texttt{spec/} layout.

%% file: appendix/D_Full_Prompt_Templates.tex
\section{Full Prompt Templates}
\label{app:prompts}

This appendix reproduces each prompt as the Python source that builds it. Each listing begins with the minimal definitions it references, such as imports, helper strings, and status enumerations, so that it can be read without the rest of the code base. The prompts in Appendices~\ref{app:prompt-synth-sys}, \ref{app:prompt-synth-user}, \ref{app:prompt-feedback}, and~\ref{app:prompt-convergence} belong to one Python module and share the definitions at the top of Appendix~\ref{app:prompt-synth-sys}. The prompts in Appendices~\ref{app:prompt-eval-sys} and~\ref{app:prompt-eval-user} belong to a second module.

\paragraph{Corrected prompts.} The listings below are corrected versions of the prompts. The original prompts contain two inconsistencies.
\begin{itemize}
\item They describe a structural-coverage stage (COV) and ask the LLM to improve line, toggle, condition, and branch coverage, although COV is disabled in the final dataset synthesis and benchmark and is not scored.
\item The dataset synthesis system prompt states that Type-N parameters should never appear in \texttt{param\_combinations}, while the same prompt instructs the LLM to use their default values in all combinations and its JSON example lists a Type-N parameter. This inconsistency affects only dataset synthesis, because the evaluation prompts do not ask the LLM for parameter combinations.
\end{itemize}
The corrected listings remove the COV instructions and state that Type-N parameters appear at their default values in every combination and are never varied. The original prompts, exactly as sent to the models, are in \texttt{vrn2/prompt.py} and \texttt{vrn2/prompt\_eval.py} of the code repository at \url{https://github.com/HierSVAAnon/HierSVACodeAndArtifacts}.

\subsection{Dataset Synthesis System Prompt}
\label{app:prompt-synth-sys}

This listing is corrected. The COV stage is removed from the list of FV runs, and the Type-N rule is aligned with the parameter-combination rules and the JSON example. The JSON example is appended to the system prompt only for the models that require it (\texttt{insert\_json=True}).

\begin{Verbatim}
import json

prompt = {}

rely_on_rtl_comments_sysprompt = "- You can rely on RTL code comments to understand design functionality and signal usage."
verified_design_sysprompt = "- The design is already verified by simulation and there are no known design bugs. Your task is to write assertions that can be formally proven to be correct based on the design functionality, without worrying about design bugs or syntax errors in the RTL code."

json_format = '''
{
    "design_bugs": [],
    "converged": false,
    "functionality": "Summary of design...",
    "verification_plan": "Plan details...",
    "special_attentions": ["Note 1..."],
    "module_declaration": "module <top_module_name> #(...) (...);",
    "bind_command": "bind <top_module_name> <top_module_name>_assertions #(<param_connections>) i_assertions (.*);",
    "signal_extractions": [
        {"signal_name": "write_enable", "signal_type": "input", "related_signals": ["full", "data_in"],
         "rtl_code_snippet": "Raw RTL code extraction...", "usage": "Description...", "behavior": "Description..."}
    ],
    "assert_explanations": [
        {"name": "data_integrity_check", "related_signals": ["write_enable", "full", "data_in"], "related_assumes": [],
         "goal": "Description of the check...", "property": "Natural language property description...",
         "sva_code_snippet": "assert property ...", "auxiliary_rtl_logic": "logic [3:0] aux_count; ..."}
    ],
    "assume_explanations": [
        {"name": "", "related_signals": ["write_enable", "full", "data_in"], "related_assert": [],
         "goal": "Description of the assume...", "property": "Natural language property description...",
         "sva_code_snippet": "assume property ...", "auxiliary_rtl_logic": "..."}
    ],
    "cover_explanations": [
        {"name": "full_cover", "related_signals": ["full"],
         "goal": "Description of the check...", "property": "Natural language property description...",
         "sva_code_snippet": "cover property ...", "auxiliary_rtl_logic": "..."}
    ],
    "assertions.v": "module <top_module_name>_assertions(\\n    input clk,\\n    input rst_n,\\n    ...\\n);\\n\\n    // Full content here\\n\\nendmodule",
    "parameter_info": {
        "width_p": {"type": 1, "default": 10},
        "clear_over_set_p": {"type": 2, "default": 0},
        "harden_p": {"type": "N", "default": 1}
    },
    "param_combinations": {
        "fv1": {"width_p": 10, "clear_over_set_p": 0, "harden_p": 1},
        "fv2": {"width_p": 10, "clear_over_set_p": 1, "harden_p": 1},
        "fv3": {"width_p": 1, "clear_over_set_p": 0, "harden_p": 1},
        "fv4": {"width_p": 1, "clear_over_set_p": 1, "harden_p": 1}
    }
}
'''

prompt["system_init"] = lambda rely_on_rtl_comments, verified_design, insert_json=False, include_bsg_interface_description=True: f'''
Role: You are an expert Digital Verification Engineer specializing in Formal Verification using SystemVerilog Assertions (SVA).

Core Competencies:
1.  Deep understanding of digital logic as well as important protocols such as AXI, AHB, APB, PCIe, USB, etc.
2.  Expertise in writing concurrent assertions, cover properties, and assume constraints.
3.  Ability to analyze EDA tool logs and counter-examples to refine and fix assertions.
4.  Rich experience in formal verification methodologies and best practices, including assertion-based verification, cover-driven verification, and verification planning.

Responsibilities: 
You are responsible for maintaining and refining a formal verification assertion set for a given RTL design across multiple iterations.
There are three types of FV runs:
    - FPV: Check if all assert/cover properties are proven to be correct. Sub-module asserts are converted to assumes; only current module's assertions are verified.
    - FTA: Check if injected faults in the CURRENT MODULE ONLY can be detected. Sub-module RTL is not perturbed.
    - FC:  Analyze formal core to identify RTL logic in the CURRENT MODULE that is outside the formal core of the top-level assertions.

Global Constraints & Coding Standards (Apply to ALL turns):
1.  **Language:** Use English for all code comments and explanations.
2.  ** Verification Scope: **
        * Do not modify the RTL design. Only modify or add SVA, assumptions, cover properties, or auxiliary assertion logic.
        * If there are multiple modules in a single RTL file, focus only on the needed module as specified by the user.
        * Do not change design intent.
        * Do NOT create assert/assume/cover for input signals. 
        * The RTL (Verilog/SystemVerilog) code you receive is guaranteed to be syntactically valid. Do NOT spend effort on syntax checking or speculate about syntax errors; focus only on behavior, interfaces, and property generation.
        {rely_on_rtl_comments_sysprompt if rely_on_rtl_comments else ""}
        {verified_design_sysprompt if verified_design else ""}
        * Add assume properties if necessary to constrain the design behavior.
        * Add auxiliary RTL logic if necessary to aid the assertions.
        * Do NOT generate assert/assume/cover properties for reset behaviour
        * Sub-module behavior is already formally verified. Their proven assertions will be converted to assumes at the top level. Do NOT re-verify sub-module internal properties. Treat sub-module behavior as guaranteed contracts.
        * When writing top-level assertions, you may reference sub-module output signals directly. Build on top of sub-module contracts rather than reimplementing them.
3.  ** Immediate vs Concurrent Assertions: **
        * Do NOT use immediate assert/assume/cover for sequential logic. Use concurrent assert/assume/cover only.
        * Do NOT use concurrent assert/assume/cover for purely combinational logic. Use immediate assert/assume/cover only. Must wrap immediate assert/assume/cover in always_comb blocks.
4.  ** SVA Syntax, Naming and Semantics: **
        * Do NOT use `ifdef` or non-standard macros.
        * Do NOT use initial blocks.
        * Do NOT declare variables with initializers inside procedural blocks within generate constructs. Formal EDA does not support static initialization of block variables in generates. Move variable declarations outside the procedural block, or declare them without initializers and assign separately. Wrong: `for (...) begin integer row = k / N;` Right: `integer row; for (...) begin row = k / N;`
        * Use generate constructs if necessary.
        * All concurrent assert/assume/cover must have unique names. Must use statement labels to name assert/assume/cover properties. The name of the property must be the same as the statement label.
        * All begin-end blocks that have assert/assume/cover statements inside must have names.
        * Do NOT change the name of assert/assume/cover properties across iterations. If you need to add new assert/assume/cover, give them new unique names.
        * All assertions should be self-contained within the generated file. 
        * All assertions should have clocking unless the design is purely combinational.
        * All assertions should have disable checking on reset unless there's no reset signal.
5 . ** Parameterized Designs: **
        * Do NOT hardcode any parameter values in the assertions. Use the parameters as-is from the top module.
        * Classify each parameter into one of three categories:
          - **Type-1 (Sizing/Structural):** Width, depth, element count, etc. These do NOT change the functional spec. Examples: width_p, els_p, segments_p.
          - **Type-2 (Behavioral/Mode):** Mode select, priority policy, operation type, scan direction, etc. These change what the module DOES. Examples: clear_over_set_p, xor_p, lo_to_hi_p, hold_on_sr_p.
          - **Type-N (Non-functional):** Synthesis hints that have NO effect on RTL behavior. Examples: harden_p, strength_p, vertical_p, debug_p. Keep every Type-N parameter at its default value in every element of param_combinations and NEVER vary it — different values do not create different logic paths, and verifying them wastes resources.
        * If any Type-2 parameter changes the functional behavior, write assertions that cover ALL values using generate blocks. Each value should have a corresponding generate branch with appropriate assertions.
        * Populate the "param_combinations" field in the JSON output following these rules:
          **Validity:**
          - Every combination MUST be elaboration-safe. Check the RTL to trace how each parameter feeds into signal widths, array sizes, and loop bounds. Do NOT include any combination that would result in zero-width signals, zero-size arrays, or out-of-range bit selects — whether directly (width_p=0) or indirectly (e.g., max_val_p=0 when ptr_width = $clog2(max_val_p+1) = 0). The minimum useful value for any parameter that determines a signal width is typically 1.
          - Do NOT include degenerate initial values that make the module non-functional (e.g., LFSR with init_val_p=0 is stuck at zero forever).
          - If the default parameters are invalid for elaboration, the first element must use the smallest valid configuration.
          - If the module only works for power-of-2 sizes (check RTL comments and logic), do NOT include non-power-of-2 sizes.
          - For mutually exclusive mode parameters (e.g., xor_p/and_p/or_p where only one can be 1), do NOT cross-product them — enumerate each mode individually instead.
          **Coverage:**
          - Each element is a dict that must list ALL parameters with their values (do NOT use {{}} for defaults — always spell out every parameter explicitly).
          - The first element should be the RTL default parameter values (if valid).
          - For Type-2 parameters: include ALL valid values. If there are multiple Type-2 parameters, include the cross-product of their values (not just one-at-a-time). Example: if op can be xor/and/or and direction can be lo_to_hi/hi_to_lo, include all 6 combinations, not just 4.
          - For Type-1 parameters: include the default value plus ONE edge case (typically the minimum valid value, e.g., width_p=1). Do NOT create many sizing variants — two (default + minimal) is sufficient.
          - For Type-N parameters: do NOT vary them. Use the default value in all combinations.
          - If related parameters must be consistent (e.g., upper_width_p must be >= ceil(log2(num_banks_p)) + some offset), ensure every combination satisfies these constraints.
          **Example** for a module with width_p (Type-1, default=10), clear_over_set_p (Type-2, default=0, values 0/1), and harden_p (Type-N, default=1):
          "parameter_info": {{"width_p": {{"type": 1, "default": 10}}, "clear_over_set_p": {{"type": 2, "default": 0}}, "harden_p": {{"type": "N", "default": 1}}}}
          "param_combinations": {{"fv1": {{"width_p": 10, "clear_over_set_p": 0, "harden_p": 1}}, "fv2": {{"width_p": 10, "clear_over_set_p": 1, "harden_p": 1}}, "fv3": {{"width_p": 1, "clear_over_set_p": 0, "harden_p": 1}}, "fv4": {{"width_p": 1, "clear_over_set_p": 1, "harden_p": 1}}}}
6. ** SVA Module: **
        * All ports of the SVA module should be declared as input ports matching the Top-Level Module signals. 
        * Ensure all internal signals needed for assert/assume/cover are properly declared as inputs.
        * Explicitly import any required SystemVerilog packages.
7.  ** User Interaction: **
        * Treat all user messages as part of the same verification task unless explicitly stated otherwise.
        * Later user messages may provide additional RTL files, simulation results, formal tool feedback, counterexamples, or failure descriptions.
        * Even when refining a small part of the logic, you must always output the **COMPLETE** content for each assert/assume/cover in the JSON. Do not use placeholders like `// ... same as before ...`.
8.  ** Design Bugs: **
        * The "design_bug" field lists design bugs identified in iterations.
        * Each string in the "design_bugs" list should be a concise description of a specific design bug, with reference to the related signals, assertions, and RTL code sections.
9.  ** Bind: **
        * The bind command must follow the format:`bind <module_name> <module_name>_assertions #(<param_connections_or_empty>) i_assertions (.*);`
        * If the top module has parameters, pass them explicitly using .PARAM(PARAM) syntax. Example: `bind fifo fifo_assertions #(.WIDTH(WIDTH), .DEPTH(DEPTH)) i_assertions (.*);`
        * If the top module has no parameters, omit the #(...) entirely. Example: `bind top top_assertions i_assertions (.*);`
        * Always use bind to module definition name, not instance path.
{"""10.  ** Interface Handshake Description: **
        * In module's handshake protocols, "helpful" means a side provides its handshake information up front, without depending on the other side's current response. "Demanding" means a side must first observe the other side's handshake signal before deciding its own.
        * Protocol semantics are as follows:
            - r&v: both sides are helpful.
            - r->v: the side providing ready is helpful; the side generating valid based on ready is demanding.
            - v->r (valid-yumi): the side providing valid is helpful; the side generating yumi/ready based on valid is demanding.
            - valid-credit: both sides are helpful.
        * The consumer/return-side signal name depends on the protocol:
            - `ready_and` for r&v
            - `ready_then` for r->v
            - `yumi` for v->r (valid-yumi)
            - `credit` for valid-credit
""" if include_bsg_interface_description else ""}

Output Contract:
- Return a STRICT JSON object. No markdown formatting outside the JSON.
- The JSON must include:
  - verification plan ("verification_plan" field)
  - design functionality summary ("functionality" field)
  - module declaration ("module_declaration" field)
  - signal usage analysis ("signal_extractions" field)
  - bind statement ("bind_command" field)
  - structured explanation for each assertion ("assert_explanations" field), each assumption ("assume_explanations" field) and each cover property ("cover_explanations" field).
  - full assertion module code ("assertions.v" field) 
  - special fields ("design_bugs", "converged" fields)
- Strictly follow the **Critical Reminders** section in every user prompt regarding "converged" fields.
- Update the verification plan in the JSON if there are any changes to the assertions, assumptions, or cover properties. The verification plan should reflect the current state of the assertions and the intended verification strategy.

{"EXAMPLE JSON FORMAT:" if insert_json else ""}
{json_format if insert_json else ""}
'''.replace("\\n", "\n")
\end{Verbatim}

\subsection{Dataset Synthesis Initial User Prompt}
\label{app:prompt-synth-user}

This listing is unchanged from the original. It uses \texttt{json} and \texttt{prompt} from Appendix~\ref{app:prompt-synth-sys}.

\begin{Verbatim}
prompt["initial"] = lambda module_name, rtl_json, submodule_context=None : f'''
I am providing the RTL source files for a new verification task. Top-level module is {module_name}.

**Input Design Data (JSON):**
```json
{json.dumps(rtl_json)}
```

{f"""**Sub-module Contexts (already formally verified):**
The following sub-modules are instantiated in {module_name}. Their interfaces and 
proven assertions are provided as contracts you can rely on. Do NOT re-verify 
these properties. Use them as assumed guarantees when writing top-level assertions.
```json
{json.dumps(submodule_context)}
```
""" if submodule_context else ""}

**Immediate Request:**
1. **Analyze:** Perform the functionality analysis and signal extraction as defined in the System Prompt.
2. **Plan:** Devise a comprehensive formal verification plan based on the design functionality.
3. **Generate:** Create the complete `assertions.v` file and the corresponding bind command.

**Critical Reminders:**
* **Output Format:** Return ONLY the strict JSON object. Do not wrap it in Markdown code blocks.
* **Completeness:** The value for the SVA file must contain the **FULL, COMPILABLE** SystemVerilog source code.
* **Plan Consistency:** Ensure the verification plan aligns with the generated assertions.
* **Design Bugs:** Set "design_bugs" to an empty list for this run.
'''
\end{Verbatim}

\subsection{Evaluation System Prompt}
\label{app:prompt-eval-sys}

This listing is corrected. The COV stage is removed from the verification pipeline description and from the scoring description.

\begin{Verbatim}
import json

json_format = '''
{
  "assertions.v": "module <DUT>_assertions(\\n    input clk,\\n    input rst_n,\\n    ...\\n);\\n  // full assertion body\\nendmodule",
  "bind_command": "bind <DUT> <DUT>_assertions #(.PARAM(PARAM)) i_assertions (.*);"
}
'''.strip()

# ---------------------------------------------------------------------------
# Auxiliary text blocks. Edit these — they will dominate the prompt's length.
# ---------------------------------------------------------------------------

SCHEMA_FIELD_GUIDE = """
## Output Schema (STRICT — return ONLY these two fields)

Return a single JSON object with the following fields and nothing else:

- **assertions.v** (string, REQUIRED): the FULL contents of the `<DUT>_assertions` module — one
  self-contained SystemVerilog file. Must be syntactically valid. Every DUT port is declared `input`.
  The assertions module MUST work for every parameter combination listed in the eval harness's
  fixed `param_info.json` (which you do NOT see and do NOT control). Use parameters symbolically
  rather than hardcoding values; this is the only way your assertions can pass across all
  Type-2 sweep variants the harness will exercise.
- **bind_command** (string, REQUIRED): one of:
    `bind <DUT> <DUT>_assertions #(.PARAM(PARAM)...) i_assertions (.*);`
    `bind <DUT> <DUT>_assertions i_assertions (.*);`     // when DUT has no parameters

Do NOT emit any other top-level fields. No `functionality`, no `verification_plan`, no
`signal_extractions`, no `assert_explanations` / `assume_explanations` / `cover_explanations`,
no `parameter_info`, no `param_combinations`, no `design_bugs`, no `converged`. Return STRICT
JSON — no markdown fences, no commentary, no chain-of-thought outside the JSON.
"""

SVA_STYLE_GUIDE = """
## SVA Style Guide

### Naming and structure
- Every concurrent assert/assume/cover MUST have a unique statement label. The Pydantic `name`
  field MUST equal the SVA statement label.
- Wrap related properties inside named `begin … end : block_name` blocks. Do NOT leave anonymous
  blocks containing assert/assume/cover.
- All assertions live in a single module named `<DUT>_assertions` whose ports mirror the DUT 1:1
  (every port declared `input`). Use `bind` to attach.
- Do NOT change a property's name across iterations. To add new properties, give them new names.

### Clocking and reset
- Concurrent properties MUST use `@(posedge clk)` (or the DUT's clock). Use `disable iff (!rst_n)`
  (or matching polarity) on every concurrent property — except modules with no reset signal.
- Pure combinational designs MUST NOT use concurrent properties at all. Wrap immediate
  assert/assume/cover inside `always_comb` blocks.
- Sequential designs MUST NOT use immediate assertions for sequential behavior.

### Operator choice
- `|->` (overlapping) checks the consequent in the SAME cycle the antecedent holds.
- `|=>` (non-overlapping) checks the consequent ONE cycle after.
- Use `##N` for explicit cycle delay, `##[a:b]` for ranges. Never use `##0` — use `|->` instead.
- Use `$past(sig, n)` for value at n cycles ago. Always pair `$past` with a guard against
  the first n cycles after reset (e.g., `$past(rst_n) && ...`).
- Use `$stable(sig)` / `$rose(sig)` / `$fell(sig)` instead of hand-rolled equivalents.
- For "always eventually" use `s_eventually` only when bounded; unbounded liveness is rarely
  provable in practice — express as a bounded `##[1:N]` instead.

### Assumptions
- Add `assume`s to constrain INPUT signals when needed (e.g., handshake protocol obligations
  on the producer side). Do NOT add assertions on input signals.
- Submodule output assertions have already been proven during the synthesis pipeline; you
  receive them as contracts (assumes). Do NOT re-verify submodule properties.

### Auxiliary logic
- If a property needs a counter or shadow register to express, declare that logic INSIDE the
  assertions module under `auxiliary_rtl_logic`. Do NOT modify the DUT.
- Auxiliary logic must follow the same clk/rst as the DUT region it tracks.

### Parameter handling
- The harness's fixed `param_info.json` decides which parameter combinations get exercised;
  you only need to make `assertions.v` correct under ALL of them.
- For Type-2 mode parameters that change which logic path is taken, write assertions that cover
  every legal value via `generate` blocks (one branch per Type-2 value), not just the default.
- For Type-1 sizing parameters, never hardcode a numeric width — always reference the parameter
  symbolically (e.g., `count_o == {width_p{1'b1}}`).
"""

COMMON_PITFALLS = """
## Common Pitfalls (verified failure modes from prior runs)

These are the recurring ways generated SVA gets rejected by the formal tool. Avoid them.

1. **Vacuous antecedents.** A property like `assert property (@(posedge clk) disable iff (!rst_n)
   (foo == 4'hF) |-> bar)` will pass vacuously if `foo == 4'hF` never holds. Always add a `cover`
   for the antecedent so vacuity is observable. Better still, choose antecedents that *are*
   exercised by the design's normal operation.

2. **Off-by-one in `|->` vs `|=>`.** If you say "when ready_o rises, valid_i must already be high",
   that's `|->`. If you say "one cycle after a request, response must come", that's `|=>` or `##1`.
   Mismatching this is the #1 source of false falsifications.

3. **Reset-window violations.** Properties referencing `$past(x)` will fail in the first cycle
   after reset because `$past(x)` is undefined there. Either guard with `$past(rst_n)` or use
   `disable iff` aggressively.

4. **Asserting on inputs.** Properties of the form `assert property (... input_signal ...)` where
   the right-hand side is purely an input combination will be falsified by any unconstrained
   waveform. Constrain inputs with `assume` (or accept that you're describing input protocol
   obligations, in which case use `assume` not `assert`).

5. **Hardcoded parameter values.** Writing `assert property (count_o == 4'd15)` when the DUT has
   `parameter width_p = 4` makes the assertion fail for every other `width_p`. Always use the
   parameter symbolically: `assert property (count_o == {width_p{1'b1}})`.

6. **Generate-block scope.** If your assertion lives inside a `generate` block, every property
   needs a unique label PER GENERATE INSTANCE. Use the `genvar` in the label or wrap each
   instance's properties in a named begin/end.

7. **Concurrent vs immediate misuse.** Combinational designs (no flip-flops) cannot use
   `@(posedge clk)` properties — there's no clock to anchor them on. Use immediate
   `always_comb begin assert (cond); end`.

8. **Cover starvation.** A passing assertion set with NO cover properties produces no signal
   about whether the assertions exercise interesting design states. Add at least one cover for
   each major interface event (handshake fire, FIFO not-empty / full transition, etc.).
"""

SVA_DESIGN_BUGS_NOTE = """
## When the RTL Has a Bug

You are evaluating a tape-out validated design. Bugs are unlikely. If a property genuinely
cannot pass under any correct interpretation of the design, write the assertion you BELIEVE
is correct (according to your design intent reading) and accept that it may be falsified —
do NOT add prose explaining the bug; the schema has no field for that.
"""

SVA_DESIGN_BUGS_NOTE_SPEC_MODE = """
## When the RTL Contradicts the Spec

The RTL provided below is a CANDIDATE implementation that may contain bugs. The
natural-language specification given in the user message is the source of truth.

If you observe ANY contradiction between the spec and the RTL — different reset
behavior, different output value, different timing, different control-flow polarity,
different parameter handling, anything — you MUST write the assertion against the
SPEC. Doing so means the assertion will FALSIFY when run on this particular RTL.
That is the correct outcome: the falsification is the bug being caught.

Specifically:
- Do NOT "fix" the assertion to match what the RTL does.
- Do NOT relax the property until it passes on the RTL shown.
- Do NOT skip writing an assertion just because the RTL implementation looks
  inconsistent with the spec on that property.
- An assertion that PASSES on the buggy RTL is a bug-detection FAILURE; an
  assertion that FAILS on the buggy RTL while being faithful to the spec is
  the GOAL.

Submodule contracts (when present) and the parameter sweep stay authoritative as
before — only the top-level RTL is potentially buggy.
"""

# ---------------------------------------------------------------------------
# system_init — the enriched 0-shot system prompt
# ---------------------------------------------------------------------------

def system_init(rely_on_rtl_comments=True, verified_design=True,
                include_bsg_interface_description=True):
    bsg_block = """
## BaseJump STL Handshake Conventions

In BaseJump's protocols, "helpful" means a side asserts its handshake info up front, without
waiting for the other side; "demanding" means a side decides only after seeing the other side.

| Protocol | Producer side | Consumer side | Consumer signal name |
|---|---|---|---|
| r&v (ready-and-valid)   | helpful   | helpful   | `ready_and`  |
| r->v (ready-then-valid) | helpful   | demanding | `ready_then` |
| v->r (valid-then-yumi)  | helpful   | demanding | `yumi`       |
| valid-credit            | helpful   | helpful   | `credit`     |
""" if include_bsg_interface_description else ""

    rely_block = (
        "- You may rely on RTL code comments to disambiguate behavior."
        if rely_on_rtl_comments else
        "- Treat RTL comments as untrusted; ground every assertion in observable signal behavior."
    )
    verified_block = (
        "- The RTL is silicon-validated and assumed bug-free. Your job is to write assertions that\n"
        "  are formally provable, not to find design bugs. Assume the design is correct."
        if verified_design else
        "- The RTL below is a CANDIDATE implementation that may contradict the spec given in the\n"
        "  user message. The SPEC IS THE SOURCE OF TRUTH. Write each assertion against the spec,\n"
        "  not against the RTL — assertions that falsify on the RTL because the RTL violates the\n"
        "  spec are the desired outcome (they detect the bug)."
    )

    # In spec/Tier-2 mode the standard 'tape-out validated, bugs unlikely' framing is
    # actively misleading. Swap in a spec-aware note that tells the LLM to keep
    # writing the spec-faithful assertion even when it falsifies on the buggy RTL.
    bugs_note_block = SVA_DESIGN_BUGS_NOTE if verified_design else SVA_DESIGN_BUGS_NOTE_SPEC_MODE

    return f"""
# Role
You are an expert Digital Verification Engineer. Your task: given a SystemVerilog/Verilog RTL
module, generate a complete SVA assertion module that can be formally proven on VC Formal.

# Verification Pipeline
After your assertions are generated, three formal tools run automatically:
  - **FPV** (formal property verification): proves each assert / inspects each cover.
  - **FTA** (formal testbench analyzer / mutation testing): mutates the RTL and checks each
    assertion catches the mutation. A high mutation-detection rate signals strong assertions.
  - **FC** (formal core): which RTL signals participate in proving each assert. Signals
    *outside* the formal core indicate the assertion is too weak.

You are scored on: assertion proof success rate (C2-P), non-vacuous proof rate (C2-NV), mutation detection
rate (C4), and formal-core coverage (C5). Trivially-true assertions hurt C2-NV; weak
assertions hurt C4 and FC.

# Hard Constraints
- The RTL is syntactically valid. Do NOT speculate about syntax errors.
- Do NOT modify the DUT. Only write the `<DUT>_assertions` module.
- Do NOT use `ifdef`, initial blocks, or non-standard macros.
- Do NOT declare variables with initializers inside procedural blocks within `generate`
  (formal does not support static init in generates).
- Do NOT generate assert/assume/cover for reset behavior or for input signals.
- Submodule properties are provided as PROVEN CONTRACTS (converted to assumes). Do NOT
  re-verify them. Build on top of them.
{rely_block}
{verified_block}

{SCHEMA_FIELD_GUIDE}

{SVA_STYLE_GUIDE}

{COMMON_PITFALLS}

{bugs_note_block}
{bsg_block}

# Output
Return a STRICT JSON object matching the schema above. No markdown, no prose outside the JSON.

EXAMPLE JSON FORMAT:
{json_format}
""".strip()
\end{Verbatim}

\subsection{Evaluation User Init Prompt}
\label{app:prompt-eval-user}

This listing is unchanged from the original. It uses \texttt{json} from Appendix~\ref{app:prompt-eval-sys}.

\begin{Verbatim}
def initial(module_name, rtl_json, submodule_context=None, spec_text=None):
    """Build the initial user message. param_combinations is NOT solicited from the LLM —
    the eval harness uses golden's param_info.json for the parameter sweep.

    The RTL JSON keys are basename-only (the directory portion of every file path is
    stripped) so the LLM cannot infer the source tree. Tier-2 / C3 must remain blind
    to whether the RTL came from a `golden`, `buggy_rtl`, or `buggy_rtl2` location.

    Optional Tier-2 input:
      spec_text: natural-language spec / intent (markdown). When present, it is inserted
                 as the source of truth and the RTL below is described as a candidate
                 implementation.
    """
    import os as _os
    rtl_basename = {_os.path.basename(k): v for k, v in (rtl_json or {}).items()}

    sub_block = (
        f"""
**Sub-module Contracts (already formally proven — treat as guaranteed):**
The sub-modules instantiated in `{module_name}` are listed below. For each, the entry
includes `module_name` and `assertions_v` — the verbatim SystemVerilog text of the
sub-module's bound assertion module. The port list and every `assert` / `assume` /
`cover` you see there is already proven; you may rely on them as assumes and you must
NOT re-verify them.
```json
{json.dumps(submodule_context)}
```
""" if submodule_context else ""
    )

    spec_block = (
        f"""
**Natural-language Specification (source of truth for intended behavior):**
The RTL below is a CANDIDATE implementation that may contain bugs. The specification
below describes the intended behavior — design every assertion against the SPEC.

**Conflict resolution rule:** if the RTL behavior contradicts the spec on any property
(reset value, output mapping, timing, control polarity, parameter handling — anything),
the SPEC WINS. Write the assertion as the spec requires, even when this means the
assertion will FALSIFY when run on the RTL shown below. A spec-faithful assertion
that fails on this RTL is the *desired* outcome — it detects the bug. Do NOT relax,
remove, or rewrite the property to match what the RTL happens to do.

```markdown
{spec_text}
```
""" if spec_text else ""
    )

    return f"""
Top-level module: `{module_name}`.
{spec_block}
**Input Design Data (JSON):**
```json
{json.dumps(rtl_basename)}
```
{sub_block}

**Task:** Generate the assertion module for `{module_name}`. Return STRICT JSON containing
ONLY `assertions.v` and `bind_command`. The `assertions.v` field must hold the FULL,
COMPILABLE SystemVerilog source.
""".strip()
\end{Verbatim}

\subsection{Feedback Prompt}
\label{app:prompt-feedback}

This listing is corrected. The COV feedback block is removed, and the formal-core feedback no longer asks for line, toggle, condition, and branch coverage. The feedback prompt is built by the deterministic loop controller (Section~\ref{sec:sp-loop}) and uses \texttt{prompt} from Appendix~\ref{app:prompt-synth-sys}. The enumerations at the top of the listing are the per-stage statuses produced by the rule-based VC~Formal log parser.

\begin{Verbatim}
from enum import Enum

class Stages(Enum):       FPV, FTA, FC = range(3)
class StageStatus(Enum):  NOT_RUN, PASSED, FAILED = range(3)
# HAS_CEX_VAC_INC_UNC: some property is falsified, vacuous, inconclusive, or uncoverable
class FPVRunStatus(Enum): FAILED, HAS_CEX_VAC_INC_UNC, ALL_PASSED = range(3)
class FTARunStatus(Enum): ALL_DETECTED, HAS_NON_DETECTED_MUTANTS = range(2)
# HAS_COMPLEMENT: some RTL logic lies outside the formal core
class FCRunStatus(Enum):  ALL_PASSED, HAS_COMPLEMENT = range(2)

# Arguments filled in by the loop controller:
#   status       one of the enumerations above for the corresponding stage
#   status_txt   per-property verdicts or per-signal results parsed from the VC Formal report of that stage
#   vcd_content  counterexample waveforms (VCD text, whitespace-normalized) of the failing properties,
#                concatenated over all parameter combinations; once the text exceeds 32,768 characters,
#                the remaining waveforms are replaced by the line
#                "Other counterexample waveforms omitted due to length limit..."

def return_fpv(status, status_txt, verified_design, vcd_content):
    if status == StageStatus.NOT_RUN:
        return ""
    if status == FPVRunStatus.ALL_PASSED:
        return ""
    elif status == FPVRunStatus.FAILED:
        return f"""
The previous FPV run failed. Logs:
{status_txt}
For the failed FPV run
1. Analyze the failure. Determine if it's due wrong syntax in assertions or failed extraction of clock and reset signals.
2. If it's due to wrong assertion syntax or failed extraction of clock and reset signals, refine the assertions or correct the signal extraction.
3. Return the FULL updated JSON.
"""
    else:
        if not status_txt.strip():
            return ""

        return f"""
The following assertions or cover properties are either falsified, vacuous, inconclusive or uncoverable. There also might be some timeout issues:
{status_txt}
Couter-Example Waveforms:
{vcd_content}
For these falsified/vacuous/inconclusive assertions, uncoverable cover properties, or timeout issues:
1a. For each falsified assertion, analyze the counter-example waveform to distinguish between design bug vs assertion bug.
1b. For each vacuous assertion, analyze the antecedent conditions.
1c. For each inconclusive assertion, analyze the related logic and conditions.
1d. For each uncoverable cover property, analyze the related logic and conditions.
1e. For each timeout issue, analyze the related logic and conditions.
2. Modify the assertions to ensure that all assertions are non-vacuously passing, unless a design bug is identified. For timeout issues, refine the assertions to avoid the timeout, or add necessary assume properties to constrain the design behavior and avoid the timeout. 
If the CEX waveform is not provided, focus on other failed assertions first.
"""

def return_fta(status, status_txt):
    if status == StageStatus.NOT_RUN:
        return ""
    if status == FTARunStatus.ALL_DETECTED:
        return ""
    
    return f"""
There are FTA issues. These may include non-detected faults and/or timeout issues:
{status_txt}
For these non-detected faults or timeout issues:
1. Analyze the assertion and the related design logic.
2. Modify the assertions or add new assertions to ensure all injected faults can be detected.
For timeout issues, analyze the related assertions and design logic, and refine the assertions or add necessary assume properties to constrain the design behavior and avoid the timeout.
"""
    
def return_fc(status, status_txt):
    if status == StageStatus.NOT_RUN:
        return ""
    if status == FCRunStatus.ALL_PASSED:
        return ""
    
    return f"""
The following RTL logic are outside the formal core of the assertions:
{status_txt}
For these RTL logic:
1. Analyze the RTL code and assertions used in the verification.
2. Modify or add assertions to ensure the RTL logic that are outside the formal core of the assertions are properly covered.
"""

def return_unified(run_type_status, run_type_status_txt, verified_design, vcd_content):
    fpv_part = return_fpv(
        run_type_status[Stages.FPV],
        run_type_status_txt[Stages.FPV],
        verified_design,
        vcd_content,
    )
    fta_part = return_fta(run_type_status[Stages.FTA], run_type_status_txt[Stages.FTA])
    fc_part  = return_fc  (run_type_status[Stages.FC], run_type_status_txt[Stages.FC])

    return f"""
{fpv_part}
{fta_part}
{fc_part}
Note that
* **Convergence:** Set "converged" field to false for this iteration.
* **Design Bugs:** If a design bug is identified from the analysis of failed assertions, add the description of the design bug to the "design_bugs" list field.
"""

prompt["unified"] = return_unified
\end{Verbatim}

\subsection{Convergence Check Prompt}
\label{app:prompt-convergence}

This listing is corrected. The convergence condition refers to formal-core coverage instead of line, toggle, condition, and branch coverage. It uses \texttt{prompt} from Appendix~\ref{app:prompt-synth-sys}.

\begin{Verbatim}
prompt["convergence"] = """
The previous verification run has successfully completed without assertion failures. Also, all RTL logic is inside the formal core of the assertions, and all injected faults are detected.

**Instructions:**
1. Analyze the RTL code and assertions used in the verification.
2. Determine if the assertions sufficiently cover the intended functionality of the design, and if there are enough cover properties to ensure thorough verification.
3a. If you believe the verification is comprehensive and the design is verified, respond with a JSON object with the field "converged" set to true.
3b. If you believe the verification is NOT comprehensive enough, respond with a JSON object with the field "converged" set to false, and add more assertions or cover properties as needed.

**Critical Reminders:**
* **Output Format:** Return ONLY the strict JSON object. Do not wrap it in Markdown code blocks.
* **Completeness:** The value for the SVA file must contain the **FULL, COMPILABLE** SystemVerilog source code.
* **Plan Consistency:** If assertions are added, ensure the verification plan is also updated to reflect the changes.
* **Design Bugs:** If a design bug is identified from the analysis of failed assertions, add the description of the design bug to the "design_bugs" list field.
"""
\end{Verbatim}

%% file: appendix/E_SVA_Primer_for_ML_Audience.tex
\section{SVA Primer for an ML Audience}
\label{app:sva-primer}

This appendix gives a self-contained introduction to SystemVerilog Assertion concepts, formal-verification workflow, and tool-specific semantics required to read the rest of this paper. It is intended for readers without prior exposure to hardware design or formal RTL verification. For a more detailed treatment, we refer the reader to Seligman et al.~\cite{the_formal_book}.

\subsection{Introduction to SystemVerilog Assertions and Hardware Verification}
\label{app:sva-intro}

A digital hardware design is described in a register-transfer-level (RTL) language such as Verilog or SystemVerilog, which expresses the design as a network of registers and combinational logic clocked by one or more periodic signals. Hardware functional verification asks whether a given RTL design behaves according to its specification and is a critical phase of pre-silicon design because once a chip is fabricated, bug fixes are prohibitively expensive.

A SystemVerilog Assertion (SVA) is a property statement written alongside the RTL that captures a piece of intended behavior, for example, "after a request is asserted, an acknowledgment must arrive within five clock cycles." SVA serves two functions in industrial verification design. In simulation-based verification, an SVA is monitored as the design is exercised by stimulus, and any violation is reported as a simulation failure. In formal property verification (FPV), an SVA is mathematically proven against the design without stimulus by exhaustively reasoning about the entire reachable state space.

\subsection{Temporal Operators and Properties}
\label{app:sva-ops}

An SVA property combines Boolean expressions over RTL signals with temporal operators that constrain when and how those expressions must hold. Common operators are the overlapped implication \texttt{|->} and the non-overlapped implication \texttt{|=>}, which encode whether, when the left-hand side holds at the current clock tick, the right-hand side must hold at the same tick (\texttt{|->}) or one tick later (\texttt{|=>})." The cycle-delay operator \texttt{\#\#k} expresses a delay of exactly k clock ticks. Sequence operators such as \texttt{[*n]} for repetition, \texttt{intersect}, and \texttt{within} compose temporal expressions across multiple clock ticks. A simple example is:
\begin{Verbatim}
(req && !ack) |-> ##[1:5] ack
\end{Verbatim}
which states that whenever \texttt{req} is asserted while \texttt{ack} is low, \texttt{ack} must arrive within one to five subsequent clock ticks.

\subsection{Formal Verification Workflow}
\label{app:sva-workflow}

A formal property verifier takes the RTL design and an SVA property and returns a verdict. Synopsys VC~Formal~\cite{vcf} is the commercial tool used in our evaluation. It exposes several application modes, two of which are central to HierSVA. \emph{Formal Property Verification} (FPV) takes each SVA property and either proves it (the property holds on every reachable trace), reports a counterexample (a concrete trace that violates the property), or returns inconclusive (the engine ran out of resources). The terminal status of an FPV goal is one of \texttt{PROVEN}, \texttt{VACUOUS}, \texttt{FALSIFIED}, or \texttt{INCONCLUSIVE}. \emph{Formal Testbench Analyzer} (FTA) systematically injects small mutations into the RTL and reports which assertions detect each fault, providing a robustness signal for the assertion set. Within FPV, the \emph{Formal Core} (FC) feature reports the minimal subset of design statements that contribute to each proof, providing a coverage signal relative to the cone-of-influence of each proven assertion rather than against the full design.

\subsection{Vacuity Detection}
\label{app:sva-vacuity}

A vacuous property is a conditional statement that is true because the antecedent cannot be satisfied. For example, the property \texttt{p |-> q} trivially passes if \texttt{p} is never true. Such a property almost always flags an issue in the environment or property.

Vacuity has a richer theory in the model-checking literature~\cite{beer2001efficient,kupferman2003vacuity}, covering cases where any subformula does not influence the truth value of the property. In practice, tool support for vacuity is more restricted. VC~Formal flags vacuity for the specific case of \texttt{|->} and \texttt{|=>} properties whose left-hand antecedent is never satisfiable on any reachable trace. When vacuity checking is enabled, VC~Formal classifies a successful proof attempt as one of two terminal statuses, namely \texttt{PROVEN} (proof succeeded and the vacuity goal is non-vacuous) or \texttt{VACUOUS} (proof succeeded but the antecedent is unsatisfiable). HierSVA uses this distinction directly in its metric framework. C2-P counts only \texttt{PROVEN} (proven non-vacuously), and C2-V counts only \texttt{VACUOUS}; the two are reported separately rather than aggregated into a single provability number.

\subsection{Mutation-Based Testbench Analysis}
\label{app:sva-mutation}

Mutation testing originated in software testing~\cite{hamlet1977analysis, budd1978design} as a way to evaluate test-suite quality by inserting small syntactic faults into the program and measuring how many of the inserted faults the test suite catches. The same idea applies to hardware assertion sets. A mutation operator perturbs the RTL in a small, well-defined way (flipping a comparison operator, deleting a statement, swapping operands, changing a constant value), and an assertion set is said to detect a fault if at least one assertion fails on the mutated RTL while passing on the original. The detection rate over a set of non-equivalent faults is a measure of how robustly the assertion set constrains the RTL it accompanies. VC~Formal's FTA app implements this systematically and reports which assertions detected each generated fault.

\subsection{Worked Example: A Toy FIFO}
\label{app:sva-fifo}

Consider a single-clock first-in-first-out (FIFO) buffer of depth eight, with input handshake signals \texttt{enq} (push request) and \texttt{full} (cannot accept), and output handshake signals \texttt{deq} (pop request) and \texttt{empty} (nothing to pop). A reasonable assertion set might include
\begin{Verbatim}
no_overflow:  full  |-> !enq
no_underflow: empty |-> !deq
count_bound:  count <= 8
\end{Verbatim}
The first property states that the FIFO never accepts a push while full; the second states the dual condition for the empty case; and the third asserts that the internal occupancy counter never exceeds the FIFO depth. Each of these compiles to an SVA property that VC~Formal can verify using FPV. On a correct FIFO, all three should reach \texttt{PROVEN} status. On a buggy FIFO that incorrectly increments the counter, \texttt{count\_bound} would reach \texttt{FALSIFIED} with a counterexample trace.

The first two properties illustrate why vacuity matters. If the formal environment assumptions make the \texttt{full} state unreachable, the first property's antecedent is never satisfied on any reachable trace, and VC~Formal would report it as \texttt{VACUOUS} rather than \texttt{PROVEN}.

\subsection{Glossary}
\label{app:sva-glossary}

\paragraph{Assertion.}
A property statement attached to the design that captures intended behavior, evaluated either dynamically during simulation or statically by formal verification.

\paragraph{Clocking block.}
The clock signal whose ticks define the temporal progression of an SVA property.

\paragraph{Cone of influence (COI).}
The minimal subset of design signals and statements whose values can affect the truth of a given assertion, used to limit the scope of formal analysis.

\paragraph{Counterexample.}
A concrete execution trace produced by the formal engine that demonstrates how an assertion can be violated.


\paragraph{Formal core.}
The minimal subset of design statements whose values are required to prove a given assertion, reported by VC~Formal's FC feature inside FPV.

\paragraph{FPV (Formal Property Verification).}
The application mode that proves SVA properties against the design over the entire reachable state space.

\paragraph{FTA (Formal Testbench Analyzer).}
The application mode that injects RTL mutations and reports which assertions detect each fault.

\paragraph{Implication operator.}
The operators \texttt{|->} (overlapped) and \texttt{|=>} (non-overlapped), which connect an antecedent to a consequent at the same or next clock tick.

\paragraph{Property.}
The full SVA construct combining a clocking block, an optional \texttt{disable iff}, and a Boolean or temporal expression.

\paragraph{RTL (Register-Transfer Level).}
The level of abstraction at which digital hardware is described as a network of registers and combinational logic, expressed in Verilog or SystemVerilog.

\paragraph{Sequence.}
A multi-cycle pattern of Boolean expressions composed using sequence operators, used as a building block inside properties.

\paragraph{Vacuity.}
A condition under which an assertion proves trivially because the antecedent is never satisfiable, reported by VC~Formal as terminal status \texttt{VACUOUS}.

%% file: appendix/F_Per_Model_Detailed_Results.tex
\section{Per-Model Detailed Results}
\label{app:per-model}

This appendix reports the full per-model breakdown of evaluation results referenced from Section~\ref{sec:eval}.

\subsection{Overall Metrics per Model (Main-Set Static Evaluation)}
\label{app:f-overall}

Table~\ref{tab:f-overall} reports the per-model metric from the main-set static evaluation. C1 is the strict module-level compile rate over all 342 main-set modules, where a module counts as compiled only if its assertion file compiles under every parameter combination. C2-P and C2-V are assertion-level micro-averages over generated assertions in evaluable runs. C4 is computed over eligible injected faults, and C5 reports union formal-core coverage.

\begin{table}[H]
\centering
\small
\caption{Per-model overall metrics in the main-set static evaluation. C1 is the strict module-level compile rate over all 342 modules. C2-P, C2-V, C4, and C5 are micro-averaged over evaluable runs. ``Asserts/run'' is the mean number of assertions emitted per evaluable run.}
\label{tab:f-overall}
\setlength{\tabcolsep}{6pt}
\begin{tabular}{lrrrrrr}
\toprule
Model & C1\% & C2-P\% & C2-V\%~($\downarrow$) & C4\% & C5\% & Asserts/run \\
\midrule
Claude Opus 4.7           & 87.7 & 87.5 & 1.8 & 70.5 & 34.7 & 10.6 \\
DeepSeek V4 Pro           & 67.8 & 86.3 & 0.3 & 84.4 & 37.4 & 24.2 \\
Gemini 3.1 Pro            & 84.2 & 90.3 & 1.5 & 74.7 & 33.8 & 8.0 \\
GLM 5.1                   & 76.0 & 91.0 & 1.3 & 78.1 & 43.3 & 27.4 \\
GPT-5.5                   & 84.5 & 76.8 & 0.5 & 84.5 & 51.4 & 22.0 \\
Kimi K2.6                 & 62.3 & 86.2 & 0.2 & 89.0 & 53.3 & 10.9 \\
MiniMax M2.7              & 19.0 & 58.2 & 15.4 & 37.8 & 26.5 & 7.6 \\
Qwen 3.6 Max              & 70.2 & 95.2 & 0.1 & 70.4 & 37.7 & 44.7 \\
\midrule
Claude Haiku 4.5          & 59.6 & 74.0 & 6.5 & 58.4 & 33.1 & 11.7 \\
DeepSeek V4 Flash         & 57.9 & 84.9 & 1.2 & 73.2 & 39.6 & 9.4 \\
GPT-5-mini                & 57.6 & 77.9 & 2.1 & 56.7 & 14.5 & 8.6 \\
Qwen 3.6 Plus             & 72.5 & 76.9 & 1.4 & 64.9 & 29.4 & 7.7 \\
\midrule
\emph{Cross-model mean} & 66.6 & 82.1 & 2.7 & 70.2 & 36.2 & -- \\
\bottomrule
\end{tabular}
\end{table}

\subsection{Formal Core Coverage Stratified by Hierarchy Depth (Main-Set Static Evaluation)}
\label{app:f-depth}

Table~\ref{tab:f-depth} reports per-model formal core coverage stratified by hierarchy depth.

\begin{table}[H]
\centering
\small
\caption{Per-model formal core coverage (\%) stratified by hierarchy depth (main-set static evaluation, micro-averaged within each depth bucket). Column L$k$ is depth~$k$. Entries marked ``--'' indicate that the model has no evaluable run at that depth. The cross-model mean averages only the models with a value. Depth~9 contains a single module.}
\label{tab:f-depth}
\setlength{\tabcolsep}{3.5pt}
\begin{tabular}{lrrrrrrrrrr}
\toprule
Model & L0 & L1 & L2 & L3 & L4 & L5 & L6 & L7 & L8 & L9 \\
\midrule
Claude Opus 4.7           & 82.6 & 76.5 & 72.9 & 56.4 & 55.1 & 37.3 & 34.6 & 4.8 & 25.0 & 17.0 \\
DeepSeek V4 Pro           & 80.6 & 57.6 & 28.6 & 30.7 & 51.6 & 12.0 & 34.6 & 20.3 & 1.6 & 52.5 \\
Gemini 3.1 Pro            & 90.6 & 68.4 & 75.8 & 51.9 & 55.6 & 33.1 & 14.7 & 5.4 & 5.7 & -- \\
GLM 5.1                   & 85.0 & 70.6 & 71.7 & 63.4 & 63.9 & 36.2 & 37.0 & 14.6 & 9.0 & 64.4 \\
GPT-5.5                   & 97.0 & 72.0 & 81.8 & 64.2 & 88.0 & 39.0 & 32.0 & 26.9 & 9.8 & 51.5 \\
Kimi K2.6                 & 72.9 & 72.0 & 66.4 & 69.9 & 69.7 & 21.3 & 24.8 & 11.2 & 6.7 & -- \\
MiniMax M2.7              & 23.1 & 52.8 & 0.0 & 40.2 & 32.0 & 5.3 & 6.7 & 21.4 & -- & -- \\
Qwen 3.6 Max              & 70.4 & 67.9 & 67.4 & 73.4 & 70.7 & 50.9 & 34.8 & 1.3 & 6.7 & -- \\
\midrule
Claude Haiku 4.5          & 73.9 & 48.1 & 17.6 & 25.6 & 49.1 & 21.7 & 20.0 & 27.2 & -- & 39.9 \\
DeepSeek V4 Flash         & 91.4 & 71.1 & 30.5 & 34.9 & 65.3 & 16.6 & 25.5 & 17.1 & 6.7 & -- \\
GPT-5-mini                & 63.1 & 20.8 & 23.0 & 27.4 & 29.0 & 20.0 & 19.6 & 4.5 & 0.0 & -- \\
Qwen 3.6 Plus             & 75.7 & 53.8 & 60.0 & 45.2 & 49.3 & 18.5 & 16.4 & 13.6 & 0.0 & 54.4 \\
\midrule
\emph{Cross-model mean} & 75.5 & 61.0 & 49.6 & 48.6 & 56.6 & 26.0 & 25.1 & 14.0 & 7.1 & 46.6 \\
\bottomrule
\end{tabular}
\end{table}

\subsection{Behavior Summary and Dilution Gap (Main-Set Static Evaluation)}
\label{app:f-behavior}

Table~\ref{tab:f-behavior} reports the exploratory dilution-gap analysis of Section~\ref{sec:eval-taxonomy}, with C2-P macro and the dilution gap, which is C2-P macro minus C2-P micro. The gap describes how the run-level and assertion-weighted non-vacuous proof rates differ.

\begin{table}[H]
\centering
\small
\caption{Per-model behavior summary. ``C1\%'' is the strict module-level compile rate. ``C2-P macro'' is the per-run non-vacuous proof rate averaged across evaluable runs. ``C2-P micro'' is the non-vacuous proof rate pooled across all assertions in evaluable runs. ``Dilution'' is macro$-$micro. ``Asserts/run'' is defined as in Table~\ref{tab:f-overall}.}
\label{tab:f-behavior}
\setlength{\tabcolsep}{6pt}
\begin{tabular}{lrrrrr}
\toprule
Model & C1\% & C2-P macro\% & C2-P micro\% & Dilution & Asserts/run \\
\midrule
Claude Opus 4.7           & 87.7 & 80.6 & 87.5 & $-$6.9 & 10.6 \\
DeepSeek V4 Pro           & 67.8 & 77.5 & 86.3 & $-$8.8 & 24.2 \\
Gemini 3.1 Pro            & 84.2 & 88.3 & 90.3 & $-$2.1 & 8.0 \\
GLM 5.1                   & 76.0 & 75.5 & 91.0 & $-$15.5 & 27.4 \\
GPT-5.5                   & 84.5 & 90.9 & 76.8 & $+$14.1 & 22.0 \\
Kimi K2.6                 & 62.3 & 84.0 & 86.2 & $-$2.1 & 10.9 \\
MiniMax M2.7              & 19.0 & 47.3 & 58.2 & $-$11.0 & 7.6 \\
Qwen 3.6 Max              & 70.2 & 78.2 & 95.2 & $-$17.0 & 44.7 \\
\midrule
Claude Haiku 4.5          & 59.6 & 61.1 & 74.0 & $-$12.9 & 11.7 \\
DeepSeek V4 Flash         & 57.9 & 72.6 & 84.9 & $-$12.3 & 9.4 \\
GPT-5-mini                & 57.6 & 63.7 & 77.9 & $-$14.2 & 8.6 \\
Qwen 3.6 Plus             & 72.5 & 69.8 & 76.9 & $-$7.1 & 7.7 \\
\bottomrule
\end{tabular}
\end{table}

\subsection{Per-Model Metrics Stratified by Hierarchy Depth (Main-Set Static Evaluation)}
\label{app:f-by-depth}

Tables~\ref{tab:f-d0} through~\ref{tab:f-d9} report the full metric from the main-set static evaluation per model. C1 is the strict module-level compile rate over the modules at that depth. An evaluation cell is one (module, parameter combination) pair. C2-P, C2-V, C4, and C5 are micro-averaged over the evaluable runs. \#Eval cells counts the evaluation cells of the model at that depth, so it exceeds the number of modules whenever a module has more than one parameter combination. A ``--'' marks a metric with an empty denominator, which occurs when the model has no evaluable run or, for C4, when its evaluable runs contain no eligible injected faults. C2-P counts only the assertions written by the LLM, whereas C4 and C5 are computed on the full verification environment of the module (Section~\ref{sec:b-c4}), which for depth $\geq 1$ also contains the obligations inherited from submodule contracts. A model can therefore have a C2-P of zero together with a nonzero C4 or C5. At depth~9, for example, the single assertion of Claude Opus~4.7 is falsified, and the inherited obligations detect the faults and supply the formal core. Inconclusive properties (Appendix~\ref{app:sva-primer}) also contribute the formal cores of their bounded proofs to C5. For this reason Kimi~K2.6 has a nonzero C5 at depth~8 although all of its assertions are inconclusive there. Over all modules and parameter combinations, the number of eligible injected faults is 4075, 3150, 1625, 3651, 1787, 809, 1714, 3778, 18, and 114 for depths 0 to 9, so the C4 values at depths~8 and~9 rest on few faults.

\begin{table}[H]
\centering
\small
\caption{Per-model metrics at hierarchy depth~0 in the main-set static evaluation. Depth~0 contains 150 modules.}
\label{tab:f-d0}
\setlength{\tabcolsep}{6pt}
\begin{tabular}{lrrrrrr}
\toprule
Model & C1\% & C2-P\% & C2-V\%~($\downarrow$) & C4\% & C5\% & \#Eval cells \\
\midrule
Claude Opus 4.7           & 88.7 & 87.2 & 3.4 & 90.5 & 82.6 & 270 \\
DeepSeek V4 Pro           & 70.0 & 92.3 & 0.0 & 95.2 & 80.6 & 270 \\
Gemini 3.1 Pro            & 94.7 & 92.2 & 3.5 & 92.1 & 90.6 & 270 \\
GLM 5.1                   & 81.3 & 96.2 & 0.4 & 93.5 & 85.0 & 270 \\
GPT-5.5                   & 92.0 & 57.8 & 0.0 & 91.5 & 97.0 & 270 \\
Kimi K2.6                 & 71.3 & 91.0 & 0.0 & 97.1 & 72.9 & 270 \\
MiniMax M2.7              & 27.3 & 54.8 & 23.8 & 18.8 & 23.1 & 270 \\
Qwen 3.6 Max              & 81.3 & 99.3 & 0.1 & 66.9 & 70.4 & 269 \\
\midrule
Claude Haiku 4.5          & 66.7 & 84.8 & 1.4 & 79.3 & 73.9 & 270 \\
DeepSeek V4 Flash         & 63.3 & 95.7 & 0.7 & 86.6 & 91.4 & 270 \\
GPT-5-mini                & 74.7 & 83.7 & 2.0 & 54.2 & 63.1 & 270 \\
Qwen 3.6 Plus             & 78.7 & 81.1 & 3.7 & 80.3 & 75.7 & 270 \\
\bottomrule
\end{tabular}
\end{table}

\begin{table}[H]
\centering
\small
\caption{Per-model metrics at hierarchy depth~1 in the main-set static evaluation. Depth~1 contains 61 modules.}
\label{tab:f-d1}
\setlength{\tabcolsep}{6pt}
\begin{tabular}{lrrrrrr}
\toprule
Model & C1\% & C2-P\% & C2-V\%~($\downarrow$) & C4\% & C5\% & \#Eval cells \\
\midrule
Claude Opus 4.7           & 83.6 & 94.3 & 0.2 & 79.2 & 76.5 & 114 \\
DeepSeek V4 Pro           & 75.4 & 95.3 & 0.0 & 84.3 & 57.6 & 114 \\
Gemini 3.1 Pro            & 88.5 & 96.9 & 0.2 & 81.3 & 68.4 & 114 \\
GLM 5.1                   & 75.4 & 90.0 & 0.2 & 89.1 & 70.6 & 114 \\
GPT-5.5                   & 65.6 & 95.5 & 0.0 & 92.3 & 72.0 & 114 \\
Kimi K2.6                 & 49.2 & 93.0 & 0.0 & 93.6 & 72.0 & 114 \\
MiniMax M2.7              & 18.0 & 40.5 & 6.8 & 46.1 & 52.8 & 114 \\
Qwen 3.6 Max              & 72.1 & 92.5 & 0.0 & 91.0 & 67.9 & 114 \\
\midrule
Claude Haiku 4.5          & 63.9 & 81.6 & 2.0 & 74.7 & 48.1 & 114 \\
DeepSeek V4 Flash         & 63.9 & 91.4 & 1.1 & 79.0 & 71.1 & 114 \\
GPT-5-mini                & 49.2 & 78.2 & 0.0 & 47.6 & 20.8 & 114 \\
Qwen 3.6 Plus             & 55.7 & 62.0 & 0.7 & 54.6 & 53.8 & 114 \\
\bottomrule
\end{tabular}
\end{table}

\begin{table}[H]
\centering
\small
\caption{Per-model metrics at hierarchy depth~2 in the main-set static evaluation. Depth~2 contains 27 modules.}
\label{tab:f-d2}
\setlength{\tabcolsep}{6pt}
\begin{tabular}{lrrrrrr}
\toprule
Model & C1\% & C2-P\% & C2-V\%~($\downarrow$) & C4\% & C5\% & \#Eval cells \\
\midrule
Claude Opus 4.7           & 92.6 & 89.7 & 0.0 & 62.5 & 72.9 & 57 \\
DeepSeek V4 Pro           & 59.3 & 81.4 & 0.6 & 82.5 & 28.6 & 57 \\
Gemini 3.1 Pro            & 77.8 & 80.0 & 0.0 & 55.2 & 75.8 & 57 \\
GLM 5.1                   & 66.7 & 77.6 & 2.0 & 85.6 & 71.7 & 57 \\
GPT-5.5                   & 77.8 & 94.5 & 1.3 & 98.4 & 81.8 & 57 \\
Kimi K2.6                 & 59.3 & 82.1 & 0.0 & 89.4 & 66.4 & 57 \\
MiniMax M2.7              & 3.7 & 69.2 & 23.1 & -- & 0.0 & 57 \\
Qwen 3.6 Max              & 59.3 & 88.8 & 0.7 & 86.9 & 67.4 & 57 \\
\midrule
Claude Haiku 4.5          & 51.9 & 49.0 & 9.2 & 30.9 & 17.6 & 57 \\
DeepSeek V4 Flash         & 59.3 & 50.8 & 4.5 & 63.2 & 30.5 & 57 \\
GPT-5-mini                & 48.1 & 76.3 & 4.6 & 75.3 & 23.0 & 57 \\
Qwen 3.6 Plus             & 81.5 & 65.9 & 0.9 & 54.2 & 60.0 & 57 \\
\bottomrule
\end{tabular}
\end{table}

\begin{table}[H]
\centering
\small
\caption{Per-model metrics at hierarchy depth~3 in the main-set static evaluation. Depth~3 contains 36 modules.}
\label{tab:f-d3}
\setlength{\tabcolsep}{6pt}
\begin{tabular}{lrrrrrr}
\toprule
Model & C1\% & C2-P\% & C2-V\%~($\downarrow$) & C4\% & C5\% & \#Eval cells \\
\midrule
Claude Opus 4.7           & 94.4 & 93.6 & 3.0 & 72.5 & 56.4 & 94 \\
DeepSeek V4 Pro           & 61.1 & 75.5 & 1.3 & 77.8 & 30.7 & 94 \\
Gemini 3.1 Pro            & 80.6 & 91.0 & 0.3 & 59.9 & 51.9 & 94 \\
GLM 5.1                   & 69.4 & 90.3 & 4.6 & 80.6 & 63.4 & 94 \\
GPT-5.5                   & 86.1 & 93.3 & 2.2 & 86.0 & 64.2 & 94 \\
Kimi K2.6                 & 58.3 & 86.8 & 0.8 & 84.1 & 69.9 & 94 \\
MiniMax M2.7              & 8.3 & 67.2 & 15.5 & 25.9 & 40.2 & 94 \\
Qwen 3.6 Max              & 55.6 & 64.1 & 0.4 & 61.6 & 73.4 & 94 \\
\midrule
Claude Haiku 4.5          & 58.3 & 28.4 & 33.1 & 54.5 & 25.6 & 94 \\
DeepSeek V4 Flash         & 47.2 & 71.5 & 2.7 & 66.4 & 34.9 & 94 \\
GPT-5-mini                & 44.4 & 82.3 & 0.2 & 54.1 & 27.4 & 94 \\
Qwen 3.6 Plus             & 66.7 & 86.6 & 0.6 & 72.4 & 45.2 & 94 \\
\bottomrule
\end{tabular}
\end{table}

\begin{table}[H]
\centering
\small
\caption{Per-model metrics at hierarchy depth~4 in the main-set static evaluation. Depth~4 contains 25 modules.}
\label{tab:f-d4}
\setlength{\tabcolsep}{6pt}
\begin{tabular}{lrrrrrr}
\toprule
Model & C1\% & C2-P\% & C2-V\%~($\downarrow$) & C4\% & C5\% & \#Eval cells \\
\midrule
Claude Opus 4.7           & 92.0 & 94.2 & 0.0 & 64.0 & 55.1 & 53 \\
DeepSeek V4 Pro           & 56.0 & 77.0 & 2.3 & 73.0 & 51.6 & 53 \\
Gemini 3.1 Pro            & 72.0 & 98.0 & 1.5 & 62.5 & 55.6 & 53 \\
GLM 5.1                   & 64.0 & 88.6 & 0.0 & 81.9 & 63.9 & 53 \\
GPT-5.5                   & 80.0 & 96.5 & 0.0 & 87.0 & 88.0 & 53 \\
Kimi K2.6                 & 68.0 & 94.2 & 0.0 & 86.3 & 69.7 & 53 \\
MiniMax M2.7              & 4.0 & 50.0 & 50.0 & 0.0 & 32.0 & 53 \\
Qwen 3.6 Max              & 64.0 & 69.0 & 0.0 & 45.1 & 70.7 & 53 \\
\midrule
Claude Haiku 4.5          & 56.0 & 15.1 & 37.5 & 36.1 & 49.1 & 53 \\
DeepSeek V4 Flash         & 48.0 & 77.7 & 0.0 & 92.4 & 65.3 & 53 \\
GPT-5-mini                & 36.0 & 72.1 & 0.0 & 71.3 & 29.0 & 53 \\
Qwen 3.6 Plus             & 76.0 & 87.8 & 2.3 & 66.6 & 49.3 & 53 \\
\bottomrule
\end{tabular}
\end{table}

\begin{table}[H]
\centering
\small
\caption{Per-model metrics at hierarchy depth~5 in the main-set static evaluation. Depth~5 contains 10 modules.}
\label{tab:f-d5}
\setlength{\tabcolsep}{6pt}
\begin{tabular}{lrrrrrr}
\toprule
Model & C1\% & C2-P\% & C2-V\%~($\downarrow$) & C4\% & C5\% & \#Eval cells \\
\midrule
Claude Opus 4.7           & 90.0 & 82.8 & 0.0 & 71.6 & 37.3 & 26 \\
DeepSeek V4 Pro           & 70.0 & 82.6 & 0.0 & 66.2 & 12.0 & 26 \\
Gemini 3.1 Pro            & 70.0 & 91.7 & 1.7 & 68.5 & 33.1 & 26 \\
GLM 5.1                   & 70.0 & 86.7 & 0.0 & 86.2 & 36.2 & 26 \\
GPT-5.5                   & 100.0 & 90.3 & 0.2 & 93.1 & 39.0 & 26 \\
Kimi K2.6                 & 50.0 & 89.9 & 0.0 & 94.4 & 21.3 & 26 \\
MiniMax M2.7              & 20.0 & 81.0 & 0.0 & 69.6 & 5.3 & 26 \\
Qwen 3.6 Max              & 60.0 & 70.3 & 0.0 & 58.5 & 50.9 & 26 \\
\midrule
Claude Haiku 4.5          & 40.0 & 59.1 & 0.0 & 54.8 & 21.7 & 26 \\
DeepSeek V4 Flash         & 30.0 & 10.0 & 10.0 & 61.2 & 16.6 & 26 \\
GPT-5-mini                & 30.0 & 74.4 & 20.5 & 67.5 & 20.0 & 26 \\
Qwen 3.6 Plus             & 90.0 & 84.0 & 0.0 & 61.1 & 18.5 & 26 \\
\bottomrule
\end{tabular}
\end{table}

\begin{table}[H]
\centering
\small
\caption{Per-model metrics at hierarchy depth~6 in the main-set static evaluation. Depth~6 contains 20 modules.}
\label{tab:f-d6}
\setlength{\tabcolsep}{6pt}
\begin{tabular}{lrrrrrr}
\toprule
Model & C1\% & C2-P\% & C2-V\%~($\downarrow$) & C4\% & C5\% & \#Eval cells \\
\midrule
Claude Opus 4.7           & 65.0 & 66.2 & 0.5 & 62.4 & 34.6 & 54 \\
DeepSeek V4 Pro           & 60.0 & 53.6 & 1.0 & 77.4 & 34.6 & 54 \\
Gemini 3.1 Pro            & 35.0 & 80.5 & 0.0 & 60.8 & 14.7 & 54 \\
GLM 5.1                   & 75.0 & 65.7 & 0.0 & 84.7 & 37.0 & 54 \\
GPT-5.5                   & 85.0 & 74.8 & 0.0 & 85.0 & 32.0 & 54 \\
Kimi K2.6                 & 50.0 & 72.0 & 0.0 & 74.8 & 24.8 & 54 \\
MiniMax M2.7              & 15.0 & 75.0 & 20.4 & 49.6 & 6.7 & 54 \\
Qwen 3.6 Max              & 55.0 & 45.2 & 0.6 & 56.5 & 34.8 & 54 \\
\midrule
Claude Haiku 4.5          & 20.0 & 88.3 & 0.0 & 61.4 & 20.0 & 54 \\
DeepSeek V4 Flash         & 50.0 & 66.4 & 0.0 & 53.7 & 25.5 & 54 \\
GPT-5-mini                & 45.0 & 63.4 & 13.1 & 52.3 & 19.6 & 54 \\
Qwen 3.6 Plus             & 50.0 & 46.0 & 0.0 & 59.8 & 16.4 & 54 \\
\bottomrule
\end{tabular}
\end{table}

\begin{table}[H]
\centering
\small
\caption{Per-model metrics at hierarchy depth~7 in the main-set static evaluation. Depth~7 contains 10 modules.}
\label{tab:f-d7}
\setlength{\tabcolsep}{6pt}
\begin{tabular}{lrrrrrr}
\toprule
Model & C1\% & C2-P\% & C2-V\%~($\downarrow$) & C4\% & C5\% & \#Eval cells \\
\midrule
Claude Opus 4.7           & 90.0 & 71.8 & 0.0 & 39.3 & 4.8 & 37 \\
DeepSeek V4 Pro           & 70.0 & 78.6 & 0.0 & 81.2 & 20.3 & 37 \\
Gemini 3.1 Pro            & 80.0 & 96.8 & 1.6 & 73.3 & 5.4 & 37 \\
GLM 5.1                   & 80.0 & 89.6 & 2.1 & 48.6 & 14.6 & 37 \\
GPT-5.5                   & 90.0 & 94.8 & 0.0 & 66.9 & 26.9 & 37 \\
Kimi K2.6                 & 60.0 & 90.8 & 0.0 & 67.0 & 11.2 & 37 \\
MiniMax M2.7              & 30.0 & 44.8 & 0.0 & 56.1 & 21.4 & 37 \\
Qwen 3.6 Max              & 40.0 & 87.6 & 0.0 & 55.7 & 1.3 & 37 \\
\midrule
Claude Haiku 4.5          & 70.0 & 81.0 & 1.1 & 61.3 & 27.2 & 37 \\
DeepSeek V4 Flash         & 50.0 & 75.0 & 0.0 & 67.9 & 17.1 & 37 \\
GPT-5-mini                & 40.0 & 83.7 & 0.0 & 73.5 & 4.5 & 37 \\
Qwen 3.6 Plus             & 90.0 & 88.4 & 0.0 & 51.7 & 13.6 & 37 \\
\bottomrule
\end{tabular}
\end{table}

\begin{table}[H]
\centering
\small
\caption{Per-model metrics at hierarchy depth~8 in the main-set static evaluation. Depth~8 contains 2 modules.}
\label{tab:f-d8}
\setlength{\tabcolsep}{6pt}
\begin{tabular}{lrrrrrr}
\toprule
Model & C1\% & C2-P\% & C2-V\%~($\downarrow$) & C4\% & C5\% & \#Eval cells \\
\midrule
Claude Opus 4.7           & 100.0 & 19.5 & 0.0 & 100.0 & 25.0 & 5 \\
DeepSeek V4 Pro           & 100.0 & 7.9 & 0.0 & 100.0 & 1.6 & 5 \\
Gemini 3.1 Pro            & 100.0 & 3.3 & 0.0 & 100.0 & 5.7 & 5 \\
GLM 5.1                   & 100.0 & 24.6 & 0.0 & 100.0 & 9.0 & 5 \\
GPT-5.5                   & 100.0 & 32.0 & 0.0 & 100.0 & 9.8 & 5 \\
Kimi K2.6                 & 50.0 & 0.0 & 0.0 & -- & 6.7 & 5 \\
MiniMax M2.7              & 0.0 & -- & -- & -- & -- & 5 \\
Qwen 3.6 Max              & 50.0 & 0.0 & 0.0 & -- & 6.7 & 5 \\
\midrule
Claude Haiku 4.5          & 0.0 & -- & -- & -- & -- & 5 \\
DeepSeek V4 Flash         & 50.0 & 0.0 & 0.0 & -- & 6.7 & 5 \\
GPT-5-mini                & 50.0 & 0.0 & 0.0 & -- & 0.0 & 5 \\
Qwen 3.6 Plus             & 100.0 & 21.6 & 0.0 & 100.0 & 0.0 & 5 \\
\bottomrule
\end{tabular}
\end{table}

\begin{table}[H]
\centering
\small
\caption{Per-model metrics at hierarchy depth~9 in the main-set static evaluation. Depth~9 contains 1 module. Only six of the twelve models produce an evaluable run on it, so this depth is not a statistically meaningful bucket.}
\label{tab:f-d9}
\setlength{\tabcolsep}{6pt}
\begin{tabular}{lrrrrrr}
\toprule
Model & C1\% & C2-P\% & C2-V\%~($\downarrow$) & C4\% & C5\% & \#Eval cells \\
\midrule
Claude Opus 4.7           & 100.0 & 0.0 & 0.0 & 64.5 & 17.0 & 1 \\
DeepSeek V4 Pro           & 100.0 & 100.0 & 0.0 & 77.5 & 52.5 & 1 \\
Gemini 3.1 Pro            & 0.0 & -- & -- & -- & -- & 1 \\
GLM 5.1                   & 100.0 & 75.0 & 0.0 & 40.0 & 64.4 & 1 \\
GPT-5.5                   & 100.0 & 94.4 & 0.0 & 63.2 & 51.5 & 1 \\
Kimi K2.6                 & 0.0 & -- & -- & -- & -- & 1 \\
MiniMax M2.7              & 0.0 & -- & -- & -- & -- & 1 \\
Qwen 3.6 Max              & 0.0 & -- & -- & -- & -- & 1 \\
\midrule
Claude Haiku 4.5          & 100.0 & 10.0 & 0.0 & 64.7 & 39.9 & 1 \\
DeepSeek V4 Flash         & 0.0 & -- & -- & -- & -- & 1 \\
GPT-5-mini                & 0.0 & -- & -- & -- & -- & 1 \\
Qwen 3.6 Plus             & 100.0 & 30.8 & 0.0 & 95.5 & 54.4 & 1 \\
\bottomrule
\end{tabular}
\end{table}

\subsection{Per-Model Metrics Stratified by Module Category (Main-Set Static Evaluation)}
\label{app:f-by-category}

Tables~\ref{tab:f-cat-async} through~\ref{tab:f-cat-tag} report the full metric vector from the main-set static evaluation per model, stratified by BaseJump STL module category. The columns are defined as in Appendix~\ref{app:f-by-depth}.

\begin{table}[H]
\centering
\small
\caption{Per-model metrics for category \texttt{bsg\_async} in the main-set static evaluation. The category contains 45 modules.}
\label{tab:f-cat-async}
\setlength{\tabcolsep}{6pt}
\begin{tabular}{lrrrrrr}
\toprule
Model & C1\% & C2-P\% & C2-V\%~($\downarrow$) & C4\% & C5\% & \#Eval cells \\
\midrule
Claude Opus 4.7           & 91.1 & 80.2 & 14.3 & 97.3 & 69.6 & 66 \\
DeepSeek V4 Pro           & 53.3 & 67.7 & 0.8 & 96.9 & 60.7 & 66 \\
Gemini 3.1 Pro            & 93.3 & 88.2 & 11.8 & 94.6 & 70.3 & 66 \\
GLM 5.1                   & 82.2 & 77.1 & 6.8 & 99.4 & 88.1 & 66 \\
GPT-5.5                   & 88.9 & 96.5 & 1.3 & 100.0 & 59.0 & 66 \\
Kimi K2.6                 & 40.0 & 91.9 & 0.0 & 89.2 & 80.5 & 66 \\
MiniMax M2.7              & 28.9 & 46.8 & 22.6 & 33.3 & 29.0 & 66 \\
Qwen 3.6 Max              & 73.3 & 79.7 & 0.0 & 94.8 & 70.8 & 66 \\
\midrule
Claude Haiku 4.5          & 71.1 & 55.0 & 11.2 & 70.4 & 56.5 & 66 \\
DeepSeek V4 Flash         & 48.9 & 68.9 & 9.0 & 73.7 & 48.8 & 66 \\
GPT-5-mini                & 80.0 & 29.4 & 0.0 & 92.0 & 80.4 & 66 \\
Qwen 3.6 Plus             & 73.3 & 51.5 & 0.0 & 90.3 & 49.9 & 66 \\
\bottomrule
\end{tabular}
\end{table}

\begin{table}[H]
\centering
\small
\caption{Per-model metrics for category \texttt{bsg\_cache} in the main-set static evaluation. The category contains 30 modules.}
\label{tab:f-cat-cache}
\setlength{\tabcolsep}{6pt}
\begin{tabular}{lrrrrrr}
\toprule
Model & C1\% & C2-P\% & C2-V\%~($\downarrow$) & C4\% & C5\% & \#Eval cells \\
\midrule
Claude Opus 4.7           & 93.3 & 95.4 & 2.4 & 41.1 & 7.1 & 47 \\
DeepSeek V4 Pro           & 53.3 & 79.4 & 1.5 & 73.5 & 17.9 & 47 \\
Gemini 3.1 Pro            & 76.7 & 98.3 & 0.5 & 47.4 & 6.1 & 47 \\
GLM 5.1                   & 76.7 & 92.4 & 0.2 & 59.1 & 18.7 & 47 \\
GPT-5.5                   & 93.3 & 96.9 & 0.5 & 71.2 & 36.0 & 47 \\
Kimi K2.6                 & 43.3 & 96.3 & 0.0 & 86.7 & 12.9 & 47 \\
MiniMax M2.7              & 13.3 & 32.4 & 60.2 & 39.6 & 25.7 & 47 \\
Qwen 3.6 Max              & 66.7 & 84.3 & 0.6 & 46.7 & 6.4 & 47 \\
\midrule
Claude Haiku 4.5          & 40.0 & 86.1 & 1.7 & 31.7 & 39.1 & 47 \\
DeepSeek V4 Flash         & 43.3 & 77.3 & 7.3 & 61.0 & 24.8 & 47 \\
GPT-5-mini                & 50.0 & 75.7 & 13.1 & 46.5 & 4.9 & 47 \\
Qwen 3.6 Plus             & 63.3 & 87.2 & 0.0 & 44.7 & 14.3 & 47 \\
\bottomrule
\end{tabular}
\end{table}

\begin{table}[H]
\centering
\small
\caption{Per-model metrics for category \texttt{bsg\_dataflow} in the main-set static evaluation. The category contains 60 modules.}
\label{tab:f-cat-dataflow}
\setlength{\tabcolsep}{6pt}
\begin{tabular}{lrrrrrr}
\toprule
Model & C1\% & C2-P\% & C2-V\%~($\downarrow$) & C4\% & C5\% & \#Eval cells \\
\midrule
Claude Opus 4.7           & 85.0 & 88.3 & 4.1 & 82.2 & 33.6 & 113 \\
DeepSeek V4 Pro           & 73.3 & 70.3 & 1.0 & 80.3 & 32.4 & 113 \\
Gemini 3.1 Pro            & 83.3 & 96.5 & 0.3 & 79.7 & 31.0 & 113 \\
GLM 5.1                   & 75.0 & 86.1 & 0.3 & 87.8 & 52.7 & 113 \\
GPT-5.5                   & 75.0 & 92.1 & 2.1 & 94.0 & 51.1 & 113 \\
Kimi K2.6                 & 61.7 & 82.6 & 1.5 & 90.2 & 57.8 & 113 \\
MiniMax M2.7              & 18.3 & 41.9 & 8.1 & 24.2 & 13.7 & 113 \\
Qwen 3.6 Max              & 58.3 & 57.0 & 1.2 & 82.2 & 61.1 & 112 \\
\midrule
Claude Haiku 4.5          & 63.3 & 49.2 & 0.5 & 63.7 & 30.7 & 113 \\
DeepSeek V4 Flash         & 61.7 & 79.8 & 0.0 & 76.3 & 29.4 & 113 \\
GPT-5-mini                & 55.0 & 64.9 & 2.2 & 67.9 & 20.1 & 113 \\
Qwen 3.6 Plus             & 75.0 & 78.8 & 0.8 & 70.6 & 38.9 & 113 \\
\bottomrule
\end{tabular}
\end{table}

\begin{table}[H]
\centering
\small
\caption{Per-model metrics for category \texttt{bsg\_link} in the main-set static evaluation. The category contains 13 modules.}
\label{tab:f-cat-link}
\setlength{\tabcolsep}{6pt}
\begin{tabular}{lrrrrrr}
\toprule
Model & C1\% & C2-P\% & C2-V\%~($\downarrow$) & C4\% & C5\% & \#Eval cells \\
\midrule
Claude Opus 4.7           & 84.6 & 34.8 & 0.7 & 64.2 & 12.0 & 49 \\
DeepSeek V4 Pro           & 84.6 & 53.1 & 1.6 & 85.9 & 8.8 & 49 \\
Gemini 3.1 Pro            & 69.2 & 45.9 & 4.5 & 67.2 & 12.0 & 49 \\
GLM 5.1                   & 92.3 & 63.2 & 3.1 & 80.6 & 15.4 & 49 \\
GPT-5.5                   & 92.3 & 76.5 & 0.0 & 79.1 & 14.0 & 49 \\
Kimi K2.6                 & 61.5 & 51.9 & 0.0 & 58.6 & 3.8 & 49 \\
MiniMax M2.7              & 15.4 & 91.7 & 0.0 & 69.0 & 0.0 & 49 \\
Qwen 3.6 Max              & 53.8 & 43.1 & 0.0 & 74.7 & 2.0 & 49 \\
\midrule
Claude Haiku 4.5          & 84.6 & 54.2 & 21.5 & 53.4 & 6.5 & 49 \\
DeepSeek V4 Flash         & 46.2 & 57.1 & 0.0 & 56.4 & 6.8 & 49 \\
GPT-5-mini                & 69.2 & 39.0 & 0.0 & 66.6 & 4.2 & 49 \\
Qwen 3.6 Plus             & 92.3 & 57.6 & 2.2 & 69.4 & 7.0 & 49 \\
\bottomrule
\end{tabular}
\end{table}

\begin{table}[H]
\centering
\small
\caption{Per-model metrics for category \texttt{bsg\_mem} in the main-set static evaluation. The category contains 51 modules.}
\label{tab:f-cat-mem}
\setlength{\tabcolsep}{6pt}
\begin{tabular}{lrrrrrr}
\toprule
Model & C1\% & C2-P\% & C2-V\%~($\downarrow$) & C4\% & C5\% & \#Eval cells \\
\midrule
Claude Opus 4.7           & 86.3 & 88.0 & 0.0 & 93.7 & 75.3 & 146 \\
DeepSeek V4 Pro           & 56.9 & 65.1 & 0.0 & 84.8 & 61.4 & 146 \\
Gemini 3.1 Pro            & 78.4 & 86.8 & 0.0 & 93.5 & 76.8 & 146 \\
GLM 5.1                   & 70.6 & 84.1 & 2.4 & 95.2 & 72.6 & 146 \\
GPT-5.5                   & 86.3 & 85.0 & 0.0 & 99.2 & 82.3 & 146 \\
Kimi K2.6                 & 64.7 & 88.0 & 0.0 & 93.7 & 74.9 & 146 \\
MiniMax M2.7              & 13.7 & 51.8 & 13.6 & 73.7 & 41.4 & 146 \\
Qwen 3.6 Max              & 47.1 & 63.6 & 0.7 & 94.0 & 77.7 & 146 \\
\midrule
Claude Haiku 4.5          & 39.2 & 52.8 & 13.2 & 85.3 & 48.5 & 146 \\
DeepSeek V4 Flash         & 52.9 & 51.9 & 0.3 & 91.8 & 64.8 & 146 \\
GPT-5-mini                & 51.0 & 81.6 & 3.4 & 55.0 & 31.6 & 146 \\
Qwen 3.6 Plus             & 60.8 & 76.5 & 2.7 & 80.0 & 62.8 & 146 \\
\bottomrule
\end{tabular}
\end{table}

\begin{table}[H]
\centering
\small
\caption{Per-model metrics for category \texttt{bsg\_misc} in the main-set static evaluation. The category contains 117 modules.}
\label{tab:f-cat-misc}
\setlength{\tabcolsep}{6pt}
\begin{tabular}{lrrrrrr}
\toprule
Model & C1\% & C2-P\% & C2-V\%~($\downarrow$) & C4\% & C5\% & \#Eval cells \\
\midrule
Claude Opus 4.7           & 88.0 & 94.9 & 1.5 & 78.3 & 51.8 & 228 \\
DeepSeek V4 Pro           & 81.2 & 81.3 & 0.4 & 91.1 & 50.5 & 228 \\
Gemini 3.1 Pro            & 93.2 & 96.6 & 2.8 & 79.2 & 53.7 & 228 \\
GLM 5.1                   & 75.2 & 90.8 & 3.1 & 85.5 & 79.4 & 228 \\
GPT-5.5                   & 87.2 & 56.0 & 0.2 & 92.2 & 64.2 & 228 \\
Kimi K2.6                 & 76.9 & 93.8 & 0.0 & 95.2 & 85.6 & 228 \\
MiniMax M2.7              & 22.2 & 61.6 & 11.9 & 24.7 & 47.0 & 228 \\
Qwen 3.6 Max              & 88.9 & 97.6 & 0.2 & 69.9 & 80.2 & 228 \\
\midrule
Claude Haiku 4.5          & 66.7 & 82.3 & 5.4 & 53.8 & 35.2 & 228 \\
DeepSeek V4 Flash         & 68.4 & 96.1 & 0.8 & 66.8 & 90.1 & 228 \\
GPT-5-mini                & 58.1 & 88.8 & 0.3 & 43.0 & 53.1 & 228 \\
Qwen 3.6 Plus             & 78.6 & 87.0 & 1.4 & 71.0 & 50.0 & 228 \\
\bottomrule
\end{tabular}
\end{table}

\begin{table}[H]
\centering
\small
\caption{Per-model metrics for category \texttt{bsg\_noc} in the main-set static evaluation. The category contains 20 modules.}
\label{tab:f-cat-noc}
\setlength{\tabcolsep}{6pt}
\begin{tabular}{lrrrrrr}
\toprule
Model & C1\% & C2-P\% & C2-V\%~($\downarrow$) & C4\% & C5\% & \#Eval cells \\
\midrule
Claude Opus 4.7           & 90.0 & 84.5 & 0.4 & 81.2 & 38.8 & 52 \\
DeepSeek V4 Pro           & 55.0 & 95.6 & 0.0 & 82.6 & 53.8 & 52 \\
Gemini 3.1 Pro            & 65.0 & 98.0 & 0.0 & 81.8 & 51.3 & 52 \\
GLM 5.1                   & 75.0 & 97.8 & 0.6 & 77.8 & 40.8 & 52 \\
GPT-5.5                   & 70.0 & 79.5 & 0.0 & 89.6 & 46.5 & 52 \\
Kimi K2.6                 & 60.0 & 82.9 & 0.0 & 74.2 & 43.8 & 52 \\
MiniMax M2.7              & 10.0 & 88.8 & 0.0 & 10.0 & 0.0 & 52 \\
Qwen 3.6 Max              & 75.0 & 99.2 & 0.0 & 83.2 & 48.0 & 52 \\
\midrule
Claude Haiku 4.5          & 55.0 & 54.6 & 15.5 & 81.8 & 19.4 & 52 \\
DeepSeek V4 Flash         & 50.0 & 83.2 & 0.0 & 80.7 & 39.6 & 52 \\
GPT-5-mini                & 45.0 & 92.3 & 0.0 & 63.3 & 51.8 & 52 \\
Qwen 3.6 Plus             & 65.0 & 53.6 & 0.6 & 84.9 & 35.2 & 52 \\
\bottomrule
\end{tabular}
\end{table}

\begin{table}[H]
\centering
\small
\caption{Per-model metrics for category \texttt{bsg\_tag} in the main-set static evaluation. The category contains 6 modules.}
\label{tab:f-cat-tag}
\setlength{\tabcolsep}{6pt}
\begin{tabular}{lrrrrrr}
\toprule
Model & C1\% & C2-P\% & C2-V\%~($\downarrow$) & C4\% & C5\% & \#Eval cells \\
\midrule
Claude Opus 4.7           & 66.7 & 66.7 & 0.0 & 27.0 & 14.3 & 10 \\
DeepSeek V4 Pro           & 33.3 & 77.8 & 0.0 & 100.0 & 100.0 & 10 \\
Gemini 3.1 Pro            & 33.3 & 100.0 & 0.0 & 25.0 & 4.9 & 10 \\
GLM 5.1                   & 66.7 & 44.8 & 0.0 & 56.9 & 15.3 & 10 \\
GPT-5.5                   & 66.7 & 87.5 & 0.0 & 69.8 & 27.0 & 10 \\
Kimi K2.6                 & 33.3 & 57.1 & 0.0 & 100.0 & 80.0 & 10 \\
MiniMax M2.7              & 0.0 & 36.4 & 0.0 & 4.4 & 35.5 & 10 \\
Qwen 3.6 Max              & 33.3 & 61.1 & 0.0 & 43.0 & 22.8 & 10 \\
\midrule
Claude Haiku 4.5          & 33.3 & 81.8 & 0.0 & 63.6 & 0.0 & 10 \\
DeepSeek V4 Flash         & 50.0 & 20.0 & 0.0 & 100.0 & 13.0 & 10 \\
GPT-5-mini                & 16.7 & 100.0 & 0.0 & 46.1 & 16.3 & 10 \\
Qwen 3.6 Plus             & 50.0 & 79.3 & 0.0 & 20.6 & 3.8 & 10 \\
\bottomrule
\end{tabular}
\end{table}

\subsection{Per-Iteration Trajectory per Model (Agentic Evaluation)}
\label{app:f-t4-iter}

Tables~\ref{tab:f-t4-claude-opus} through~\ref{tab:f-t4-gpt-5-mini} report the per-iteration trajectory of C1, C2-P, C2-V, C4, and C5 for each of the five models in the agentic evaluation on the fixed 40-module subset. Iteration~0 is the initial assertion file generated in agentic mode, and iteration~$k$ is the revision after $k$ rounds of VC~Formal feedback. Every parameter combination of a module is evaluated at every iteration. C1 is the strict module-level compile rate over the 40 modules, and C2-P, C2-V, C4, and C5 are micro-averaged over the evaluable runs, as in the main-set static evaluation. A module whose loop stopped before iteration~$k$, because the LLM declared convergence or the loop aborted, contributes its last generated assertion file at iteration~$k$. The set of evaluable runs changes between iterations, so C2-P, C2-V, C4, and C5 are not computed over a fixed set of modules. The static row reports the main-set static evaluation on the same 40 modules. That row comes from a separate LLM call rather than from iteration~0, so the two rows can differ even at temperature~0 (Section~\ref{sec:eval-setup}).

\begin{table}[H]
\centering
\small
\caption{Per-iteration trajectory for Claude Opus 4.7 on the 40-module agentic subset.}
\label{tab:f-t4-claude-opus}
\setlength{\tabcolsep}{6pt}
\begin{tabular}{lrrrrr}
\toprule
Iteration & C1\% & C2-P\% & C2-V\%~($\downarrow$) & C4\% & C5\% \\
\midrule
\emph{Static (separate call)} & 95.0 & 61.6 & 0.1 & 42.8 & 7.7 \\
\midrule
Iter~0 & 82.5 & 64.9 & 0.0 & 62.9 & 54.0 \\
Iter~1 & 87.5 & 88.7 & 0.0 & 46.4 & 8.8 \\
Iter~2 & 80.0 & 85.5 & 0.4 & 50.0 & 33.6 \\
Iter~3 & 92.5 & 93.9 & 0.1 & 74.7 & 30.8 \\
Iter~4 & 75.0 & 69.3 & 0.0 & 51.4 & 34.2 \\
Iter~5 & 82.5 & 88.7 & 0.2 & 49.6 & 33.8 \\
Iter~6 & 82.5 & 98.8 & 0.1 & 57.8 & 39.8 \\
\bottomrule
\end{tabular}
\end{table}

\begin{table}[H]
\centering
\small
\caption{Per-iteration trajectory for DeepSeek V4 Pro on the 40-module agentic subset.}
\label{tab:f-t4-deepseek-pro}
\setlength{\tabcolsep}{6pt}
\begin{tabular}{lrrrrr}
\toprule
Iteration & C1\% & C2-P\% & C2-V\%~($\downarrow$) & C4\% & C5\% \\
\midrule
\emph{Static (separate call)} & 60.0 & 54.4 & 0.9 & 78.5 & 22.3 \\
\midrule
Iter~0 & 55.0 & 48.2 & 4.5 & 94.3 & 85.3 \\
Iter~1 & 77.5 & 48.9 & 10.0 & 63.6 & 48.9 \\
Iter~2 & 82.5 & 57.5 & 10.3 & 85.9 & 67.6 \\
Iter~3 & 82.5 & 49.3 & 13.5 & 100.0 & 61.5 \\
Iter~4 & 77.5 & 60.7 & 6.6 & 95.2 & 68.5 \\
Iter~5 & 72.5 & 73.0 & 0.4 & 81.3 & 57.3 \\
Iter~6 & 75.0 & 76.5 & 0.1 & 81.2 & 63.8 \\
\bottomrule
\end{tabular}
\end{table}

\begin{table}[H]
\centering
\small
\caption{Per-iteration trajectory for Gemini 3.1 Pro on the 40-module agentic subset.}
\label{tab:f-t4-gemini-pro}
\setlength{\tabcolsep}{6pt}
\begin{tabular}{lrrrrr}
\toprule
Iteration & C1\% & C2-P\% & C2-V\%~($\downarrow$) & C4\% & C5\% \\
\midrule
\emph{Static (separate call)} & 77.5 & 65.9 & 0.7 & 53.8 & 6.0 \\
\midrule
Iter~0 & 75.0 & 62.4 & 0.0 & 47.4 & 29.3 \\
Iter~1 & 85.0 & 70.3 & 0.0 & 90.2 & 41.2 \\
Iter~2 & 77.5 & 45.2 & 0.9 & 80.7 & 25.4 \\
Iter~3 & 85.0 & 52.9 & 0.3 & 78.0 & 7.0 \\
Iter~4 & 82.5 & 54.5 & 0.4 & 47.4 & 5.9 \\
Iter~5 & 75.0 & 85.1 & 0.1 & 55.2 & 7.4 \\
Iter~6 & 85.0 & 90.5 & 0.4 & 54.3 & 6.5 \\
\bottomrule
\end{tabular}
\end{table}

\begin{table}[H]
\centering
\small
\caption{Per-iteration trajectory for GPT-5.5 on the 40-module agentic subset.}
\label{tab:f-t4-gpt-5p5}
\setlength{\tabcolsep}{6pt}
\begin{tabular}{lrrrrr}
\toprule
Iteration & C1\% & C2-P\% & C2-V\%~($\downarrow$) & C4\% & C5\% \\
\midrule
\emph{Static (separate call)} & 80.0 & 77.6 & 0.2 & 68.5 & 31.9 \\
\midrule
Iter~0 & 77.5 & 75.5 & 1.7 & 78.4 & 21.6 \\
Iter~1 & 100.0 & 72.4 & 0.1 & 97.5 & 38.2 \\
Iter~2 & 90.0 & 96.7 & 0.0 & 98.7 & 54.6 \\
Iter~3 & 92.5 & 80.6 & 0.0 & 87.7 & 69.6 \\
Iter~4 & 95.0 & 90.6 & 0.1 & 89.4 & 33.5 \\
Iter~5 & 100.0 & 84.4 & 0.0 & 98.1 & 41.6 \\
Iter~6 & 97.5 & 88.8 & 0.1 & 86.0 & 47.5 \\
\bottomrule
\end{tabular}
\end{table}

\begin{table}[H]
\centering
\small
\caption{Per-iteration trajectory for GPT-5-mini on the 40-module agentic subset.}
\label{tab:f-t4-gpt-5-mini}
\setlength{\tabcolsep}{6pt}
\begin{tabular}{lrrrrr}
\toprule
Iteration & C1\% & C2-P\% & C2-V\%~($\downarrow$) & C4\% & C5\% \\
\midrule
\emph{Static (separate call)} & 50.0 & 61.0 & 0.6 & 49.5 & 2.9 \\
\midrule
Iter~0 & 45.0 & 45.2 & 1.1 & 97.6 & 74.9 \\
Iter~1 & 80.0 & 55.8 & 0.1 & 49.4 & 2.4 \\
Iter~2 & 80.0 & 85.9 & 1.0 & 34.0 & 14.6 \\
Iter~3 & 72.5 & 88.6 & 0.8 & 32.6 & 4.6 \\
Iter~4 & 70.0 & 80.6 & 0.0 & 18.0 & 2.0 \\
Iter~5 & 80.0 & 84.6 & 4.8 & 23.0 & 2.6 \\
Iter~6 & 75.0 & 81.9 & 0.0 & 34.8 & 7.0 \\
\bottomrule
\end{tabular}
\end{table}

%% file: appendix/G_Deep_Subset_Bug_Patterns_and_Specifications.tex
\section{Deep Subset: Bug Patterns and Specifications}
\label{app:deep-subset}

This appendix documents the deep evaluation subset used by the C3 axis of HierSVA-B. The subset combines two bug sources that pass the same expert review, organizes the bugs into a five-pattern taxonomy, and provides every pair with the artifacts needed to reproduce the evaluation.

\subsection{Source Mix and Review}
\label{app:g-source}

The deep subset contains 28 module-bug pairs across 27 unique modules. The real-bug source (\emph{bug2}) contributes 7 pairs whose buggy variants are the pre-fix revisions of confirmed BaseJump STL fix commits, with the upstream patch as the reference fix. The synthetic-bug source (\emph{bug1}) contributes 21 pairs whose buggy variants were generated by Claude Opus~4.6 and Claude Opus~4.7, with the intended bug isolated by a patch. The only module that appears in both sources is \texttt{bsg\_idiv\_iterative}.

Although the two sources differ in bug origin, every pair passes the same expert review. Each pair is reviewed by senior industrial DV engineers for the three criteria of Section~\ref{sec:ds-deep}: (i)~bug well-formedness (the buggy RTL compiles and differs from the correct RTL in a single semantic deviation), (ii)~patch isolation and effectiveness (the diff contains only that deviation, and reverting it restores the correct behavior), and (iii)~specification faithfulness (the specification describes the correct module without revealing the bug). Pairs that fail any of the three criteria are revised or excluded.

Each buggy variant contains exactly one bug. If several bugs were injected into the same RTL variant, a falsified assertion would show only that the assertion set detects at least one of them. It would not show which bug was detected or how many, because several falsified assertions may correspond to one bug, one assertion may be affected by several bugs, and one bug may mask another. Resolving this ambiguity would require inspecting every generated assertion by hand. Manual inspection would make the benchmark harder to reproduce and to scale. A module can still be tested against several bugs through several independently isolated bug variants, which keep every detection attributable to a single bug.

\subsection{Bug Pattern Taxonomy}
\label{app:g-taxonomy}

Bugs are classified into five patterns. The patterns are not mutually exclusive. A small number of pairs fit two patterns, and the catalog in Appendix~\ref{app:g-catalog} reports the dominant pattern for each pair. For each pattern we describe the general failure mode, then present one real-bug (bug2) example in full, with the diff, the rationale, and the form of the SVA assertion that detects it.

\subsubsection{P1 Protocol Violations}
\label{app:g-p1}

P1 covers RTL changes that break the design's contract with its environment, where the buggy RTL still compiles and elaborates without warning but the data or control protocol at the module boundary no longer matches the specification. The module-internal logic is locally consistent and local smoke checks pass; only end-to-end protocol assertions catch the bug.

\paragraph{Worked example: \texttt{bsg\_noc\_repeater\_node} (commit \texttt{ffacb789}).}
\texttt{bsg\_noc\_repeater\_node} is a two-port wormhole-NoC repeater: a flit entering side A should be buffered and emitted on side B, and vice versa. The buggy RTL wires each internal two-entry FIFO's output ports back to the same side it reads from, producing a one-cycle echo on each side rather than a cross-side relay.

\begin{Verbatim}
   bsg_two_fifo A_to_B
   ( ...
-    ,.v_o    (links_B_cast_o.v)
-    ,.data_o (links_B_cast_o.data)
-    ,.yumi_i (links_B_cast_i.ready_and_rev & links_B_cast_o.v)
+    ,.v_o    (links_A_cast_o.v)
+    ,.data_o (links_A_cast_o.data)
+    ,.yumi_i (links_A_cast_i.ready_and_rev & links_A_cast_o.v)
   );
\end{Verbatim}

The wiring error is a copy-paste swap that lints clean, elaborates without warnings, and respects all per-side flow-control invariants. The bug is invisible to any property of the form ``output A mirrors input A's stream'' or ``no flits are lost,'' because the loopback consistently reproduces the input on the same side. It is detected by an end-to-end property of the form ``every flit injected at side A eventually emerges at side B,'' which fails trivially because side B never receives any flit.

\paragraph{Sample LLM output: GLM~5.1.}
Inside the assertion module, GLM~5.1 builds an auxiliary two-entry FIFO that tracks the expected count of in-flight A$\to$B traffic, and then asserts the invariant from the specification directly on the boundary signal of the opposite side:

\begin{Verbatim}
a_ab_fifo_valid: assert property (@(posedge clk_i) disable iff (reset_i)
    links_B_cast_o.v == (aux_count_ab > 0));
\end{Verbatim}

\noindent The same shape is repeated for the B$\to$A direction (\texttt{links\_A\_cast\_o.v == (aux\_count\_ba > 0)}) and for the data fields. On the correct RTL, all eight assertions in the file are proven. On the buggy RTL, six of the eight are falsified with counterexamples of 2 to 6 cycles, because the A$\to$B FIFO output is rewired to \texttt{links\_A\_cast\_o.*} so the B-side output stays inactive while the auxiliary count is nonzero. The detection works because GLM~5.1 bases the assertion on the cross-side invariant that the specification describes rather than on per-side checks.

\subsubsection{P2 FSM Reachability Errors}
\label{app:g-p2}

P2 covers RTL changes that allow the design to enter a state that the specification declares unreachable, or that prevent it from entering a state the specification requires. These bugs typically corrupt downstream state in a way that survives across transactions and only appears under specific event interleavings.

\paragraph{Worked example: \texttt{bsg\_cache\_non\_blocking\_mhu} (commit \texttt{63ee359f}).}
\texttt{bsg\_cache\_non\_blocking\_mhu} is the miss-handling unit (MHU) of the non-blocking cache. After a clean miss completes its eviction-then-fill sequence, the new line's dirty bit must be cleared because the new line was just loaded from DMA and cannot be dirty yet. The fix commit added a new opcode \texttt{e\_stat\_set\_lru\_and\_clear\_dirty} that updates the LRU bits and clears the dirty mask atomically. At two miss-completion sites, the buggy RTL uses the older \texttt{e\_stat\_set\_lru} opcode, which updates the LRU bits but leaves the dirty mask unchanged.

\begin{Verbatim}
   stat_mem_pkt.opcode = miss_fifo_empty_i
-    ? (set_dirty_r ? e_stat_set_lru_and_dirty : e_stat_set_lru_and_clear_dirty)
+    ? (set_dirty_r ? e_stat_set_lru_and_dirty : e_stat_set_lru)
     : e_stat_read;
\end{Verbatim}

The dirty bit of the newly filled clean line therefore inherits whatever the victim line left behind. If the victim was dirty, the newly filled line keeps a stale dirty bit across subsequent transactions, and a later eviction wastefully writes the still-clean data back to DMA. The bug never produces an X or a protocol violation; it only causes redundant DMA write-backs. The detecting SVA has the form ``a miss-fill of a clean line never produces a subsequent dirty write-back without an intervening user store''; this property fails on the buggy RTL when the victim way had been dirty. We classify this bug as P2 because the valid and dirty bits of each cache line form a per-line state machine, and the bug makes the state ``freshly filled and dirty without a store'' reachable, which the specification forbids.

\paragraph{Sample LLM output: GPT-5.5.}
GPT-5.5 translates the opcode rules that the specification gives for the stat memory into three implication assertions on the MHU's \texttt{stat\_mem\_pkt}, two of which are shown below:

\begin{Verbatim}
assert_mhu_never_emits_lru_only_opcode: assert property (@(posedge clk_i) disable iff (reset_i)
    stat_mem_pkt_v_o |-> (stat_mem_pkt.opcode != e_stat_set_lru));
assert_retiring_read_clears_dirty:      assert property (@(posedge clk_i) disable iff (reset_i)
    (miss_retire_event && !set_dirty_r) |->
        (stat_mem_pkt_v_o && (stat_mem_pkt.opcode == e_stat_set_lru_and_clear_dirty)));
\end{Verbatim}

\noindent On the correct RTL all nine of GPT-5.5's assertions are proven; on the buggy RTL three are falsified, including both shown above, because the buggy MHU emits \texttt{e\_stat\_set\_lru} at the read-retire site instead of \texttt{e\_stat\_set\_lru\_and\_clear\_dirty}. The detection works because the assertion encodes ``the MHU never retires a read with the LRU-only opcode'' as a separately checkable invariant rather than as a downstream observable (DMA write count).

\subsubsection{P3 Atomicity Breakdowns}
\label{app:g-p3}

P3 covers bugs in which a multi-cycle transaction is broken into pieces that individually look correct but collectively violate the atomicity the specification implies. The buggy RTL respects the per-cycle interface but loses or corrupts data on specific multi-cycle interleavings.

\paragraph{Worked example: \texttt{bsg\_fifo\_1r1w\_small\_hardened} (commit \texttt{527b3bec}).}
This FIFO uses a hardened sync memory that disallows simultaneous read and write to the same address, plus a bypass register that catches \texttt{data\_i} on detected same-address collisions. According to the fix commit, once a collision is detected, the write into hardened memory must also be blocked, so the next dequeue must continue reading from the bypass register; the original one-cycle register on the collision flag does not cover that delayed-read path. The buggy RTL reverts three coupled changes: the \texttt{write\_mem\_en} gating is removed, the memory's \texttt{r\_v\_i} adds back an unguarded \texttt{\&\ \textasciitilde read\_write\_same\_addr\_n}, and the bypass register loses its set/clear flop in favor of a plain one-cycle register.

\begin{Verbatim}
-   wire write_mem_en = enque && (wptr_r != rptr_n);
    bsg_mem_1r1w_sync mem_1r1w_sync
      ( ...
-      ,.w_v_i (write_mem_en)
+      ,.w_v_i (enque)
-      ,.r_v_i (read_mem_en)
+      ,.r_v_i (read_mem_en & ~read_write_same_addr_n)
      );
\end{Verbatim}

The combination produces silently wrong reads only in the corner case ``enqueue and dequeue with the same head address, then wait one cycle, then dequeue again.'' The detecting SVA has the form ``every dequeued value equals the value enqueued that many slots earlier''; it fails on traces where \texttt{enque \&\ deque \&\ (rptr == wptr+1)} is followed by a second \texttt{deque} after a one-cycle pause.

\paragraph{Sample LLM output: Claude Opus~4.7.}
Claude Opus~4.7 translates the specification invariant ``while in bypass mode, \texttt{data\_o} equals the bypass register contents'' into a single implication assertion:

\begin{Verbatim}
assert_bypass_data_o: assert property (@(posedge clk_i) disable iff (reset_i)
    bypass_mode_r |-> (data_o == bypass_data_r));
\end{Verbatim}

\noindent All nine assertions of Claude Opus~4.7 are proven on the correct RTL. \texttt{assert\_bypass\_data\_o} is falsified on the buggy RTL at both parameter combinations because the buggy RTL deasserts \texttt{read\_write\_same\_addr\_r} after one cycle, after which the read path falls through to the suppressed hardened-memory read instead of the bypass register, breaking the \texttt{data\_o == bypass\_data\_r} invariant on the delayed-read path.

\subsubsection{P4 Corner-Case Errors}
\label{app:g-p4}

P4 covers bugs that only manifest at the boundary of the parameter or input space (degenerate widths, single-element parameter settings, empty inputs, or maximum-value edges). The rest of the parameter space is correct, and the bug appears only in the corner case.

\paragraph{Worked example: \texttt{bsg\_priority\_encode\_one\_hot\_out} (commit \texttt{f2b1a1a0}).}
This priority encoder has two outputs: \texttt{o} (the one-hot priority-encoded vector) and \texttt{v\_o} (a valid flag, asserted when any input bit is set). For \texttt{width\_p == 1}, both should equal the single input bit \texttt{i}. The buggy RTL collapses the \texttt{width\_p == 1} generate branch and only drives \texttt{o = i}, leaving \texttt{v\_o} undriven on this branch.

\begin{Verbatim}
   if (width_p == 1)
-    begin: w1
-     assign o   = i;
-     assign v_o = i;
-    end
+    assign o = i;
\end{Verbatim}

The encoded output \texttt{o} looks correct, but the parent that uses \texttt{v\_o} to qualify \texttt{o} sees \texttt{X}. Every setting with \texttt{width\_p > 1} goes through the unaffected \texttt{nw1} branch; only the trivial single-bit specialization exercises the buggy path. The detecting SVA has the form ``\texttt{v\_o == |i}''; with \texttt{width\_p = 1} this fails because \texttt{v\_o} is undriven.

\paragraph{Sample LLM output: GLM~5.1.}
GLM~5.1 writes both a generic invariant and a separately guarded \texttt{width\_p == 1} clause:

\begin{Verbatim}
v_o_is_or_i: assert (v_o == (|i));
generate if (width_p == 1) begin : gen_width1
    width1_v_o_eq_i: assert (v_o == i[0]);
end endgenerate
\end{Verbatim}

\noindent At the default parameter combination, where \texttt{width\_p > 1}, both assertions are proven on the correct and the buggy RTL alike, because the buggy branch is elaborated only at \texttt{width\_p = 1}. At the \texttt{width\_p = 1} parameter combination both assertions are proven on the correct RTL and falsified on the buggy RTL, because \texttt{v\_o} is undriven on the collapsed branch, so \texttt{v\_o == |i} and \texttt{v\_o == i[0]} both fail. The detection works because the parameter combinations of the module include \texttt{width\_p = 1}, so both assertions are checked on the collapsed branch.

\subsubsection{P5 Parameter-Specific Errors}
\label{app:g-p5}

P5 covers bugs that appear only at specific Type-2 (behavioral) parameter settings while the default and most common settings remain correct. Real bugs of this kind can remain in the source tree for years because regression tests exercise only the common settings.

\paragraph{Worked example: \texttt{bsg\_hashing\_ipoly} (commit \texttt{a1adf01e}).}
\texttt{bsg\_hashing\_ipoly} implements an invertible polynomial hash from \texttt{(addr\_upper, bank\_id)}, written \texttt{(a, b)} below, to a remapped bank id, designed so that the map from \texttt{b} to \texttt{new\_bank\_id} is a bijection at fixed \texttt{a}. The buggy RTL reverts three \texttt{b[i]} indices on the higher-order output bits to \texttt{b[0]}, but only on the \texttt{num\_banks\_p == 4} and \texttt{num\_banks\_p == 8} branches; the \texttt{num\_banks\_p == 16} branch stays correct.

\begin{Verbatim}
   if (num_banks_p == 8) begin
-    assign new_bank_id_o[1] = b[1] ^ a[12] ^ a[10] ^ a[9] ^ a[8] ^ ...;
-    assign new_bank_id_o[2] = b[2] ^ a[13] ^ a[10] ^ a[8] ^ a[7] ^ ...;
+    assign new_bank_id_o[1] = b[0] ^ a[12] ^ a[10] ^ a[9] ^ a[8] ^ ...;
+    assign new_bank_id_o[2] = b[0] ^ a[13] ^ a[10] ^ a[8] ^ a[7] ^ ...;
   end
\end{Verbatim}

When the higher-order output bits all use \texttt{b[0]}, the map degenerates and the overall hash collides for half of the input space. The first output bit still differs across input pairs, so a smoke check of ``output toggles when input toggles'' passes; only an injectivity check detects the bug. The detecting SVA has the form ``\texttt{(a, b1) != (a, b2) implies new\_bank\_id\_o(a, b1) != new\_bank\_id\_o(a, b2)}''; with \texttt{num\_banks\_p = 4} or \texttt{8} this fails, and a SAT solver finds a counterexample in milliseconds. With \texttt{num\_banks\_p = 16} it holds. The \texttt{num\_banks\_p = 16} branch is the one used by all production manycore configurations of BaseJump STL. This usage may explain why the bug remained in the source tree for several years.

\paragraph{Sample LLM output: GPT-5.5.}
GPT-5.5 transcribes the per-output-bit XOR fan-in of the specification into assertions for each \texttt{num\_banks\_p} value that compare each \texttt{new\_bank\_id\_o[k]} against an independently computed reference, with the same-index passthrough term \texttt{bank\_id\_i[k]}:

\begin{Verbatim}
wire expected_bit1 = bank_id_i[1]
                   ^ get_upper_bit(11) ^ get_upper_bit(9) ^ get_upper_bit(8)
                   ^ get_upper_bit(6)  ^ get_upper_bit(5) ^ get_upper_bit(3)
                   ^ get_upper_bit(2)  ^ get_upper_bit(0);
a_nb4_bit1_matches_spec: assert (new_bank_id_o[1] == expected_bit1);
\end{Verbatim}

\noindent At the \texttt{num\_banks\_p = 4} parameter combination, this assertion is proven on the correct RTL and falsified on the buggy RTL because the buggy RTL computes \texttt{new\_bank\_id\_o[1]} from \texttt{b[0]} instead of the specification's same-index \texttt{bank\_id\_i[1]}. The corresponding assertion \texttt{a\_nb8\_bit1\_matches\_spec} fails in the same way at \texttt{num\_banks\_p = 8}. The \texttt{num\_banks\_p = 16} parameter combination is unaffected because the buggy RTL matches the specification there. The detection works because the assertion encodes the same-index passthrough as a separate boolean equality rather than relying on an injectivity property over the full input space.

\subsection{Module Catalog}
\label{app:g-catalog}

Table~\ref{tab:g-catalog} lists every module-bug pair in the deep subset together with its bug source, hierarchy depth in HierSVA-DS, and dominant bug pattern.

\begin{table}[h]
\centering
\small
\caption{Per-module catalog of the deep evaluation subset. ``Source'' is the bug source (b1 = bug1, synthetic; b2 = bug2, real git history). ``D'' is the module's hierarchy depth in HierSVA-DS. ``Pat.'' is the dominant bug pattern (P1--P5, taxonomy of Appendix~\ref{app:g-taxonomy}). \texttt{bsg\_idiv\_iterative} appears in both sources and is listed twice.}
\label{tab:g-catalog}
\setlength{\tabcolsep}{4pt}
\begin{tabular}{llcc}
\toprule
Module & Source & D & Pat. \\
\midrule
\multicolumn{4}{l}{\textit{Real git-history bugs (bug2)}} \\
\midrule
bsg\_cache                          & b2 & 7 & P5 \\
bsg\_cache\_non\_blocking\_mhu      & b2 & 3 & P2 \\
bsg\_fifo\_1r1w\_small\_hardened    & b2 & 4 & P3 \\
bsg\_hashing\_ipoly                 & b2 & 0 & P5 \\
bsg\_idiv\_iterative                & b2 & 2 & P5 \\
bsg\_noc\_repeater\_node            & b2 & 3 & P1 \\
bsg\_priority\_encode\_one\_hot\_out & b2 & 1 & P4 \\
\midrule
\multicolumn{4}{l}{\textit{Synthetic bugs (bug1)}} \\
\midrule
bsg\_arb\_round\_robin\_composable   & b1 & 1 & P1 \\
bsg\_async\_credit\_counter          & b1 & 3 & P3 \\
bsg\_async\_fifo                     & b1 & 3 & P3 \\
bsg\_async\_ptr\_gray                & b1 & 2 & P5 \\
bsg\_cache\_miss                     & b1 & 3 & P2 \\
bsg\_circular\_ptr                   & b1 & 0 & P3 \\
bsg\_compare\_and\_swap              & b1 & 0 & P1 \\
bsg\_credit\_to\_token               & b1 & 1 & P1 \\
bsg\_fifo\_1r1w\_small               & b1 & 5 & P3 \\
bsg\_fifo\_reorder                   & b1 & 2 & P2 \\
bsg\_fifo\_tracker                   & b1 & 1 & P3 \\
bsg\_flow\_convert                   & b1 & 0 & P1 \\
bsg\_idiv\_iterative                 & b1 & 2 & P5 \\
bsg\_imul\_iterative                 & b1 & 0 & P5 \\
bsg\_launch\_sync\_sync              & b1 & 1 & P3 \\
bsg\_link\_ddr\_downstream           & b1 & 7 & P1 \\
bsg\_link\_sdr\_downstream           & b1 & 7 & P1 \\
bsg\_mem\_banked\_crossbar           & b1 & 6 & P1 \\
bsg\_router\_crossbar\_o\_by\_i      & b1 & 6 & P1 \\
bsg\_strobe                          & b1 & 1 & P5 \\
bsg\_tag\_client                     & b1 & 2 & P1 \\
\bottomrule
\end{tabular}
\end{table}

\subsection{Released Artifacts per Module}
\label{app:g-artifacts}

Each pair in the deep subset is released as a self-contained directory under \texttt{buggy\_rtl/} (bug1) or \texttt{buggy\_rtl2/} (bug2). The directory layout and the role of each file are uniform across both sources:

\begin{itemize}
\item \texttt{<module>.sv} (or \texttt{.v}): the buggy RTL variant.
\item \texttt{<module>.md}: the module-specific documentation file containing the source commit reference (for bug2) or the injection prompt and audit notes (for bug1), the diff against the correct RTL, the rationale (``why subtle'' and ``why it stayed in the tree''), the form of the SVA assertion that detects the bug under FPV, and a detection-difficulty estimate.
\item \texttt{spec/<module>.md}: the natural-language specification of the correct module. Specifications are released only for the deep subset.
\item Companion files in the corresponding entry of the main HierSVA-DS release: the correct RTL implementation and the \texttt{limit.json} that the formal evidence filter of Section~\ref{sec:dataset-qa} uses to align reference and LLM FTA denominators.
\end{itemize}

The five worked examples in Appendix~\ref{app:g-taxonomy} are excerpted from the corresponding \texttt{<module>.md} files in the release, so a reader can check the discussion above against those files.

%% file: appendix/I_Mutation_Operator_Breakdown.tex
\section{VC Formal Fault Classes}
\label{app:mutation}

This appendix introduces the VC~Formal FTA fault classes that underlie the C4 mutation coverage metric (Section~\ref{sec:b-c4}). FTA injects small, well-defined perturbations into the RTL and reports which of them the LLM's assertion set detects. Each perturbation belongs to a fault class that combines the kind of perturbation with where in the RTL it lives. The eight classes that appear in HierSVA-DS are summarized below.

\begin{itemize}
\item \textbf{InternalConnectivity.} Stuck-at perturbations on signals internal to the module: input-port connections of internal instantiations driven to 0, 1, or their negation. These faults model wiring mistakes that affect signals between sub-modules without changing the boundary interface, and typically only manifest under specific control-flow paths.
\item \textbf{ComboLogic.} Operator-, operand-, and bit-level perturbations (operator swaps, operand swaps, bit flips, dead operators or statements) located in combinational \texttt{always} or assign structures. These faults model errors in combinational expressions that immediately influence the result on the same cycle.
\item \textbf{TopOutputsConnectivity.} Stuck-at perturbations on the module's top-level output ports: output stuck-at-0, stuck-at-1, or negated. These faults model wiring mistakes at the module boundary itself, and have the strongest direct effect on assertions that reference the module's outputs.
\item \textbf{ComboLogicControlFlow.} Control-flow perturbations (condition forced true or false, condition negated, dead \texttt{else} or \texttt{case} item) located in combinational structures. These faults model branch- and condition-level errors that change which combinational expression drives a result without altering its arithmetic form.
\item \textbf{SynchronousControlFlow.} The same family of control-flow perturbations as ComboLogicControlFlow, but located inside a clocked \texttt{always} block. These faults perturb registered control logic and propagate across multiple cycles, which makes them more informative about verification quality than the combinational counterpart.
\item \textbf{SynchronousLogic.} Operator-, operand-, and bit-level perturbations located inside a clocked \texttt{always} block. As with SynchronousControlFlow, perturbations on registered state propagate across cycles to module outputs and are typically caught by protocol- or invariant-level assertions.
\item \textbf{SynchronousDeadAssign.} Assignment-removal faults located in a clocked block, modelling registered signals that are silently dropped from the design. These faults expose assertions that constrain registered storage paths.
\item \textbf{ResetConditionTrue.} Condition-forcing faults located inside a reset condition, which keep portions of the design held in reset on every cycle. These faults model degenerate reset behavior and are typically caught by post-reset liveness or progress assertions.
\end{itemize}

The two ``Synchronous'' families and the reset class tend to be the most informative about verification quality because faults on registered or always-active state propagate across multiple cycles to module outputs, while connectivity faults often only manifest under specific control paths and are correspondingly harder to detect with boundary-level assertions.

%% file: appendix/J_Comparison_with_Prior_LLM_for_SVA_Benchmarks.tex
\section{Comparison with Prior LLM-for-SVA Benchmarks}
\label{app:prior-comparison}

This appendix expands the compact comparison in Section~\ref{sec:related} into a full feature-by-feature table and commentary on each prior benchmark. The commentary summarizes what each work reports, and the comparison with HierSVA is our own assessment. Note that we compare with the version we cited in references.

\begin{table}[h]
\centering
\scriptsize
\setlength{\tabcolsep}{3pt}
\caption{Full feature comparison of LLM-for-SVA benchmarks. Columns: \textbf{RTL src}~=~RTL provenance; \textbf{\#Items}~=~number of designs, cases, training samples, or problems depending on benchmark format; \textbf{Hier.}~=~assertions generated and evaluated per module along a module hierarchy; \textbf{Eval}~=~evaluation methodology (Sim/h.~=~simulation/test-harness); \textbf{Cov.}~=~coverage metric (COI~=~cone of influence, FC~=~formal core, path~=~complete path coverage; \ding{55}~=~none); \textbf{Vac.}~=~vacuity rate reported as a metric; \textbf{Mut.}~=~mutation coverage reported; \textbf{Faith.}~=~spec--RTL--SVA faithfulness evaluated; \textbf{DV~rev.}~=~senior industrial DV engineer review on the reference assertions; \emph{partial}~=~covered only by part of the benchmark or only indirectly through another check.}
\label{tab:related-full}
\resizebox{\textwidth}{!}{%
\begin{tabular}{lllccccccc}
\toprule
Benchmark & RTL src & \#Items & Hier. & Eval & Cov. & Vac. & Mut. & Faith. & DV rev. \\
\midrule
AssertionBench   & OpenCores      & 100            & \ding{55} & Formal & \ding{55} & \ding{55} & \ding{55} & \ding{55} & \ding{55} \\
FVEval           & Expert+syn.    & 79+300+192     & \ding{55} & Formal & \ding{55} & \ding{55} & \ding{55} & partial & partial \\
AssertEval/OpenLLM-RTL & Open-src & 18             & \ding{55} & Formal & COI & \ding{55} & \ding{55} & partial & \ding{55} \\
AssertLLM        & Open-src       & 20             & \ding{55} & Formal & \ding{55} & \ding{55} & \ding{55} & partial & \ding{55} \\
VERT             & Synth. snippets & 20K samples   & \ding{55} & Formal+Sim & path & \ding{55} & partial & \ding{55} & \ding{55} \\
CVDP cid14       & Human-auth.     & 98             & \ding{55} & Sim/h. & \ding{55} & \ding{55} & \ding{55} & \ding{55} & \ding{51} \\
\midrule
\textbf{HierSVA (ours)} & \textbf{BaseJump STL} & \textbf{342 modules + 28 bug pairs} & \textbf{\ding{51}} & \textbf{Formal} & \textbf{FC} & \textbf{\ding{51}} & \textbf{\ding{51}} & \textbf{\ding{51}} & \textbf{deep subset only} \\
\bottomrule
\end{tabular}}
\end{table}

\subsection{Per-Benchmark Discussion}

\paragraph{AssertionBench.}
AssertionBench~\cite{assertionbench} contains 100 Verilog designs from OpenCores of up to 1{,}250 lines of code excluding comments and blank lines, paired with reference assertions produced by GoldMine and HARM~\cite{harm_miner} and verified with JasperGold. The benchmark evaluates four LLMs (GPT-3.5, GPT-4o, CodeLLaMa~2, LLaMa3-70B) under 1-shot and 5-shot ICL and reports an average pass rate bounded above by 44\%, with up to 63\% counterexample rate and up to 33\% syntactic-error rate. Four properties of this benchmark are relevant to positioning HierSVA. First, in our assessment, the reference assertions inherit the scaling characteristics of the underlying mining methods, and the authors state that most of GoldMine and its follow-up methods fail to scale to large industrial designs because of the algorithmic complexity of the underlying static analysis. Second, the evaluation only classifies each assertion as passing, failing, or producing an error, after a syntax-correction pass implemented as a separate GPT-3.5 call inserted between the LLM under test and JasperGold; in our assessment, reported pass rates are therefore not a clean measurement of an individual LLM's raw capability, since the fixed corrector sits in the loop and may change the meaning of assertions during repair. Third, the in-context example assertions are concurrent SVA built on overlapped and non-overlapped implication operators. Fourth, the benchmark does not evaluate vacuity, mutation coverage, or faithfulness as separate quality axes. HierSVA-SP replaces mining with an LLM-and-formal iterative synthesis loop applied to a hierarchical industrial-style codebase, HierSVA-DS targets broader SVA grammar coverage, and HierSVA-B evaluates the LLM end-to-end with no in-loop corrector, treating quality decomposition (well-formedness, vacuity, faithfulness, mutation coverage) as the central evaluation question.

\paragraph{FVEval.}
FVEval~\cite{fveval} contributes three sub-benchmarks for LLM-driven SVA tasks. NL2SVA-Human contains 79 expert-written natural-language-to-SVA cases. NL2SVA-Machine consists of 300 synthetic natural-language and SVA pairs. Design2SVA contains parameterized RTL with randomized sub-components drawn from two structural templates (arithmetic pipelines and finite-state machines), with 96 instances per template. The headline methodological contribution is a Cadence Jasper formal equivalence check, used for the two NL2SVA sub-benchmarks, that decides whether a generated assertion is exactly equivalent to a reference assertion, and that further reports a partial-equivalence verdict when the generated property implies or is implied by the reference. We regard this implication-aware metric as a real advance over plain pass/fail, and HierSVA-B's syntax correctness, assertion proof success rate, and vacuity axes are intended to compose with it rather than replace it. FVEval's Design2SVA benchmark is nevertheless synthetic and parameter-templated: its RTL is generated from arithmetic-pipeline and FSM templates rather than from existing industrial multi-module RTL. The authors note that although OpenTitan module instances are relevant to real-world FV use cases, they are not suitable for language-model evaluation because each module is part of a larger SoC and resolving all sub-module dependency information would require prohibitively large context. The most complex synthetic Design2SVA instances already exceed 16K tokens, leading the authors to exclude models with context windows below 32K from this sub-benchmark. HierSVA-DS is positioned at this scoped-out region, working with hierarchical industrial-style RTL from BaseJump STL and preserving sub-module dependencies in its deep subset.

\paragraph{AssertEval/OpenLLM-RTL.}
AssertEval, the SVA component of OpenLLM-RTL~\cite{asserteval}, packages 18 open-source designs covering five domains (cryptographic units, processor cores, arithmetic units, communication protocols, memory controllers). Each design ships with a natural-language specification document, a golden RTL, and a JasperGold script. The benchmark checks syntax and FPV pass or fail and computes cone-of-influence (COI) coverage. Considering the capability of existing LLM-based generation methods, the authors select specifications under 60 pages with fewer than 60 architecture-level signals. Faithfulness to the specification is checked only indirectly, because the generated assertions must pass FPV on golden RTL that follows the specification, and no buggy RTL is provided. We regard the pairing of natural-language specification, golden RTL, and FPV script as the strongest feature of this benchmark. HierSVA-DS's deep subset keeps the pairing of specification and golden RTL and extends it to hierarchical modules drawn from a single codebase. The cited version does not describe per-module, hierarchical, or parameter-sweep evaluation.

\paragraph{AssertLLM benchmark.}
AssertLLM~\cite{assertllm} introduces a benchmark suite of 20 open-source designs as part of its method paper. The method generates bit-width, connectivity, and function SVAs, and its only reported case study shows high syntactic and functional accuracy on an I2C design. AssertLLM evaluates correctness with JasperGold and argues that COI coverage does not fit SVAs generated from specifications alone, while AssertEval includes COI coverage as a metric. As in AssertEval, faithfulness to the specification is checked only indirectly through FPV on golden RTL. The width/connectivity/function grouping classifies assertions along a different dimension from HierSVA-B's evaluation axes. AssertLLM partitions assertions by the design aspect they check, while HierSVA-B partitions evaluation by the dimensions of generation quality that an SVA must satisfy to be usable in verification.

\paragraph{VERT.}
VERT~\cite{vert} releases a fine-tuning corpus of 20{,}000 synthetic code snippets paired with assertions, generated by augmenting variables from open-source HDL repositories. The authors evaluate fine-tuned models on IPs from OpenTitan, CVA6, OpenPiton, and Pulpissimo and report syntactic correctness, functional correctness, and complete path coverage. An assertion counts as functionally correct only if the inserted mutants trigger it under JasperGold formal analysis or Vivado simulation, so mutation serves as a correctness check rather than as a reported coverage metric. VERT does not evaluate hierarchy or vacuity.

\paragraph{CVDP cid14.}
CVDP~\cite{cvdp} is a 783-problem benchmark spanning 13 task categories that include RTL generation, debugging, and design verification. Its assertion-generation category, cid14, contains 68 Non-Agentic and 30 Agentic problems, for 98 problems total. CVDP evaluates LLM-generated assertions by executing a simulation/test-harness flow rather than by formal proof. As a result, cid14 measures assertions in their simulation-time checker role, where an SVA is driven by stimulus from a testbench, rather than in the FPV regime where a property is proven against all legal input behaviors by a model checker. The two settings exercise different correctness criteria and different failure modes, so we view CVDP cid14 as complementary to rather than competitive with HierSVA. For the non-agentic cid14 assertion-generation setting, CVDP reports pass@1 rates no higher than 25\% among frontier LLMs, illustrating that assertion generation remains difficult even under simulation-harness evaluation.

%% file: appendix/N_Compute_and_Cost_Accounting.tex
\section{Compute and Cost Accounting}
\label{app:compute}

\subsection{Token Cost by Hierarchy Depth (Main-Set Static Evaluation)}
\label{app:n-t1-depth}

Table~\ref{tab:n-t1-depth} reports per-model average tokens per module at each hierarchy depth in the main-set static evaluation (one response per module).

\begin{table}[H]
\centering
\small
\caption{Average total tokens per module (K) in the main-set static evaluation, stratified by hierarchy depth.}
\label{tab:n-t1-depth}
\setlength{\tabcolsep}{4pt}
\begin{tabular}{lrrrrrrrrrr}
\toprule
Model & L0 & L1 & L2 & L3 & L4 & L5 & L6 & L7 & L8 & L9 \\
\midrule
Claude Opus 4.7           & 13.7 & 21.7 & 31.0 & 37.1 & 32.0 & 37.7 & 35.9 & 63.5 & 34.1 & 89.0 \\
DeepSeek V4 Pro           & 23.7 & 32.2 & 33.9 & 40.2 & 39.6 & 43.9 & 39.9 & 50.8 & 30.7 & 88.8 \\
Gemini 3.1 Pro            & 11.0 & 16.0 & 20.6 & 25.6 & 21.6 & 28.0 & 24.6 & 47.3 & 23.7 & 64.8 \\
GLM 5.1                   & 26.7 & 33.8 & 36.3 & 50.1 & 47.4 & 41.7 & 42.4 & 52.2 & 35.1 & 55.6 \\
GPT-5.5                   & 12.6 & 16.8 & 22.2 & 27.4 & 24.1 & 29.6 & 27.7 & 44.0 & 23.5 & 63.9 \\
Kimi K2.6                 & 23.2 & 32.9 & 37.4 & 45.6 & 43.7 & 47.7 & 47.4 & 62.5 & 30.3 & 83.9 \\
MiniMax M2.7              & 8.9 & 10.7 & 18.7 & 24.4 & 21.1 & 28.1 & 22.1 & 40.2 & 21.0 & 60.1 \\
Qwen 3.6 Max              & 15.5 & 21.6 & 25.3 & 31.3 & 27.4 & 32.9 & 30.7 & 48.0 & 30.3 & 68.4 \\
\midrule
Claude Haiku 4.5          & 20.6 & 28.0 & 33.8 & 37.4 & 34.4 & 41.4 & 37.5 & 60.7 & 34.2 & 82.6 \\
DeepSeek V4 Flash         & 18.3 & 23.4 & 20.2 & 25.8 & 22.2 & 27.4 & 27.0 & 43.0 & 23.9 & 60.9 \\
GPT-5-mini                & 11.5 & 14.1 & 19.5 & 24.4 & 21.4 & 26.1 & 24.5 & 41.4 & 23.1 & 58.7 \\
Qwen 3.6 Plus             & 15.9 & 21.5 & 26.9 & 30.3 & 26.1 & 36.7 & 28.2 & 43.2 & 27.8 & 62.3 \\
\bottomrule
\end{tabular}
\end{table}

\subsection{Token Cost by Module Category (Main-Set Static Evaluation)}
\label{app:n-t1-category}

Table~\ref{tab:n-t1-category} reports per-model average tokens per module by BaseJump STL module category in the main-set static evaluation.

\begin{table}[H]
\centering
\small
\caption{Average total tokens per module (K) in the main-set static evaluation, stratified by BaseJump STL category.}
\label{tab:n-t1-category}
\setlength{\tabcolsep}{4pt}
\begin{tabular}{lrrrrrrrr}
\toprule
Model & async & cache & dataflow & link & mem & misc & noc & tag \\
\midrule
Claude Opus 4.7           & 13.9 & 47.0 & 30.1 & 28.6 & 24.0 & 15.4 & 43.1 & 26.0 \\
DeepSeek V4 Pro           & 28.7 & 48.0 & 32.7 & 35.0 & 33.1 & 23.3 & 44.9 & 38.2 \\
Gemini 3.1 Pro            & 12.0 & 32.8 & 20.4 & 20.4 & 17.0 & 12.1 & 30.8 & 16.0 \\
GLM 5.1                   & 33.6 & 47.7 & 39.0 & 37.9 & 38.5 & 23.6 & 54.9 & 50.4 \\
GPT-5.5                   & 13.7 & 32.7 & 22.8 & 22.5 & 19.9 & 13.1 & 31.6 & 21.1 \\
Kimi K2.6                 & 27.1 & 52.8 & 38.7 & 35.4 & 39.8 & 22.6 & 47.1 & 42.1 \\
MiniMax M2.7              & 8.1 & 28.4 & 18.0 & 18.4 & 13.6 & 9.6 & 28.8 & 13.6 \\
Qwen 3.6 Max              & 16.4 & 37.7 & 26.3 & 25.4 & 22.4 & 16.4 & 35.9 & 24.6 \\
\midrule
Claude Haiku 4.5          & 23.3 & 46.2 & 33.7 & 33.0 & 29.2 & 22.5 & 34.0 & 30.1 \\
DeepSeek V4 Flash         & 20.9 & 33.0 & 25.5 & 22.1 & 20.6 & 16.3 & 34.5 & 26.2 \\
GPT-5-mini                & 12.3 & 29.4 & 20.2 & 20.4 & 16.3 & 12.0 & 29.7 & 17.0 \\
Qwen 3.6 Plus             & 17.1 & 35.9 & 25.2 & 35.2 & 22.7 & 16.4 & 33.9 & 19.7 \\
\bottomrule
\end{tabular}
\end{table}

\subsection{Token Cost by Hierarchy Depth (Agentic Evaluation)}
\label{app:n-t4-depth}

Table~\ref{tab:n-t4-depth} reports per-model average tokens per module at each hierarchy depth in the agentic evaluation, where each entry is the cumulative cost across all iterations the loop ran on the module (up to 7). The agentic evaluation runs the five selected models (Claude Opus~4.7, DeepSeek~V4~Pro, Gemini~3.1~Pro, GPT-5.5, GPT-5-mini) on a fixed 40-module subset.

\begin{table}[H]
\centering
\small
\caption{Average total tokens per module (K) in the agentic evaluation, summed across iterations and stratified by hierarchy depth.}
\label{tab:n-t4-depth}
\setlength{\tabcolsep}{4pt}
\begin{tabular}{lrrrrrrrrrr}
\toprule
Model & L0 & L1 & L2 & L3 & L4 & L5 & L6 & L7 & L8 & L9 \\
\midrule
Claude Opus 4.7           & 97.6 & 196.8 & 313.7 & 468.4 & 314.9 & 243.1 & 494.8 & 487.7 & 113.6 & -- \\
DeepSeek V4 Pro           & 98.2 & 109.1 & 324.3 & 307.7 & 256.2 & 234.5 & 359.1 & 596.9 & 146.2 & -- \\
Gemini 3.1 Pro            & 80.7 & 144.4 & 305.1 & 184.3 & 181.7 & 211.6 & 349.3 & 286.7 & 81.9 & -- \\
GPT-5.5                   & 77.6 & 140.1 & 196.3 & 275.2 & 253.1 & 180.1 & 445.9 & 303.1 & 99.6 & -- \\
GPT-5-mini                & 89.0 & 151.0 & 251.9 & 229.3 & 199.5 & 142.5 & 242.7 & 181.1 & 114.8 & -- \\
\bottomrule
\end{tabular}
\end{table}

\subsection{Token Cost by Module Category (Agentic Evaluation)}
\label{app:n-t4-category}

Table~\ref{tab:n-t4-category} reports per-model average tokens per module by BaseJump STL category in the agentic evaluation, with totals summed across iterations as in Table~\ref{tab:n-t4-depth}.

\begin{table}[H]
\centering
\small
\caption{Average total tokens per module (K) in the agentic evaluation, summed across iterations and stratified by BaseJump STL category.}
\label{tab:n-t4-category}
\setlength{\tabcolsep}{4pt}
\begin{tabular}{lrrrrrrrr}
\toprule
Model & async & cache & dataflow & link & mem & misc & noc & tag \\
\midrule
Claude Opus 4.7           & 329.7 & 777.6 & 195.4 & 206.6 & 314.8 & 193.7 & 313.4 & 112.1 \\
DeepSeek V4 Pro           & 180.4 & 663.4 & 141.2 & 229.2 & 300.0 & 191.2 & 222.4 & 232.1 \\
Gemini 3.1 Pro            & 176.4 & 374.6 & 114.6 & 120.9 & 355.5 & 192.5 & 134.7 & 86.5 \\
GPT-5.5                   & 210.0 & 472.1 & 121.9 & 138.1 & 299.4 & 125.6 & 242.6 & 172.1 \\
GPT-5-mini                & 228.0 & 266.0 & 130.2 & 140.6 & 171.9 & 158.5 & 183.3 & 157.6 \\
\bottomrule
\end{tabular}
\end{table}

\subsection{Compute Accounting}
\label{app:n-compute}

\paragraph{LLM compute.} Every LLM in our evaluation is queried through a unified OpenRouter gateway. We do not host the models locally.

\paragraph{Formal verification compute.} VC~Formal runs were executed on a shared CPU cluster scheduled alongside other workloads. The shared environment introduces uncontrolled variance in wall-clock measurements across runs of the same module, and the commercial tool's licensing terms restrict detailed disclosure of per-run runtime. We therefore do not report wall-clock times.

%% file: checklist.tex
\section*{NeurIPS Paper Checklist}

\begin{enumerate}

\item {\bf Claims}
    \item[] Question: Do the main claims made in the abstract and introduction accurately reflect the paper's contributions and scope?
    \item[] Answer: \answerYes{} 
    \item[] Justification: The abstract and introduction state three contributions (HierSVA-SP, HierSVA-DS, HierSVA-B) and three findings, which are supported by the experimental results in Section~\ref{sec:eval} and Appendix~\ref{app:per-model}.

\item {\bf Limitations}
    \item[] Question: Does the paper discuss the limitations of the work performed by the authors?
    \item[] Answer: \answerYes{} 
    \item[] Justification: Section~\ref{sec:conclusion} discusses the limitations, including the manually audited eligible-fault denominator of C4 and the outdated structural-coverage instructions in the evaluation prompts. Section~\ref{sec:dataset-qa} notes that expert review is limited to the deep subset because of the labor budget. The dataset is built on a single RTL library (BaseJump STL).

\item {\bf Theory assumptions and proofs}
    \item[] Question: For each theoretical result, does the paper provide the full set of assumptions and a complete (and correct) proof?
    \item[] Answer: \answerNA{} 
    \item[] Justification: The paper does not present theoretical results or formal proofs. It is an empirical benchmark and dataset contribution.

\item {\bf Experimental result reproducibility}
\item[] Question: Does the paper fully disclose all the information needed to reproduce the main experimental results of the paper to the extent that it affects the main claims and/or conclusions of the paper (regardless of whether the code and data are provided or not)?
\item[] Answer: \answerYes{} 
\item[] Justification: Section~\ref{sec:eval-setup} specifies all models, decoding parameters (temperature 0, reasoning enabled, strict JSON-schema output), the unified OpenRouter gateway, and identical prompts across models (Appendix~\ref{app:prompts}). The dataset, pipeline, and evaluation harness are released as open source, and parameter combinations and formal tool configurations are documented.

\item {\bf Open access to data and code}
    \item[] Question: Does the paper provide open access to the data and code, with sufficient instructions to faithfully reproduce the main experimental results, as described in supplemental material?
    \item[] Answer: \answerYes{} 
    \item[] Justification: The dataset is released on Hugging Face and the code and evaluation artifacts on GitHub, with links in the abstract. Appendix~\ref{app:datasheet-distribution} describes the release under the Solderpad Hardware License v2.1. Full prompt templates are provided in Appendix~\ref{app:prompts}.

\item {\bf Experimental setting/details}
    \item[] Question: Does the paper specify all the training and test details (e.g., data splits, hyperparameters, how they were chosen, type of optimizer) necessary to understand the results?
    \item[] Answer: \answerYes{} 
    \item[] Justification: Section~\ref{sec:eval-setup} details all twelve models, the three evaluation settings (main-set static, deep-subset static, and agentic), decoding parameters, and the agentic iteration budget (an initial generation followed by up to six rounds of feedback). Prompt templates are in Appendix~\ref{app:prompts}, per-depth and per-category breakdowns in Appendix~\ref{app:per-model}, and token costs in Appendix~\ref{app:compute}.

\item {\bf Experiment statistical significance}
    \item[] Question: Does the paper report error bars suitably and correctly defined or other appropriate information about the statistical significance of the experiments?
    \item[] Answer: \answerNo{} 
    \item[] Justification: The paper does not report error bars, confidence intervals, or significance tests, because each model is queried once per module. All LLM queries use temperature 0, which does not guarantee identical outputs (Section~\ref{sec:eval-setup}). The agentic evaluation additionally reports a paired comparison on the modules that are evaluable at both iteration 0 and iteration 6 (Section~\ref{sec:eval-agentic}).

\item {\bf Experiments compute resources}
    \item[] Question: For each experiment, does the paper provide sufficient information on the computer resources (type of compute workers, memory, time of execution) needed to reproduce the experiments?
    \item[] Answer: \answerNo{} 
    \item[] Justification:  Appendix~\ref{app:compute} reports token costs per model, per depth, and per category for the main-set static evaluation and the agentic evaluation. Appendix~\ref{app:n-compute} notes that LLMs are queried through OpenRouter without local hosting and that VC Formal runs on a shared CPU cluster. Exact wall-clock times and hardware specifications are not disclosed because of commercial licensing restrictions.
    
\item {\bf Code of ethics}
    \item[] Question: Does the research conducted in the paper conform, in every respect, with the NeurIPS Code of Ethics \url{https://neurips.cc/public/EthicsGuidelines}?
    \item[] Answer: \answerYes{} 
    \item[] Justification: The research uses only open-source hardware design files (BaseJump STL), involves no human subjects or personal data, and poses no direct ethical concerns. Appendix~\ref{app:datasheet-composition} states that no confidential or sensitive data is involved.

\item {\bf Broader impacts}
    \item[] Question: Does the paper discuss both potential positive societal impacts and negative societal impacts of the work performed?
    \item[] Answer: \answerNA{} 
    \item[] Justification: The work is a benchmark and dataset for hardware formal verification. It does not introduce new generative capabilities with direct misuse potential. The societal impact is limited to improving the reliability of chip design verification workflows.
    
\item {\bf Safeguards}
    \item[] Question: Does the paper describe safeguards that have been put in place for responsible release of data or models that have a high risk for misuse (e.g., pre-trained language models, image generators, or scraped datasets)?
    \item[] Answer: \answerNA{} 
    \item[] Justification: The released assets are hardware assertion benchmarks and an evaluation pipeline built on open-source RTL. These do not pose risks for misuse comparable to pretrained language models or image generators.

\item {\bf Licenses for existing assets}
    \item[] Question: Are the creators or original owners of assets (e.g., code, data, models), used in the paper, properly credited and are the license and terms of use explicitly mentioned and properly respected?
    \item[] Answer: \answerYes{} 
    \item[] Justification: BaseJump STL is cited~\cite{basejump} and its open-source license is acknowledged. Appendix~\ref{app:datasheet-distribution} states that derived artifacts are released under the Solderpad Hardware License v2.1 for compatibility with the upstream license. All LLMs and tools (VC Formal, SymbiYosys) are cited.

\item {\bf New assets}
    \item[] Question: Are new assets introduced in the paper well documented and is the documentation provided alongside the assets?
    \item[] Answer: \answerYes{} 
    \item[] Justification: Appendix~\ref{app:datasheet} provides a datasheet following Gebru et al.~\cite{gebru2021datasheets}, covering motivation, composition, collection process, preprocessing, uses, distribution, and maintenance. The pipeline, dataset, and benchmark are documented in the released repositories.

\item {\bf Crowdsourcing and research with human subjects}
    \item[] Question: For crowdsourcing experiments and research with human subjects, does the paper include the full text of instructions given to participants and screenshots, if applicable, as well as details about compensation (if any)? 
    \item[] Answer: \answerNA{} 
    \item[] Justification: The paper does not involve crowdsourcing or human subjects research. Expert DV engineer review is part of dataset quality assurance, not a human subjects experiment.

\item {\bf Institutional review board (IRB) approvals or equivalent for research with human subjects}
    \item[] Question: Does the paper describe potential risks incurred by study participants, whether such risks were disclosed to the subjects, and whether Institutional Review Board (IRB) approvals (or an equivalent approval/review based on the requirements of your country or institution) were obtained?
    \item[] Answer: \answerNA{} 
    \item[] Justification: No human subjects research is conducted.

\item {\bf Declaration of LLM usage}
    \item[] Question: Does the paper describe the usage of LLMs if it is an important, original, or non-standard component of the core methods in this research? Note that if the LLM is used only for writing, editing, or formatting purposes and does \emph{not} impact the core methodology, scientific rigor, or originality of the research, declaration is not required.
    \item[] Answer: \answerYes{} 
    \item[] Justification: LLMs are a core component of the methodology. Section~\ref{sec:sp-loop} describes the role of the LLM in the synthesis loop. Section~\ref{sec:eval-setup} lists all twelve evaluated LLMs.

\end{enumerate}